\RequirePackage{fix-cm}
\documentclass{svjour3}                     
\usepackage{longtable}
\usepackage{graphicx}
\usepackage{latexsym}
\usepackage{amssymb}
\usepackage{lscape}
\usepackage{txfonts}
\smartqed  
\usepackage{graphicx}

\begin{document}

\title{On the origin of the faint-end of the red sequence in high density environments.
}


\author{Alessandro Boselli         \and
        Giuseppe Gavazzi 
}


\institute{Alessandro Boselli \at
              Laboratoire d'Astrophysique de Marseille - LAM, Universit\'e d'Aix-Marseille \& CNRS, UMR7326, 38 rue F. Joliot-Curie, F-13388 Marseille Cedex 13, France \\
              Tel.: +33-491056976\\
              Fax: +33-491621190\\
              \email{alessandro.boselli@lam.fr}           
           \and
           Giuseppe Gavazzi \at
              Universit\`a di Milano - Bicoocca - Piazza della scienza 3, Milano - Italy\\
	      \email{gavazzi@mib.infn.it}         
}

\date{Received: 10/08/2014 / Accepted: 27/08/2014}
\maketitle
\begin{abstract}
%
%
%
%
%
%
%
%
%
%

With the advent of the next generation
wide-field cameras it became possible to survey in an unbiased mode
galaxies spanning a variety of local densities, from the core of
rich clusters, to compact and loose groups, down to filaments and
voids. The sensitivity reached by these instruments allowed to
extend the observation to dwarf galaxies, the most ``fragile''
objects in the universe. At the
same time  models and simulations have been tailored to quantify the
different effects of the environment on the evolution of galaxies.
Simulations, models, and observations consistently indicate that
star-forming dwarf galaxies entering high-density environments for
the first time can be rapidly stripped from their interstellar
medium. The lack of gas quenches the activity of star formation,
producing on timescales of ${\sim}$1~Gyr quiescent galaxies with
spectro-photometric, chemical, structural, and kinematical
properties similar to those observed in dwarf early-type galaxies
inhabiting rich clusters and loose groups. Simulations and
observations consistently identify ram pressure stripping as the
major effect responsible for the quenching of the star-formation
activity in rich clusters. Gravitational interactions (galaxy
harassment) can also be important in groups or in clusters whenever
galaxies have been members since early epochs. The observation of
clusters at different redshifts combined with the present high
infalling rate of galaxies onto clusters indicate that the quenching
of the star-formation activity in dwarf systems and the formation of
the faint end of the red sequence is a very recent phenomenon.

\end{abstract}

\section{Introduction}\label{s1}

In 2006, the present authors reviewed the ``Environmental effects on
late-type galaxies (LTGs) in nearby clusters'' (Boselli \& Gavazzi 2006). Is
there a compelling urgency for a new review on such a short time
lag? The answer is yes, because the DR7 release of the Sloan Digital
Sky Survey (SDSS, Abazajian et al. 2009) which disclosed the complete
northern sky to photometric and spectroscopic observations to limits
as faint as 17.7 ($r$) mag, was yet to come in 2006. Beside the
SDSS, several panoramic multifrequency surveys of large stretches of
the local Universe became available after 2006 (we will review them
in Sect.~\ref{surveys}). The crucial novelty of these surveys is
that they allowed for the first time to sample clusters of galaxies
{\it at large}, embedded in surrounding regions of relatively low
galactic density, making it possible to contrast directly, not only
by statistical means, the galaxy properties in ambients of
significantly different density: cluster cores, cluster outskirts,
loose groups, filaments, and voids.

Indeed, many other significant progresses in this field were
achieved after 2006, concerning both the observations and the
simulations which brought a deeper conviction that the
environment---{\it nurture}---plays a relevant role on the evolution
of galaxies, especially at the low-mass end. It is on the new
evidences that the present review is focused. As in Boselli \& Gavazzi (2006),
Virgo, Coma, and A1367 and their surrounding superclusters, being
among the best studied clusters, will dominate the discussion, and
will be treated as representative of the local Universe.

The key motivation for environmental issues on galaxy evolution is
that, no matter if galaxies spend most of their life in relatively
low-density regions of space (the filaments of the cosmic web), a
large fraction of them is sooner or later processed through denser
(groups) and denser (clusters) environments. This picture clearly
emerges from numerical simulations in the $\Lambda {CDM}$ cosmology
(e.g. the Millennium simulation, Springel et al. 2005), where satellite
galaxies continuously feed regions of higher density (central
galaxies, groups, clusters) at the intersection of multiple
filaments. Even at $z=0$ there is substantial evidence for
high-velocity infall of ``healthy'' (star forming) galaxies into
rich clusters, which also arises from the anisotropy of the velocity
distribution of LTGs compared to early-type galaxies (ETGs) (see
Fig. 14 of Boselli \& Gavazzi 2006). Whereas ETGs obey to a
Gaussian distribution, LTGs tend to populate the wings of the
distribution at high and low velocity. Considering the time scales
for the various transformations, Boselli et al. (2008a) and
Gavazzi et al. (2013a,b) were able to estimate that infall on the
Virgo (Coma) cluster occurred at a rate of approximately 300--400
(100) galaxies with mass $M_{\rm star} \gtrsim
10^9\;\hbox{M}_{\odot}$ per Gyr in the last 2 (7.5)~Gyrs.

\begin{figure}
\includegraphics[width=1\textwidth]{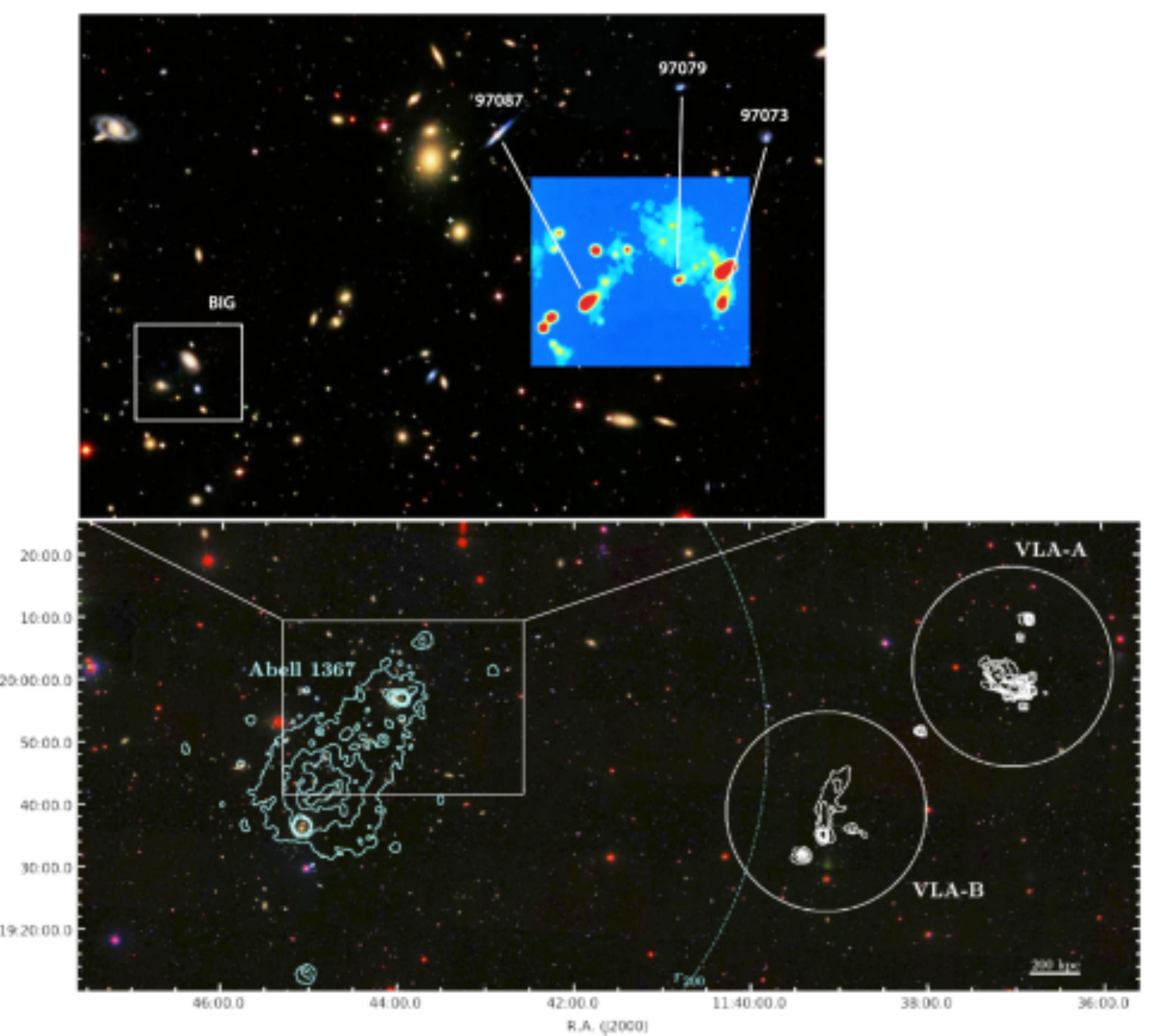}
\caption{({\it Top}) The SDSS distribution of galaxies in
the central region ($40\times 25 $~arcmin) of A1367. {\it Boxes}
highlight galaxies showing evidence of high-velocity first pass
through the cluster IGM. The {\it color inset} contains a 1.4~GHz
VLA continuum map showing extended trailing emission behind three
galaxies CGCG 97073, 97079 and 97087 (adapted from
Gavazzi et al. 1995). The {\it rectangle labeled} BIG lies close to the
X-ray cluster center. It contains a compact group of
galaxies infalling onto the cluster (Cortese et al. 2006a, b).
({\it Bottom}) The head-tail galaxy FGC1287 (VLA-B) and CGCG 97026
(VLA-A) are found just outside one virial radius of A1367 (adapted
from Scott et al. 2012). X-ray emission (ROSAT) from the hot ICM is
indicated with cyan contours. Reproduced with permission of Oxford University Press} 
\label{f1}
\end{figure}

\begin{figure}
\includegraphics[width=1\textwidth]{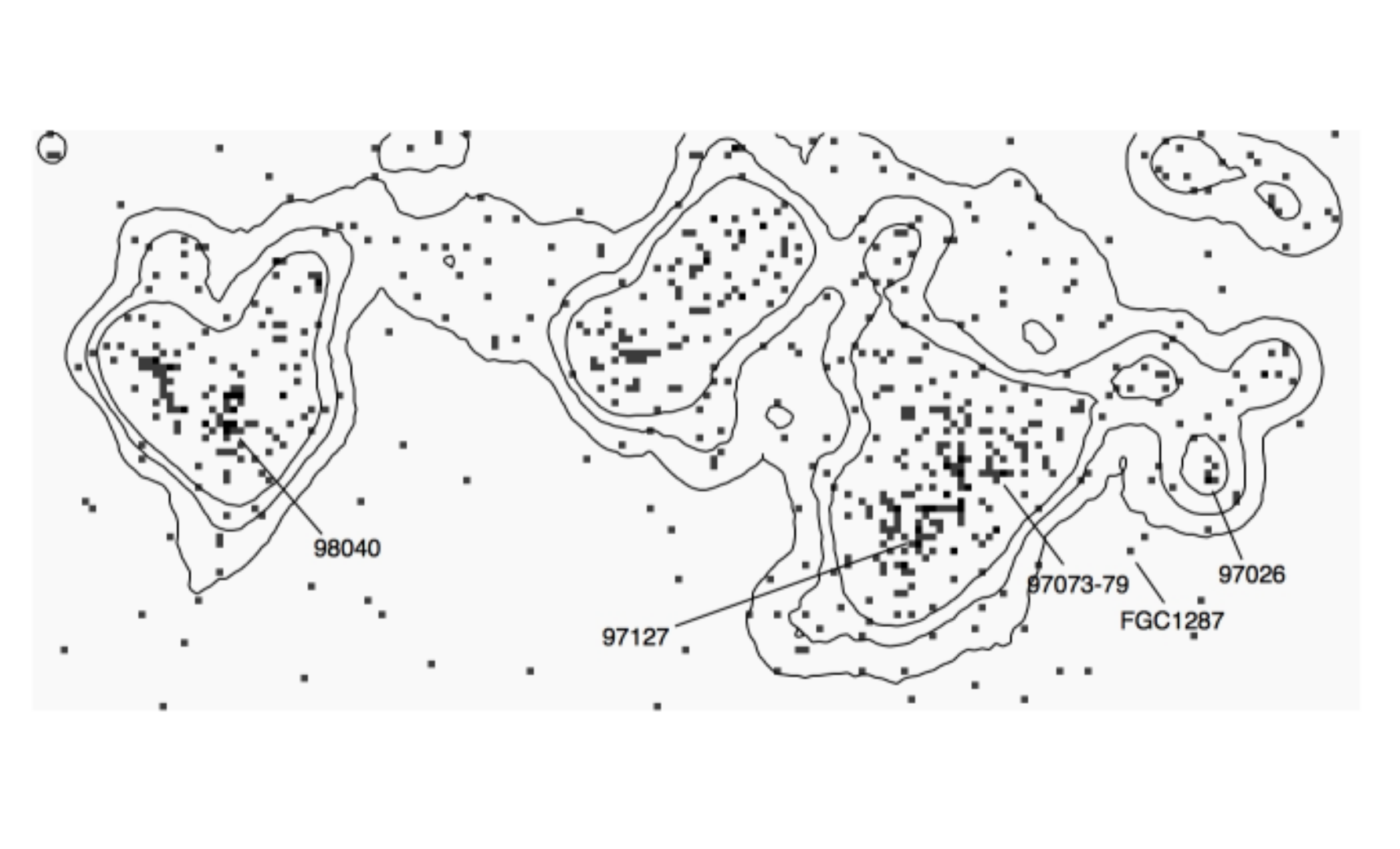}
\caption{Contours of the smoothed density distribution of
SDSS galaxies in a $9\times 3.5^\circ$ region around A1367. The
large-scale distribution of galaxies in the Coma supercluster form a
connected structure elongated in the E--W direction. Galaxies 97073
and 97079 (with radio and $\hbox{H}\alpha$ tails) and the head--tail
radio galaxy 97127 near the center of the main cluster are
highlighted. The location of the newly discovered HI head--tail
galaxies FGC 1287 and CGCG 97026 at the W periphery of the cluster
are also marked, as well as the wide-angle tail CGCG 98040 (in the
NGC 4065 group at the E of the cluster) }
\label{f3}
\end{figure}

We wish to begin and to close this review with Abell 1367, the
prototypical laboratory for the study of galaxies under the influence
of hydrodynamical processes. Figure~\ref{f1} shows the distribution
of its central galaxies from the SDSS. Also marked are two faint
$(M_{\rm star} \sim 10^{9.5}\;\hbox{M}_{\odot}$) CGCG galaxies
97073, 97079 next to 97087 that have projected distances less than
one Mpc from the cluster X-ray center (well inside the virial radius
of A1367 that was estimated  2.1 Mpc by Boselli \& Gavazzi 2006). Gavazzi \& Jaffe 1985
discovered their ``head--tail'' radio continuum emission (see color
inset from Gavazzi et al. 1995) and Gavazzi et al. (2001) detected
$\hbox{H}\alpha$ emission trailing behind them for ${\sim}$75~Mpc,
an evident sign of high-velocity interaction with the ICM. The inset
labeled ``BIG'' contains a compact group of blue galaxies infalling
at $2{,}500\,\hbox{km}\,\hbox{s}^{-1}$ into the main cluster
(Sakai et al. 2002; Gavazzi et al. 2003b; Cortese et al. 2006a,b). Next to BIG, the
elliptical galaxy NGC3862 (CGCG 97127), the brightest cluster
member, harbors the head--tail radio galaxy 3C 264 (not shown in
Fig.\,\ref{f1}).

The striking new feature of A1367 is the discovery by Scott et al. (2012)
of two ``HI head-tail'' galaxies FGC1287 and CGCG 97026 at the
cluster periphery (just outside the virial radius) reproduced in
Fig.\,\ref{f1}. This is perhaps not too surprising, as recent
hydrodynamical simulations (Tonnesen \& Bryan 2009; Bahe et al. 2013; Cen et al. 2014) claim that
ram-pressure is effective out to 2--3 times the virial radius of a
cluster. Moreover Book \& Benson (2010), confirmed by observational studies of
satellite galaxy SFR versus cluster-centric radius, suggest that
quenching of the SFR relative to the field takes place  at similarly
large clustercentric projected distances
(Balogh et al. 2000; Verdugo et al. 2008; Braglia et al. 2009). Moreover, there is evidence that
ram pressure becomes effective at lower density than previously
assumed (Bekki 2009). See, for example, Freeland et al. (2010) who studied the effects of the IGM
in the poor group NGC4065 belonging to the Coma supercluster,
visible just $4^\circ$ to the E of A1367 in Fig.\,\ref{f3}. FGC1287
($M_{\rm star} \sim 10^{9.9}\;\hbox{M}_{\odot}$) lies a little over
one virial radius away from the cluster center, i.e. approximately
one Mpc away from the hot gas in A1367, as mapped by {\it ROSAT}. We
remind, however, that the map reproduced in Fig. \ref{f1}
underestimates the real extent of the X-ray emission from A1367, as
a deeper {\it XMM}-Newton observation reveals (Finoguenov private
communication). Environmental effects (gas deficiency and
star-formation quenching) in the Virgo cluster are detected more
than one virial radius away from M87 (Gavazzi et al. 2012).

It is difficult to disentangle whether the HI-tail of FGC1287 is
triggered by the IGM associated with A1367 or with its hosting
group. Looking at the smoothed distribution of SDSS galaxies shown
in Fig.\,\ref{f3} it appears that galaxies around A1367 form a
continuum structure with contiguous groups, elongated in the  E--W
direction generally traced by the Coma Supercluster as a whole. One
group to the W contains 97026, the newly discovered HI tail, and
another group to the E (NGC4065) contains a well-known wide-angle
radio galaxy associated with 98040 (Jaffe \& Gavazzi 1986).

What is perhaps more surprising is that even evolved clusters like
Coma, as soon as their galaxies are observed with sufficiently long
exposures, reveal the presence of trailing emission (star-forming
trails or ionized gas) (Yagi et al. 2010; Yoshida et al. 2012; Fossati et al. 2012),
witnessing profound ongoing environmental transformations affecting
many dwarf and some massive LTGs.

Summarizing, adding the aforementioned cases to other well-known
examples in the Virgo cluster (e.g. Vollmer et al. 2001) there is
multiple evidence that today's clusters harbor many actively
star-forming galaxies in their first-time high-velocity pass through
the dense and hot IGM. Our aim is to show that they are being
quickly transformed into passive systems under the influence of the
dynamical pressure. We notice that most of them have stellar masses
$M_{\rm star} \lesssim 10^{10}\;\hbox{M}_{\odot}$, which is
tentatively assumed hereafter as an empirical separation between
dwarf and giant galaxies.

\begin{figure}
   \includegraphics[width=0.6\textwidth]{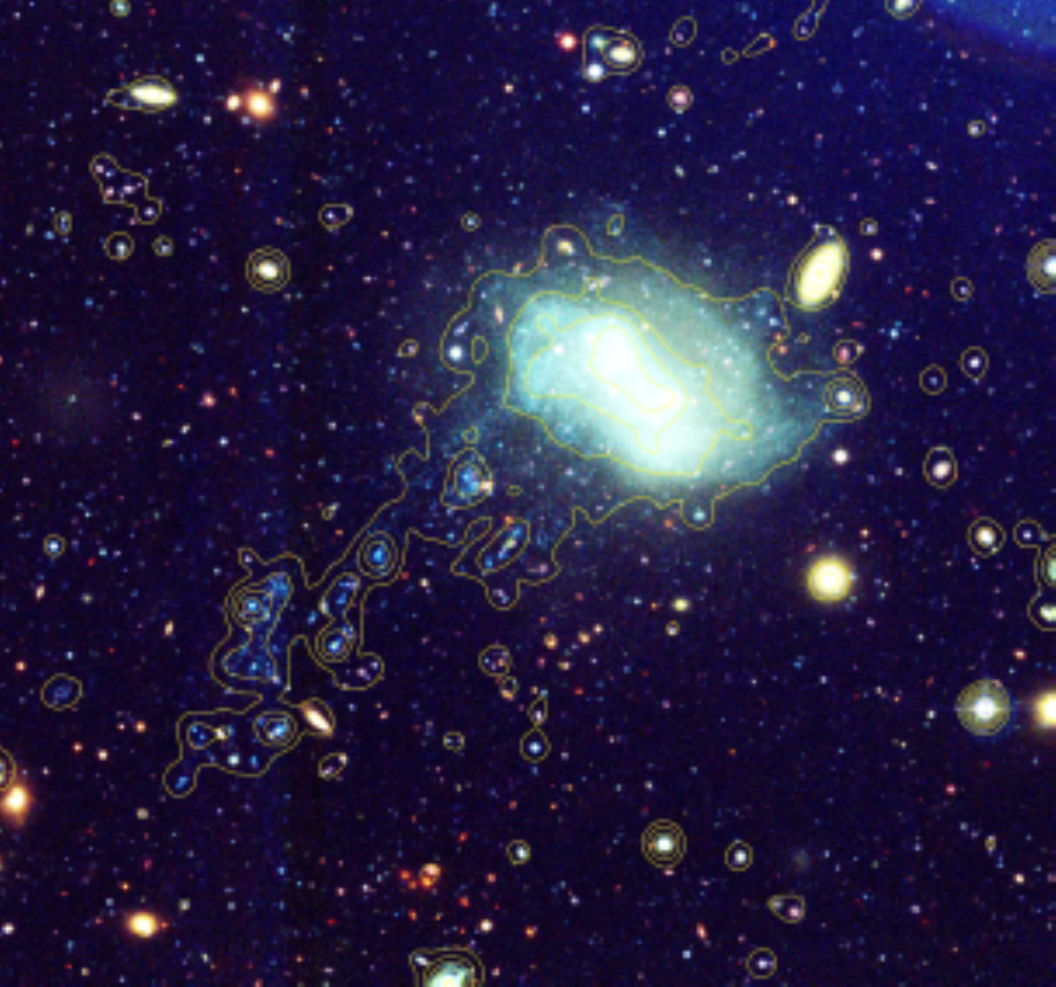}
   \includegraphics[width=0.4\textwidth]{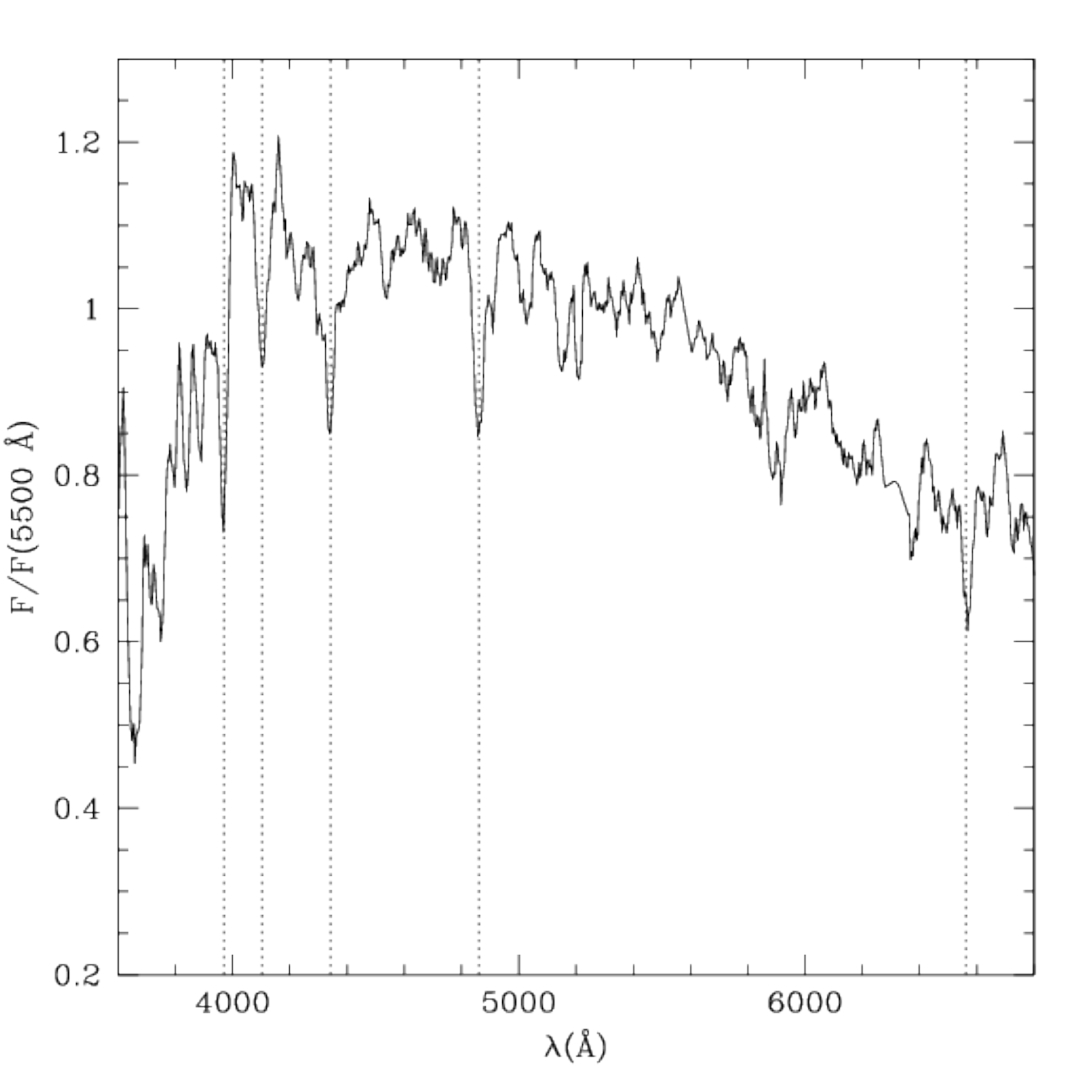}
   \caption{{\it Left panel} RGB image of VCC1217 in the Virgo
cluster obtained with $u,g,i$ NGVS images with superposed contours
obtained combining two long \textit{GALEX} NUV exposures of 16,000
and 4,500~s, respectively. The integrated spectrum of the galaxy
({\it right panel}) is typical of a post-starburst} 
\label{f1217}
\end{figure}

Dwarf galaxies, the most common objects in the universe, have a very
important role for understanding the processes that gave birth to
the local evolved stellar systems. Models of galaxy evolution
consistently indicate that they are the building blocks of massive
objects, formed by subsequent merging events. Only recently these
objects became accessible to systematic observations outside the
Local Group. Given their shallow  potential wells, these ``fragile''
stellar systems provide us with a sensitive probe of their
environment. Up to approximately 10~years ago, dwarf galaxies were
supposed to belong to two main sequences: Magellanic Irregular (Im)
and blue-compact-dwarf (BCD), actively star forming, dominating the
field on one hand, and quiescent dwarf elliptical (dE) and
spheroidal (dS0), abundant in clusters on the other. The advent of
large panchromatic and kinematic surveys allowed us to realize that
some Im found in clusters are in fact completely quiescent. An
example is VCC1217 (IC1318) (see Fig.\,\ref{f1217}), located
approximately $1.5^\circ$ south of M87 in the Virgo cluster. In this
dwarf ($M_{\rm star} \sim 10^{8.9}\;\hbox{M}_{\odot}$) irregular LSB
galaxy the star formation is absent from the disk, being probably
truncated recently, as testified by its PSB-like spectrum.
This object received recent attention due to the
presence of a long tail of star-forming blobs first revealed by
\textit{GALEX}, and trailing behind it, a clear signature of an
ongoing ram-pressure stripping event
(Hester et al. 2010; Fumagalli et al. 2011; Kenney et al. 2014).

\begin{figure}
   \includegraphics[width=0.6\textwidth]{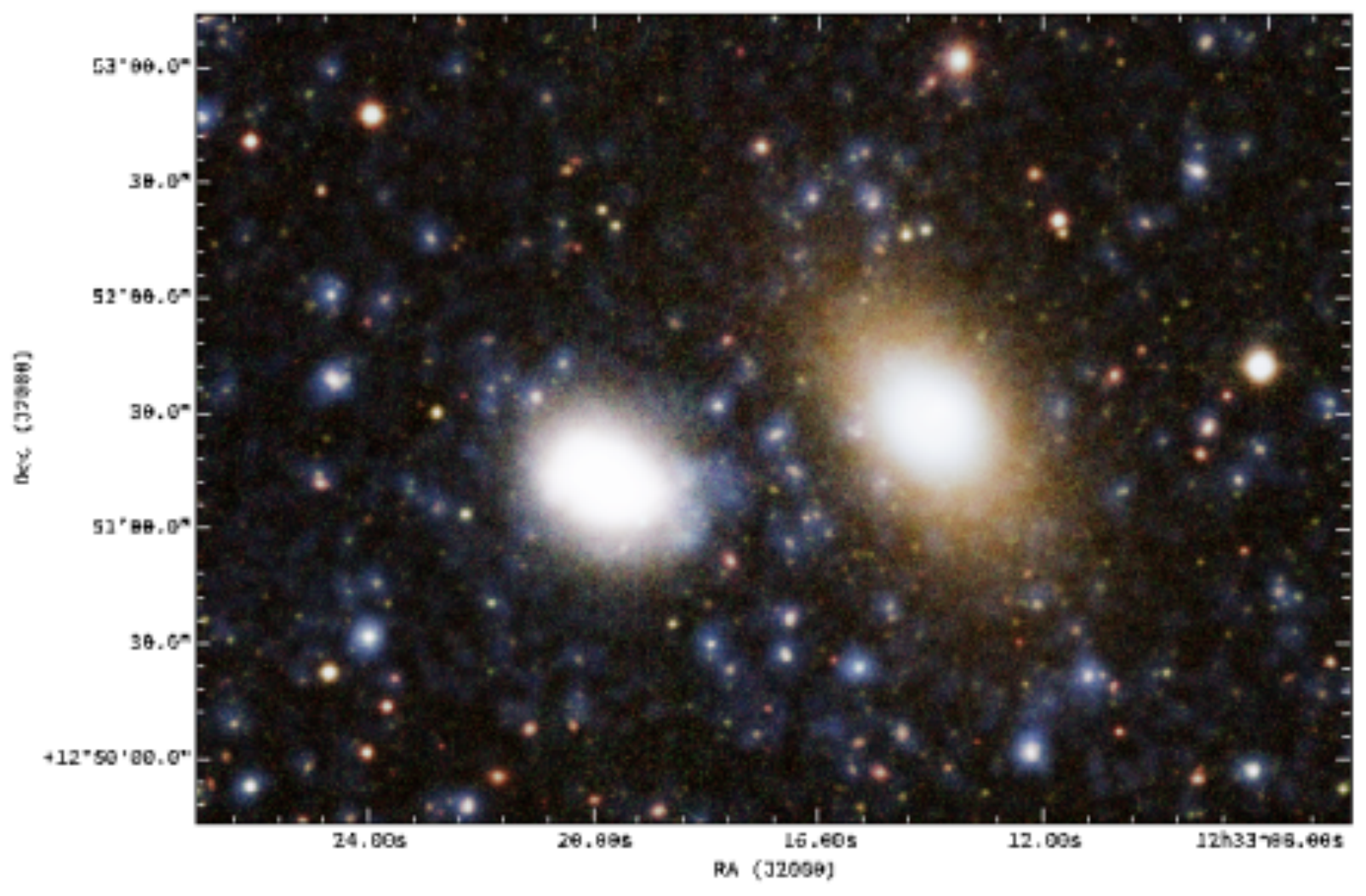}
   \includegraphics[width=0.4\textwidth]{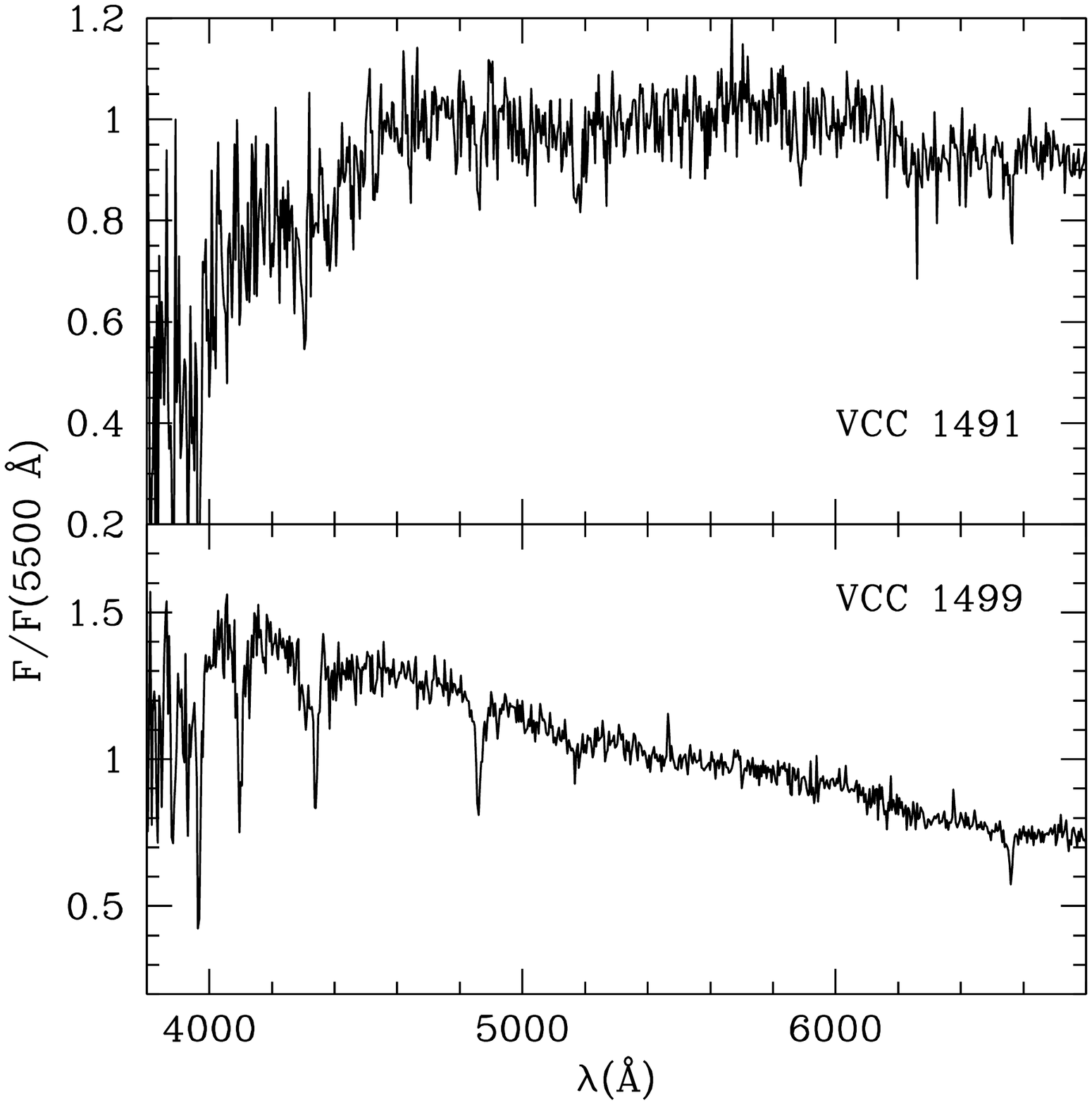}
   \caption{{\it Left panel} RGB image of VCC1491 ({\it
right-red}) and VCC1499 ({\it left-blue}) in the Virgo cluster
obtained combining UV and optical NGVS images. Their integrated
spectra ({\it right panel}) are characteristic of a red, passive
galaxy (VCC1491, {\it top}) and of a PSB (VCC1499, {\it bottom})}
\label{e1499}
\end{figure}

On the other hand, only a minority of dEs can be recognized as the
low-mass counterparts of giant elliptical galaxies. Some have a
complex morphology (e.g. pseudobulges, inner disks, spiral arms),
some rotate (Toloba et al 2009; 2011; 2012), some contain dust,
gas, and star formation in their center. Recent observations
(Boselli et al. 2008a,b; Gavazzi et al. 2010) suggest that dEs, and ultimately
the faint-end of the red sequence, can result from the recent
migration of faint star-forming galaxies through the
``green-valley''. This is sketched for the Virgo cluster on the
color--stellar mass relation shown in Fig.\,\ref{modellisam} using
the most recent spectrophotometric models for the evolution of
galaxies in rich environments. Two such cases are offered by VCC1491
and VCC1499 reproduced in Fig.\,\ref{e1499}. They are located within
$1^\circ$ projected distance from M87 in the Virgo cluster. Seen on
the exquisite B-band photographic plate taken by Binggeli et al. (1985) for
the construction of the Virgo Cluster Catalog (VCC) they look
morphologically identical and they were both classified as dEs. In
spite of their similar stellar mass [$M_{\rm star}(1491) =
10^{8.62}\; \hbox{M}_{\odot}, M_{\rm star}(1499) =
10^{8.23}\;\hbox{M}_{\odot}$], when seen on multiband CCD images
they appear dramatically different in color: $(g-i)_{1491}=1.06;
(g-i)_{1499}=0.59$, and spectroscopically (see Fig.\,\ref{e1499},
right panel). While VCC1491 has a spectrum typical of a passive
galaxy, VCC1499 has a post-star-burst (PSB) spectrum (both spectra
are integrated over the whole galaxy). Unfortunately for neither
galaxies kinematical measurements are available, but we would not be
surprised if VCC1499 was a fast rotator (Cappellari et al. 2011b), i.e. a
LTG recently converted into a ETG.

Boselli et al. (2008a) and Gavazzi et al. (2010) argued that PSB (or $k+a$) galaxies
(Poggianti et al. 2004) showing the characteristic blue continuum and
strong Balmer lines in absorption might consist of galaxies
undergoing the fast transition across the ``green valley'' due to an
abrupt truncation of the SFR by ram-pressure. As remarked by these
authors they come exclusively under the form of dwarfs (i.e. with
stellar masses $M_{\rm star} \lesssim 10^{10}\;\hbox{M}_{\odot}$) in
the outskirts of local rich clusters of galaxies. The present paper
is conceived for reviewing the works done in the past decade on
galaxy evolution in relation to the environment (we apologize for
the missing references). We hope this primarily observational review
will contribute at convincing the reader that environmental
transformations should be taken into higher consideration as drivers
of galaxy evolution. The advent of large-scale cosmological
simulations including ``gastrophysics'' will allow to constrain the
environmental effects in detail from a theoretical point of view as
well, complementing the more acknowledged stellar and AGN feedback
processes. Here we will skip a detailed discussion on the physical
processes, as they were extensively treated in Boselli \& Gavazzi (2006). We
will focus instead mainly on observations (Sects.~\ref{s2},
\ref{s3}) leaving some room for comparison with models
(Sects.~\ref{s4}, \ref{s5}) and concluding with evidences of
evolution as a function of lookback time and density.

\section{Recent blind and pointed surveys of nearby clusters}\label{surveys}\label{s2}

\subsection{Large-scale surveys}

The study of environmental effects on the evolution of galaxies took
advantage from several recent multifrequency surveys covering large
portions of the sky.\footnote{Most large-scale surveys are currently
available for the northern hemisphere. However, several large survey
of the southern sky are under way, e.g. the ESO/VST and the DES
(Dark Energy Survey) at NOAO.} Among these the one that had
certainly the major impact is the SDSS (York et al. 2000). Thanks to
its photometric and spectroscopic mode in the optical domain, the
SDSS allowed the observation of millions of galaxies in a vast
luminosity interval, located in regions spanning a wide range of
environments, from local voids to the core of the richest clusters.

Early SDSS releases have been used to study the dependence of the
structural, spectrophotometric, and star-formation properties of
galaxies as a function of galaxy density, resulting, however, in
controversial results on the role of the environment on galaxy
evolution. On the one hand Kauffmann et al. (2004) found that ``For galaxies
in the range $10^{10} - 3\times 10^{10} M_\odot$ the median specific
star-formation rate decreases by more than factor of 10 as the
population shifts from predominantly star-forming at low density to
predominantly inactive at high densities''. Conversely
Hogg et al. (2004), beside confirming the morphology-density effect, do
not find any further environmental difference once ETGs are
separated from LTGs on the basis of their Sersic index.\footnote{The
absence of a significant environmental effect in Hogg et al. (2004) is
due to a sensitivity bias: their analysis includes galaxies brighter
than $M_i={-}20$, while significant environmental issues affect
galaxies fainter than $M_i={-}19$.} Similarly, Balogh et al. (2004) do not
see a progressive reddening of galaxies as a function of local
density. In other words, they do not find evidence for objects
crossing the green valley in their way from the blue to the red
sequence.

The DR7 of the SDSS was released in 2009
(Abazajian et al. 2009) and included the complete photometry in the
northern galactic cap; the DR9 came out in 2012 (Ahn et al. 2012)
based on a new photometric pipeline; the DR10 in 2013
(Ahn et al. 2014), based on a renewed spectral pipeline. We remind
that the SDSS spectral database is complete to $r=17.77$ mag
(Strauss et al. 2002), except for ``shredding'' and fiber conflict
effects (Blanton et al. 2005a,b). To avoid large galaxies whose
photometry is uncertain because of these problems, most statistical
studies that came out from the SDSS are limited to $z>0.05$. Other
authors made a different use of the SDSS data: by directly analyzing
the images and extracting magnitudes by hand, independently of the
SDSS pipeline. With this approach Gavazzi et al. (2010, 2012, 2013a) took
advantage of the SDSS superior material to directly compare the
properties of galaxies in the centers of Coma, A1367, and Virgo with
their isolated counterparts taken in the outskirts of these
clusters.

The \textit{GALEX} mission (Martin et al. 2005) covered the entire
sky in the far (FUV, $\lambda_{\rm eff}$ 1,539~\AA) and near
ultraviolet (NUV, $\lambda_{\rm eff}$ 2,316~\AA) down to a limit of
$\simeq$21~mag with an angular resolution of $\sim$5~arcsec.
Sensitive to the emission of the youngest stars, \textit{GALEX}
provided the census of the star-formation activity in galaxies in
the local universe. The sensitivity of the instrument in targeted
observations combined with its large field of view (${\sim}
1\,\hbox{deg}^2$) allowed the detection of low surface brightness
systems such as dwarf galaxies and tidal streams in several nearby
clusters, making it an ideal instrument for the study of the effects
of the environment in the local universe
(Gil de Paz et al. 2007; Boselli et al. 2014a,b).

The gaseous component, primary feeder of the star-formation process,
was observed thanks to the Arecibo Legacy Fast Arecibo L-band Feed
Array
(ALFALFA),\footnote{http://egg.astro.cornell.edu/index.php/.}
first announced by Giovanelli et al. (2005) and completed in 2012. This survey
has mapped ${\sim}7{,}000\,{\rm deg}^2$ of the high galactic
latitude sky visible from Arecibo (i.e in the declination strip
0$^\circ$--30$^\circ$), providing a HI line spectral database covering the
redshift range between $-1{,}600\,\hbox{km}\,\hbox{s}^{-1}$ and
18,000~km~$\hbox{s}^{-1}$ with 5~km~$\hbox{s}^{-1}$ resolution at a
sensitivity of 2.3~mJy. Exploiting Arecibo's large collecting area
and relatively small beam size ($3.5'$), ALFALFA was specifically
designed to probe the faint end of the HI mass function in the local
universe down to $M({\rm HI}) \simeq 10^7\;\hbox{M}_{\odot}$
(Martin et al. 2010). Haynes et al. (2011) presented the current catalog of
21 cm H I line sources extracted from ALFALFA over
$\sim$2,800~deg$^2$ of sky: the $\alpha.40$ catalog. Covering 40\,\%
of the final survey area, the $\alpha.40$ catalog contains 15,855
sources in the spring sky, in two declination strips: 4$^\circ$--16$^\circ$ and
24$^\circ$--28$^\circ$. These include large stretches of the Local and of the
Coma superclusters.

\begin{figure}
\includegraphics[width=0.8\textwidth]{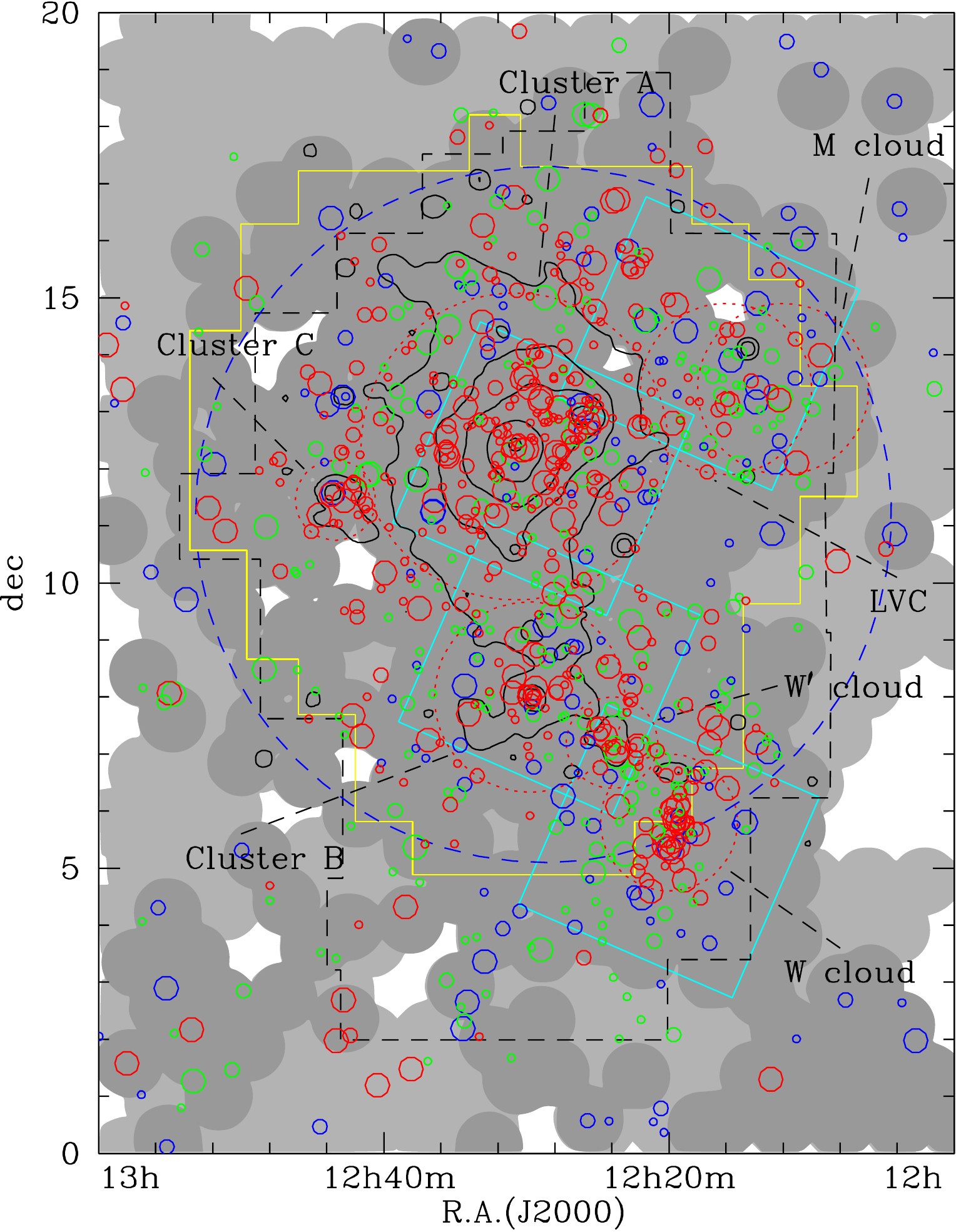}
\caption{Coverage of different surveys of the Virgo cluster
region in between $12~{\rm h} < {\rm R.A.}({\rm J}2000) < 13~{\rm
h}$ and $0^\circ < \hbox{dec} < 20^\circ$. The footprint of the
Virgo Cluster Catalogue (VCC, Binggeli et al. 1985) is indicated by the
{\it black dashed line}, the Next Generation Virgo Cluster Survey
(NGVS, Ferrarese et al. 2012) by the {\it yellow solid line}, and the
\textit{Herschel} Virgo Cluster Survey (HeViCS, Davies et al. 2010) by
the {\it cyan solid line}. The {\it light and dark gray areas}
indicate the regions observed during the \textit{GALEX} Ultraviolet
Virgo Cluster Survey (GUViCS, Boselli et al. 2011) in the NUV
(2,316~\AA) band at two different depths (adapted from
Boselli et al. 2014a). {\it Red, green, and blue empty circles}
indicate Virgo cluster galaxies belonging to the red sequence, green
valley, and blue cloud. The {\it size of the symbols} is
proportional to the galaxy stellar mass. The {\it black contours}
indicate the X-ray diffuse emission of the cluster, from
B\"ohringer et al. (1994). Courtesy of ESO}
\label{GUVICS}
\end{figure}

In the infrared domain, the all sky surveys undertaken with the
\textit{AKARI} (Murakami et al. 2007) and \textit{WISE}
(Wright et al. 2010) space missions provided photometric data for
hundreds of thousands of galaxies in the spectral range
2--180~$\mu\hbox{m}$. With a significantly better sensitivity and
angular resolution than \textit{IRAS}, \textit{AKARI} and
\textit{WISE} are crucial for quantifying the contribution of the
old stellar populations dominating the emission of galaxies for
$\lambda \lesssim 5~\mu\hbox{m}$ and that of the warm and cold dust
components at longer wavelengths. Dust emission is critical for
quantifying the attenuation of the stellar light in the UV and
optical domain and is thus crucial for a correct determination of
the present and past star-formation activity of galaxies.

\subsection{Blind panoramic surveys of clusters at $z=0$ and their host superclusters}

Because of the limited
sensitivity of the instruments used in all sky surveys, the study of
the dwarf galaxy population was mainly restricted to the very nearby
universe. Different representative regions have been the target of
dedicated studies. Among these, the most studied are certainly the
Virgo and the Coma clusters. The Virgo cluster is the largest
concentration of galaxies within 35~Mpc. Virgo is one of the closest
rich clusters, whose distance (of only 16.5~Mpc,
Gavazzi et al. 1999; Mei et al. 2007) allows us to study galaxies spanning a wide
range in morphology and luminosity, from giant spirals and
ellipticals down to dwarf irregulars, BCDs, dEs, and dS0.
Furthermore, Virgo is still in the process of being assembled so
that a wide range of processes (ram-pressure stripping, tidal
interactions, harassment, and pre-processing) are still taking
place. The Virgo cluster has been the target of several blind
multifrequency surveys covering the whole range of the
electromagnetic spectrum. The \textit{GALEX} UV Virgo Cluster Survey
(GUViCS; Boselli et al. 2011)\footnote{http://galex.lam.fr/guvics/}.
covered ${\simeq}120\,\hbox{deg}^2$ centered on the cluster in the
FUV and NUV bands providing UV data for more than 1,200,000 sources,
out of which $\sim$850 identified as cluster members
(Voyer et al. 2014; see Fig. \ref{GUVICS}). The sensitivity of the
survey, reached typically with $\sim$1,500~s (one orbit) exposures,
allowed the detection of low surface brightness features of
27.5--28~mag~$\hbox{arcsec}^{{-}2}$ such as dwarf galaxies,
including quiescent systems (Boselli et al. 2005a) and tidal features
produced during the mutual interaction of galaxies
(Boselli et al. 2005b; Arrigoni Battaia et al. 2012). The UV observations are of paramount
importance for a large number of studies. In star-forming galaxies,
the present day star-formation activity can be measured from the UV
flux emitted by the youngest stellar population
(Kennicutt 1998; Boselli et al. 2001, 2009), provided that dust extinction can be
accurately determined (e.g. using the far-IR to UV flux ratio,
Cortese et al. 2006a, 2008a; Hao et al. 2011). In quiescent galaxies, the UV
emission can help to date the last generation of stars (on a few
100~Myr timescale) or is associated to very old populations (UV
upturn; O'Connell 1999; Boselli et al. 2005a).

The Virgo cluster has been mapped in four optical bands ($u^*, g',
i', z'$) by the Next Generation Virgo cluster Survey (NGVS,
Ferrarese et al. 2012).\footnote{https://www.astrosci.ca/NGVS/The\_Next\_Generation\_Virgo\_Cluster\_Survey/Home.html}.
Carried out with the 1~deg$^2$ MegaCam instrument on the Canadian
French Hawaii Telescope, the survey covered 104~deg$^2$ of the Virgo
cluster, from the dense core out to the virial radius (see Fig.
\ref{GUVICS}). Designed to study at the same time point-like and
extended, low surface brightness sources, it has a sensitivity of
25.9~$g$ magnitudes for point-sources and a surface brightness limit
of $\mu_{g}$ 29~mag~$\hbox{arcsec}^{-2}$. The survey has
detected $\sim 3 \times 10^7$ sources, including hundreds of low
surface brightness Virgo cluster members down to absolute magnitudes
of $M_{\rm g} \sim {-}6$, and thousands of globular clusters
associated with the massive galaxies. Sensitive to the stellar
emission, the survey has been designed to study the luminosity and
mass function of cluster galaxies, as well as the color--magnitude
and the most important structural and photometric scaling relations
down to this absolute magnitude limit.

The Virgo cluster has been observed in the far infrared by
\textit{Spitzer} and \textit{Herschel}. The VIRGOFIR program (Fadda
et al., in prep.) mapped  $30\,\hbox{deg}^2$ of the Virgo cluster at
24 and $70\,\mu\hbox{m}$ with \textit{Spitzer}, while the
\textit{Herschel} Virgo Cluster Survey (HeViCS;
Davies et al. 2010, 2012)\footnote{http://wiki.arcetri.astro.it/bin/view/HeViCS/WebHome}.
$64\,{\rm deg}^2$ with PACS (100, 160~$\mu\hbox{m}$) and SPIRE
(250, 350, 500~$\mu\hbox{m}$) on board of \textit{Herschel} (see
Fig. \ref{GUVICS}). The HeViCS survey is confusion limited at
$250\;\mu\hbox{m}$ (${\sim}1\,\hbox{MJy\,sr}^{-1}$) and has an
angular resolution spanning from $6''$ at $100\,\mu\hbox{m}$ up to
$36''$ at $500\,\mu\hbox{m}$ (diffraction limited). It is thus
perfectly suited for studying the cold dust properties, one of the
most important phases of the ISM in galaxies. Far infrared data are
fundamental for accurately correcting for dust attenuation the UV
and optical emission of galaxies and are thus crucial for
quantifying and studying the present day star-formation activity,
and the past star-formation history of cluster galaxies.

At the distance of the Virgo cluster the ALFALFA survey provided HI
masses for galaxies down to $M({\rm HI}) \simeq
10^{7.5}\;\hbox{M}_{\odot}$ (Giovanelli et al. 2007; Kent et al. 2008; Haynes et al. 2008).
Slightly deeper HI data ($M({\rm HI}) \simeq
10^{7}\;\hbox{M}_{\odot}$) have been obtained by the Arecibo Galaxy
Environment Survey (AGES; Taylor et al. 2012, 2013) on the cluster
along two radial strips covering M49 $(10 \times 2\,\hbox{deg}^2$)
and east of M87 $(5\,\hbox{deg}^2)$. Finally, X-ray data relative to
the emission of the hot ICM are available thanks to \textit{ROSAT}
(B\"ohringer et al. 1994) and \textit{ASCA} (Shibata et al. 2001), including
spectroscopy from \textit{XMM}-Newton (Urban et al. 2011).

\begin{figure}
\includegraphics[width=1\textwidth]{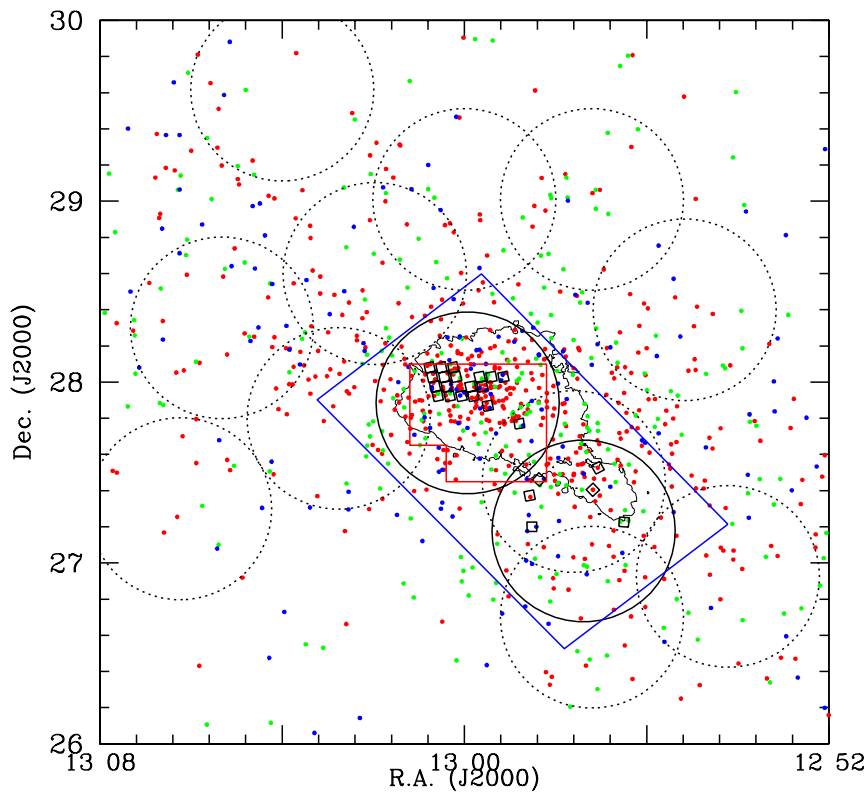}
\caption{Coverage of available surveys in the Coma cluster region. The deep
{\it GALEX} fields centered on the core of the cluster (Smith et al. 2010) and on NGC 4839 (Hammer et al. 2010b) 
are indicated with large
(1 deg diameter) {\it solid circles}, the shallower one obtained by
Cortese et al. (2008b) with {\it dotted circles}. The {\it blue
rectangle} indicates the area covered with {\it Herschel} by
Hickinbottom et al. (2014). The {\it red polynomial} region centered
on the core of the cluster shows the area covered by deep
$\hbox{H}_{\alpha}$ observations of Yagi et al. (2010). The
footprints of the ACS Coma survey (Carter et al. 2008a) are
indicated by {\it small squares}. {\it Red}, {\it green}, and {\it
blue dots} indicate SDSS galaxies belonging to the red sequence,
green valley, and blue cloud. The lowest contour of the X-ray
emission from XMM is given. The patchy black contur in the center of
the image shows the X-ray emission of the hot IGM (Briel et al. 2001)}
\label{coma}
\end{figure}

The Coma cluster has also been the target of several multifrequency
blind surveys ({see Fig.} \ref{coma}). Located at a distance of
$\sim$96~Mpc, Coma is a relaxed, spiral poor cluster characterized
by a strong X-ray emission (Sarazin 1986; Briel et al. 2001). Nine
square degree of the cluster, corresponding to ${\sim}
25\,\hbox{Mpc}^2$, have been observed in the UV with \textit{GALEX}
by Cortese et al. (2008b), and one extra field centered on the infalling
region 1.6~Mpc south-west from the cluster core with a deep
observation by Hammer et al. (2010a), and another one by Smith et al. (2010)
centered on the cluster core. The Coma cluster has been also the
target of a dedicated \textit{Hubble}/ACS blind survey designed to
cover $740\,\hbox{arcmin}^2$ but unfortunately not completed because
of the failure of the ACS camera
(Carter et al. 2008a; Hammer et al. 2010b).\footnote{http://astronomy.swin.edu.au/coma/index-mm.htm}.
The central 4~deg$^2$ of the Coma cluster have been observed in the
infrared domain by \textit{Spitzer} at 24 and 70~$\mu$m
(Bai et al. 2006). The Coma cluster is also included in the
H-ATLAS survey
(Eales et al. 2010)\footnote{http://www.h-atlas.org/}. and has
thus been observed by \textit{Herschel} at low sensitivity
($5\sigma$ sensitivity for point-like sources $\simeq$50 to 100~mJy)
in the PACS and SPIRE bands at 100, 160, 250, 350, and
$500\,\mu\hbox{m}$. The cluster has been also the target of a
deeper, dedicated observation with \textit{Herschel}/PACS at 70,
100, and 160~$\mu$m on an area of $1.75\times1.0^\circ$
encompassing the core and the southwest infalling region
(Hickinbottom et al. 2014). Spectroscopic observations of dwarf galaxies in
the Coma cluster have been also obtained by Smith et al. (2008, 2009).
Worth mentioning are also the impressive narrow band
$\hbox{H}\alpha$ imaging observations of large portions of the Coma
cluster done with the Suprime-Cam on the Subaru telescope by
Yagi et al. (2007, 2010) and Yoshida et al. (2008) and the following
spectroscopic observations of Yoshida et al. (2012). We remind that X-ray
data of the cluster are available thanks to \textit{ROSAT}
(Briel et al. 1992), \textit{XMM}
(Briel et al. 2001; Arnaud et al. 2001; Neumann et al. 2001; Finoguenov et al. 2004a), \textit{Chandra}
(Vikhlinin et al. 2001; Churazov et al. 2012; Andrade-Santos et al. 2013), and \textit{INTEGRAL}
(Renaud et al. 2006; Eckert et al. 2007).

Other recent blind surveys of very nearby clusters worth mentioning
are those of A1367 in the UV and HI bands
(Cortese et al. 2005, 2008a,b), of the Shapley supercluster in
the optical (Mercurio et al. 2006), near-infrared (Merluzzi et al. 2010),
far infrared and UV bands (Haines et al. 2011), including optical
spectroscopy (Smith et al. 2007), and X-ray
(Bonamente et al. 2001; Akimoto et al. 2003; Finoguenov et al. 2004b), and of the Fornax cluster in the
PACS and SPIRE \textit{Herschel} bands (Davies et al. 2013), and in
X-ray by \textit{XMM} (Murakami et al. 2011), while a ultra deep optical
survey (FOCUS) is under way.

\subsection{Pointed observations}

In the recent years a growing effort has been also devoted in
targeted multifrequency observations of nearby cluster and field
galaxies, including dwarfs, with the specific purpose of
understanding the physical processes at the origin of the red
sequence in high-density regions. A particular attention has been
paid in gathering high-resolution spectroscopy data necessary to
constrain the kinematical properties of the observed galaxies. The
SMAKCED (Stellar content, Mass and Kinematics of Cluster Early-type
Dwarfs) project
(Janz et al. 2012)\footnote{http://smakced.net/}. has been
designed to obtain medium resolution ($R=3{,}800$), long slit
spectroscopy and deep near infrared photometry of $\sim$100 dwarf
elliptical galaxies in the Virgo cluster. This project is a
continuation of the study of the kinematic and spectrophotometric
properties of dE started a few years before within the MAGPOP
collaboration by Toloba and collaborators
(Toloba et al. 2009, 2011, 2012). The $\hbox{ATLAS}^{\rm 3D}$ survey
(Cappellari et al. 2011a),\footnote{http://www-astro.physics.ox.ac.uk/atlas3d/}.
although limited to relatively massive objects ($M_{\rm star}
\gtrsim 6 \times 10^9\;\hbox{M}_{\odot}$), was designed to observe
with SAURON on the William Herschel Telescope a volume limited
sample ($D < 42$~Mpc) of 260 ETGs in the local universe using
kinematic and spectrophotometric data at different frequencies. The
$\hbox{ATLAS}^{\rm 3D}$ survey, which was originally defined to
study 2D intermediate resolution spectroscopy data, extended in data
quality and in statistics the SAURON project (Bacon et al. 2001).
Another recent survey of nearby galaxies based on 2D-integral field
spectroscopy at intermediate/low resolution ($R$ 850 and 1,650 in
the spectral range 3,700--7,000~\AA) data is CALIFA (Calar Alto
Legacy Integral Field Area survey;
Sanchez et al. 2012),\footnote{http://califa.caha.es/}. a project
designed to observe with PPAK at the 3.5-m telescope of Calar Alto
and study $\sim$600 galaxies in the local universe $(0.005< z <
0.03)$. Although mainly limited to massive galaxies, the sample
includes galaxies spanning a wide range in morphological type and
environment, including Coma and A1367, and is thus perfectly suited
to study any possible physical process able to transform star
forming into quiescent systems in high-density regions.

High-quality optical images of dwarf elliptical galaxies in the
Virgo cluster have been obtained with the ACS camera on the
\textit{HST} during the ACS Virgo cluster survey (ACSVCS;
Cote et al. 2004).\footnote{https://www.astrosci.ca/users/VCSFCS/Home.html}.
The survey covered 100 ETGs, out of which 35~dE, dEN or dS0, in
the F475W and F850LP bandpasses, which roughly correspond to the SDSS $g$
and $z$ bands. The same team has also undertaken the ACS Fornax
cluster survey (ACSFCS; Jordan et al. 2007a,b), a similar survey
of 43 ETGs in the Fornax cluster. In the optical domain it is worth mentioning $\hbox{H}\alpha 3$, a
narrow-band $\hbox{H}\alpha$ imaging survey of dwarf, HI-detected
star forming galaxies located in the surrounding regions of the
Virgo and Coma/A1367 clusters (Gavazzi et al. 2012, 2013a,b). This
survey extended previous observations of the clusters (Gavazzi et al. 1998, 2002, 2006a; Koopmann et al. 2001; 
Boselli et al. 2002; Boselli \& Gavazzi 2002) to objects in low density regions at
similar distances. Concerning the
gaseous component we should mention the VLA survey of Virgo galaxies
(VIVA) of Chung et al. (2009) that, although targeting bright spirals,
allowed the detection of extraplanar HI gas stripped during the
interaction with the hot ICM (Chung et al. 2007) where local episodes
of star formation can give birth to new dwarf systems (see
Sect.~\ref{INDIVIDUAL}). We must also mention the \textit{Herschel}
Reference Survey (HRS;
Boselli et al. 2010),\footnote{http://hedam.lam.fr/HRS/} a
project aimed at studying among other scientific topics, the effects
of the cluster environment on the properties of the ISM in galaxies.
Multifrequency data covering the whole electromagnetic spectrum,
including new \textit{Herschel} data, have been collected in the
literature or thanks to dedicated observations. The sample is
ideally designed for this purpose since it is a complete
volume-limited $(15 < {\it Dist} < 25 \hbox{ Mpc})$, K-band-selected
sample of galaxies spanning a wide range in morphological type,
including at the same time field objects and Virgo cluster galaxies
selected using consistent criteria. As selected, the sample includes
spiral galaxies down to $M_{\rm star} \simeq 3 \times
10^8\;\hbox{M}_{\odot}$ and thus covers the relatively bright dwarf
systems.

\section{Main observational results}\label{s3}

\subsection{Luminosity functions}

The availability of large-scale multifrequency surveys such as SDSS
and \textit{GALEX} allowed the determination of the luminosity
function in different photometric bands and the study of the
variation of their characteristic parameters as a function of galaxy
density (e.g. Blanton \& Moustakas 2009). With respect to previous works for
the major part based on the observations of the central regions of a
few well-known clusters, these surveys have the advantage of
extending the study to the periphery of the cluster, not always
mapped by pointed observations. These regions are of great
importance since they define the prevailing conditions first
encountered by galaxies infalling in high-density regions (e.g.
Boselli \& Gavazzi 2006). Owing to the unprecedented statistical significance
of datasets extracted from the SDSS, the role of the environments on
the evolution of galaxies belonging to regions of different density,
from the field to loose and compact groups up to massive clusters,
has been significantly clarified.

\begin{figure}
\includegraphics[width=0.7\textwidth]{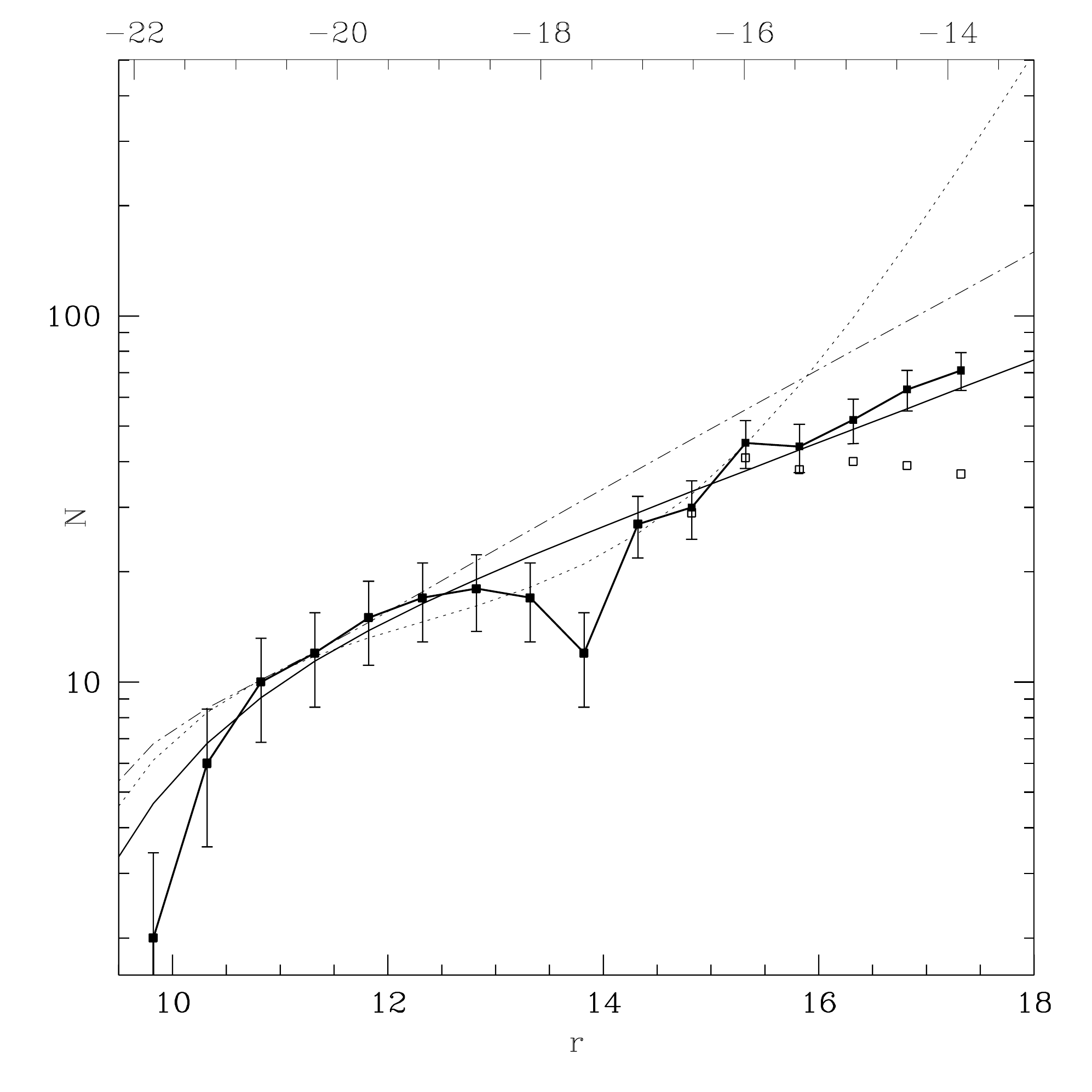}
\caption{The $r$ band luminosity function of the Virgo
cluster within 1~Mpc from M87, from Rines \& Geller (2008) ({\it filled
squares}) ({\it upper} ({\it lower}) scale $r$ absolute (apparent)
magnitude). The fitted Schechter function ({\it black solid line})
is compared to the field luminosity function from the SDSS of
Blanton et al. (2005b) ({\it dotted--dashed line}) and to the composite
luminosity function of clusters of Popesso et al. (2006) ({\it dotted
line}). The optical luminosity function of Virgo has a slope
comparable to that of the field once low surface brightness objects
are considered. \copyright\ AAS. Reproduced with permission}
\label{LFottica}
\end{figure}

By combining SDSS data of 130 X-ray selected clusters at $z \simeq
0.15$ (Popesso et al. 2005) determined and studied the properties of
the composite luminosity functions of nearby clusters. This work has
shown that the luminosity function has a bimodal behavior, with an
upturn and an evident steepening in the faint magnitude range in any
SDSS band. Both the bright and the faint end can be fitted with a
Schechter function, the former with a slope $\alpha \simeq {-}1.25$,
the latter with ${-}2.1 \leq \alpha \leq {-}1.6$ (see
Fig.~\ref{LFottica}). The same authors were also able to separate
the contribution of early- and late-type galaxies using color
indices (Popesso et al. 2006). They have shown that while the
luminosity function of late-type systems can be fitted with a single
Schechter function of slope $\alpha = {-}2.0$ in the $r$-band, the
early-types require two Schechter functions, the faint one being
responsible for the faint-end upturn observed in the global
composite luminosity function of the cluster. The shape of the
bright-end tail of the early-type luminosity function does not
depend on the local galaxy density, while the faint-end shows a
significant and continuous variation with the environment, with a
clear flattening near the core of the cluster. A steepening of the
faint-end of the luminosity function at the periphery of nearby
clusters and a flattening in the core has been also observed by
Barkhouse et al. (2007, 2009) using data on 57 low-redshift Abell clusters
observed with the KPNO 0.9~m telescope. All these teams interpreted
these results as an evidence of a combined effect of galaxy
transformation from star forming to quiescent systems through
harassment in the periphery and dwarf tidal disruption in the core
(Popesso et al. 2006; Barkhouse et al. 2007; de Filippis et al. 2011). The presence of an upturn at faint
magnitudes in the luminosity function, however, was questioned by
de Filippis et al. (2011), while a significantly flatter faint end slope
($\alpha \sim {-}1.1$) has been found in the Hydra I and the
Centaurus clusters down to $M_V \sim {-}10$ by Misgeld et al. (2008, 2009).
Hansen et al. (2009) have shown that the shape of the luminosity function
of satellites does not change with clustercentric distance. They
have also shown that the luminosity function of both red and blue
satellites is only weakly dependent on richness. Their ratio,
however, dramatically changes. The average color of satellites is
redder near the cluster centers.

The properties of the SDSS luminosity function of cluster galaxies
was extended to groups by Zandivarez et al. (2006), Zandivarez \& Martinez (2011) and
Robotham et al. (2010). Zandivarez et al. (2006) and Zandivarez \& Martinez (2011) have shown a steepening
of the faint end slope and a brightening of the characteristic
magnitude as the mass of the system increases, while Robotham et al. (2010)
concluded that the steepening at the faint end is mainly due to an
increase of the quiescent galaxy population.

Studies of very nearby clusters such as Virgo have the advantage of
both detecting the dwarf galaxy population down to much fainter
limits and determine cluster memberships according to different and
independent criteria such as colors, morphology, surface brightness,
unpractical at larger distances. Recently, Lieder et al. (2012) determined
the $V$ and $I$ luminosity function of the central $4\,{\rm deg}^2$
of the Virgo cluster, deriving a faint end slope $\alpha = {-}1.50$.
Using new spectra of galaxies in the direction of the Virgo cluster,
Rines \& Geller (2008) have shown that the faint-end of the $r$-band luminosity
function has a slope consistent with that of the field
$(\alpha={-}1.28)$ down to $M_r \simeq {-}13.5 \simeq M^* + 8$, thus
significantly flatter than that generally determined for clusters
using SDSS data (e.g. Popesso et al. 2005, 2006; see Fig.
\ref{LFottica}). The analysis of Rines \& Geller (2008) has indicated that the
difference is primarily due to the use of a statistical subtraction
for the correction of the background contamination. They have shown
that a simple separation in apparent magnitude versus surface
brightness, originally proposed by Sandage and collaborators
(Sandage et al. 1985; Binggeli et al. 1985) and later adopted by Boselli et al. (2011) in the
UV bands, provides a powerful membership classification. They thus
opened new ways for identifying cluster members, allowing the
determination of the optical luminosity function in five photometric
bands down to $M_g \simeq {-}6 (M^* + 15)$ on a region centered on
the Virgo cluster as large as $104\,{\rm deg}^2$ using the NGVS data
(Ferrarese et al. 2012). The preliminary results obtained using the NGVS
data of the central $4\,\hbox{deg}^2$ of Virgo seem to indicate that
the optical luminosity function of Virgo has a slope of $\alpha
\simeq {-}1.4$ down to this absolute magnitude limit (Ferrarese,
private communication). We recall that this value, which perfectly
matches that obtained in the eighties by Sandage and collaborators
(1985) using photographic plate material, is very
similar to the one obtained for the field using the SDSS once low
surface brightness objects as those detected in Virgo are considered
(Blanton et al. 2005b).

\begin{figure}
\includegraphics[width=1.0\textwidth]{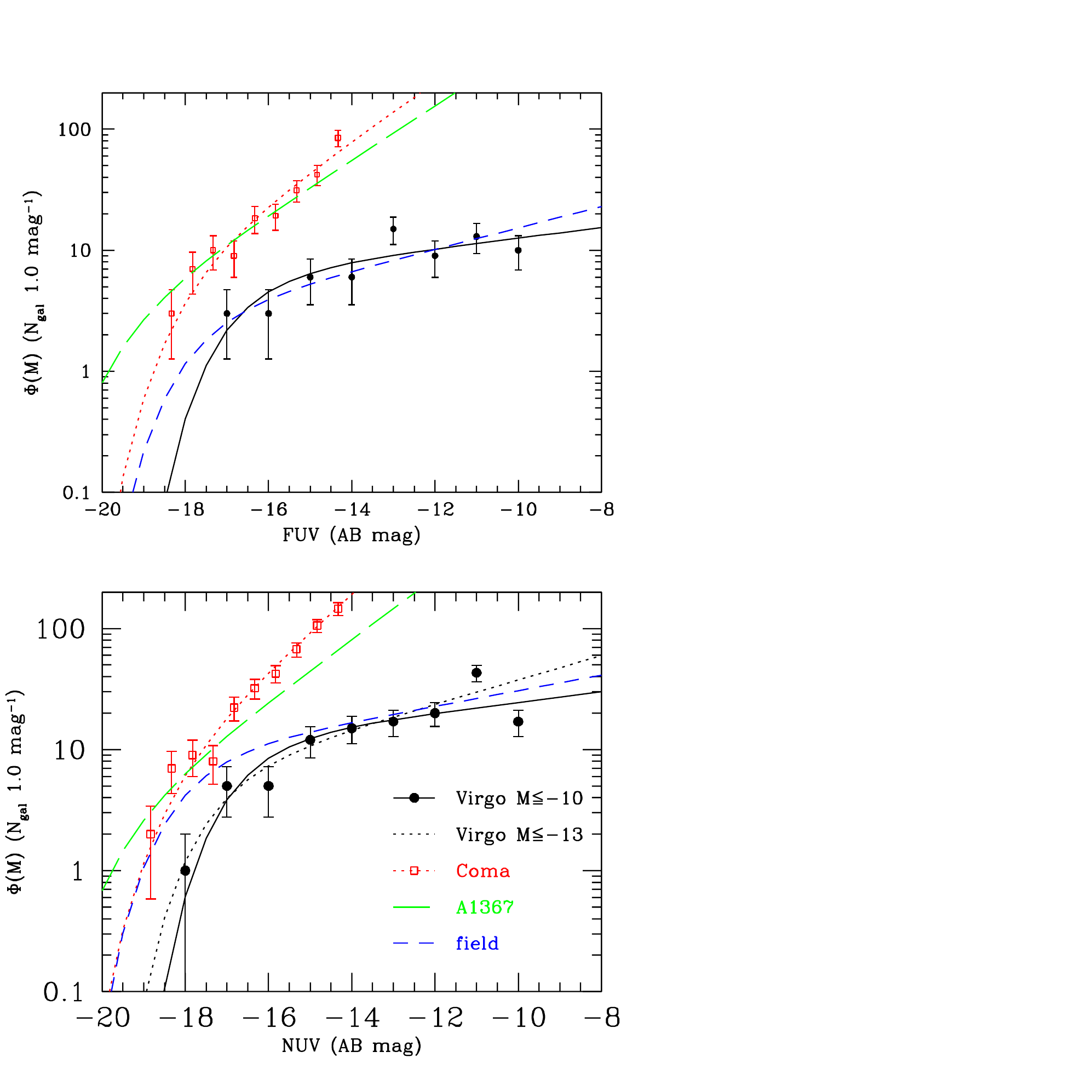}
\caption{The FUV ({\it upper panel}) and NUV ({\it lower
panel}) luminosity functions of the Virgo cluster ({\it black
symbols and solid line}) compared to that of the Coma cluster ({\it
red}), A1367 ({\it green}), and the field ({\it blue}), from
Boselli et al. (2011). The UV luminosity function of the Virgo cluster is
similar to that of the field. Courtesy of ESO}
\label{LFUV}\end{figure}

Pointed observations with the \textit{GALEX} satellite allowed the
determination of the UV luminosity function of well-known nearby
clusters such as Coma (Cortese et al. 2008b; Hammer et al. 2012), A1367
(Cortese et al. 2005), the Shapley supercluster (Haines et al. 2011), and
the Virgo cluster (Boselli et al. 2011) in two photometric bands (FUV,
$\lambda_{\rm eff} = 1{,}539$~\AA; NUV, $\lambda_{\rm eff} =
2{,}316$~\AA) (see Fig. \ref{LFUV}). Steep slopes of the fitted
Schechter functions ($\alpha \simeq {-}1.5/{-}1.6$) both in the NUV
and FUV bands have been observed in Coma (Cortese et al. 2008b), A1367
(Cortese et al. 2005), and in the Shapley supercluster
(Heines et al. 2011) down to UV absolute magnitudes of
${\simeq}{-}14$. These values of $\alpha$ are significantly steeper
than those observed in the field by 
(Wyder et al. 2005; $\alpha \simeq {-}1.2$). By dividing galaxies
according to their morphological type, these authors have also shown
that the steepening of the observed UV luminosity function is due to
the contribution of ETGs, becoming important at faint luminosities.
We recall that in these quiescent objects the UV emission is not
only related to star-forming events as in LTGs  (e.g.
Boselli et al. 2009), but rather to evolved stellar populations (UV
upturn; e.g. O'Connell 1999; Boselli et al. 2005a). Slightly flatter slopes
$(\alpha = {-}1.39)$ have been observed in the infalling region
centered on NGC 4839 in Coma by Hammer et al. (2012) using a deep
\textit{GALEX} field including galaxies down to UV magnitudes of
$-$10.5.

The GUViCS survey allowed the determination of the UV luminosity
function down to UV magnitudes of $-$10 ($M_{UV}^*+ 7$) in the
central $12\,{\rm deg}^2$ of the Virgo cluster (Boselli et al. 2011,
Fig. \ref{LFUV}). The faint-end slope of the fitted Schechter
function determined in this work ($\alpha \simeq {-}1.1/{-}1.2$) is
significantly flatter than the one observed in other clusters and
very close to the one determined for the field. Boselli et al. (2011)
interpreted the difference between the faint-end slope of the
luminosity function of Virgo and other nearby clusters as due to two
possible effects. The first is related to the small volume sampled
by their study and thus to the small number of luminous objects used
to constrain the shape of their luminosity function. They have also
shown, however, that because of the higher distance of the other
clusters, the determination of their luminosity functions suffers
from significantly narrower dynamic range in luminosity (down to
$-$14 vs. $-$10 in Virgo) and by the quite uncertain statistical
corrections for the background contamination. This conclusion is
consistent with the results of Rines \& Geller (2008) obtained in the optical
bands. The mild steepening of the faint-end of the luminosity
function is primarily due to the population of evolved galaxies
(E-S0-dE; Boselli et al. 2011). Given the similarity of the field and
cluster luminosity functions, Boselli et al. (2011) concluded that their
results are consistent with a transformation of star-forming dwarf
galaxies in quiescent systems due to a ram pressure stripping event
removing their gas content once galaxies fall into the cluster.

\begin{table*}[t]
\caption{The faint-end slope of the luminosity function.}
\centering
\begin{tabular}{c l l l}
\hline\hline
  Band  &	       &	Reference      &	$\alpha$	 \\
\hline 
OPT     & 130 clusters     & Popesso et al. (2005)        &    $-$2.0   \\
OPT     & Hydra, Centaurus & Misgeld et al. (2008, 2009)  &    $-$1.1   \\
OPT     & Virgo            & Lieder et al. (2012)         &    $-$1.5   \\
OPT     & Virgo            & Rines \& Geller (2008)       &    $-$1.28   \\
OPT     & Virgo            & Ferrarese et al. (2012)      &    $-$1.4   \\
UV      & Coma/A1367       & Cortese et al. (2005)        &    $-$1.5/$-$1.6   \\
UV      & Shapley          & Haines et al. (2011)         &    $-$1.5  \\
UV      & field            & Wyder et al. (2005)          &    $-$1.2  \\
UV      & Coma (N4839)     & Hammer et al. (2012)         &    $-$1.39  \\
UV      & Virgo            & Boselli et al. (2011)        &    $-$1.1/$-$1.2  \\
FIR     & Coma             & Bai et al. (2006, 2009)      &    $-$1.4/$-$1.5 \\
FIR     & Shapley          & Haines et al. (2011)         &    $-$1.4/$-$1.5 \\
\hline
\end{tabular}
\label{tab1}
\end{table*}

The study of the luminosity function of cluster galaxies has been
extended to the far-infrared domain thanks to \textit{Spitzer} and
\textit{Herschel}. Bai et al. (2006, 2009), using 24, 70, and
160~$\mu{\rm m}$ MIPS data determined the luminosity function of
the two clusters Coma and A3266 $(z=0.06)$ down to $L_{\rm IR} \sim
10^{42}\,\hbox{erg\,s}^{-1}$ and $L_{\rm IR} \sim
10^{43}\,\hbox{erg\,s}^{-1}$, respectively. These works have shown
that the far-infrared luminosity function of these cluster galaxies
$(\alpha={-}1.4/{-}1.5$ and $L^* \sim\,10.5 \hbox{L}_{\odot}$) is
comparable to that observed in the field. They have also shown that
both $L^*$ and $\alpha$ fade close to the cluster core. Similar
results have been obtained in the 24 and $70\,\mu\hbox{m}$ bands by
Haines et al. (2011) in the Shapley supercluster. Haines et al. (2011) interpreted
these results as an evidence that the LTG population, responsible
for the far-infrared emission, has been only recently accreted in
clusters. More recently Davies et al. (2012) determined the first
far-infrared luminosity distribution in the \textit{Herschel} PACS
(100--160\,$\mu\hbox{m}$) and SPIRE (250--350--500\,$\mu\hbox{m}$)
bands for the central $64\,{\rm deg}^2$ of the Virgo cluster. They
have shown that optically selected galaxies have a far-infrared
luminosity distribution peaked at intermediate luminosities, showing
a lack of both bright and low-luminosity systems. The faint-end
slope of the various luminosity functions is summarized in Table
\ref{tab1}.

\subsection{Gas content}

\begin{figure}
\includegraphics[width=1\textwidth]{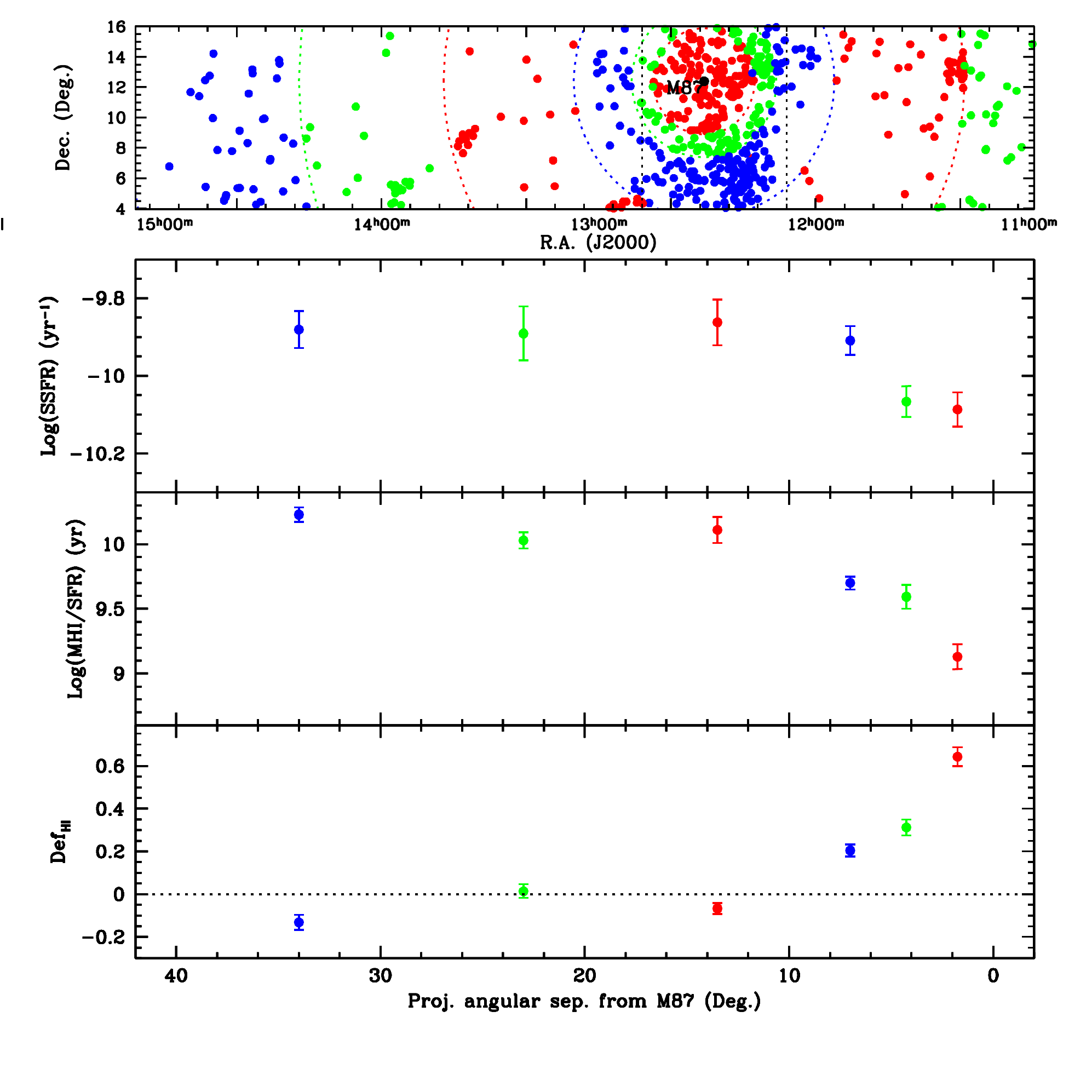}
\caption{{\it Upper panel}: the sky distribution of late-type galaxies within
seven annuli of increasing radius from M87. Galaxies in each ring
are given with a {\it different color}. Variation of the specific
star formation rate {\it SSFR} ({\it middle upper}), of the HI mass
per unit star formation rate {\it SFR} per unit HI mass ({\it middle
lower}), and of the HI-deficiency parameter $\hbox{Def}_{{\rm HI}}$
({\it lower}) as function of the projected angular separation from
M87 (from Gavazzi et al. 2013a). Courtesy of ESO}
\label{HaHI}
\end{figure}

The availability of HI blind surveys on large regions of the sky
such as ALFALFA (Giovanelli et al. 2005) and AGES
(Taylor et al. 2012, 2013) allowed the acquisition of homogeneous
sets of data down to a well-defined sensitivity limit for
extragalactic sources belonging to a large variety of environments.
Indeed ALFALFA and AGES included nearby clusters such as Virgo
(Giovanelli et al. 2007; Kent et al. 2008; Haynes et al. 2011), Coma and A1367
(Cortese et al. 2008c). These surveys  are generally less deep than
previous pointed observations of selected objects but cover
simultaneously and without any a priori selection all kind of
extragalactic sources. They mainly confirmed  that the atomic gas
content of galaxies decreases in high-density regions (e.g.
Cayatte et al. 1990; Solanes et al. 2001; Gavazzi et al. 2005, 2006b). They also allowed the
detection of a minority of ETGs, including dwarf systems, inside the
Virgo cluster with gas contents of the order of a few
$10^7\,\hbox{M}_{\odot}$ (di Serego Alighieri et al. 2007). These surveys led to
the first robust determination of the mean structural and
spectrophotometrical properties of HI-selected galaxies in different
environments and to compare them with those of optically selected
samples (Gavazzi et al. 2008; Cortese et al. 2008c). The main contribution of these
surveys to the study of the role of the environment on galaxy
evolution, however, comes from the combination of HI data with
optical and UV data from SDSS and \textit{GALEX}. Using a sample of
$\sim$10,000 objects with multifrequency data, Huang et al. (2012) have
shown that, for stellar masses below $M_{\rm star} \lesssim
10^{9.5}\,\hbox{M}_{\odot}$, galaxies follow a sequence along the
color--magnitude relation (CMR) or the {\it SSFR} (star formation per
unit stellar mass or specific star formation) vs. $M_{star}$ relation
regulated by the available quantity of HI gas, which becomes the
dominant barionic component in low mass systems. By studying this
effect in different environments, from the field to the core of
nearby rich clusters and including groups, Gavazzi et al. (2013a,b)
confirmed that the distribution of galaxies along the CMR below
$M_{\rm star} = 10^{9.5}\,\hbox{M}_{\odot}$ is regulated by their
atomic gas content. Late-type galaxies located on the blue sequence
have larger gas fractions than spirals in the green valley, or post
starburst galaxies characterized by red colors (Boselli et al. 2014a,b). The galaxies mostly
devoid of gas are the red early-type systems forming the red
sequence. Gavazzi et al. (2003a, 2013b) have also shown that this sequence
in colors or gas fraction is also related to the mean density of the
environment where galaxies reside, with redder colors and gas-poor
systems dominating high-density regions (see Fig. \ref{HaHI}). These
results, that extend previous analysis mainly based on massive
galaxies (Hughes \& Cortese 2009; Cortese \& Hughes 2009), have been interpreted as a clear
evidence that gas removal due to ram pressure stripping events
quenches the activity of star formation, transforming gas-rich,
star-forming systems into quiescent objects, as first proposed by
Boselli et al. (2008a,b). The work of Gavazzi et al. (2013b), combined with
that of Fabello et al. (2012) based on stacking analysis of HI data of a
large sample of nearby low-mass galaxies and that of Catinella et al. (2013)
for massive systems, however, have shown that gas removal and the
relative quenching of the star-formation activity is already present
in intermediate-density regions such as groups with halo masses $M >
10^{13}\,\hbox{M}_{\odot}$.

Interestingly, all these evidences questioned the general assumption
that in massive galaxies the quenching of the activity of star
formation is mainly due to AGN feedback (e.g.
Martin et al. 2007; Schawinski et al. 2009, see, however, Schawinski et al. 2014), proposing
environmental effects as an alternative process for the origin of
the green valley (Hughes \& Cortese 2009; Cortese \& Hughes 2009). The most recent CO surveys
of cluster and field objects support this scenario. The analysis of
a large K-band-selected, volume-limited sample of nearby field and
cluster galaxies, the \textit{Herschel} Reference Survey
(Boselli et al. 2010), as well as the detailed study of a small sample
of CO mapped galaxies, have both shown that also the molecular gas
phase is stripped by the interaction of galaxies with the hot
intergalactic medium permeating rich clusters such as Virgo
(Fumagalli et al. 2009; Boselli et al. 2014b). The stripping process, however, is less
efficient than that on the atomic phase because the molecular
hydrogen is mainly located in the central regions of galaxies, where
the gravitational restoring force is at its maximum. This evidence supports the idea that the responsible process for gas removal is ram pressure
stripping (Fumagalli et al. 2009; Scott et al. 2013; Boselli et al. 2014b). All these works have
been mostly limited to massive galaxies since the CO observation of
metal-poor, low-luminosity objects is still challenging. We can
expect, however, that given the shallower potential well of dwarf
galaxies with respect to massive systems, the stripping of the
gaseous component in all its phases is even more efficient in
low-mass objects.

\begin{figure}
\includegraphics[width=0.5\textwidth]{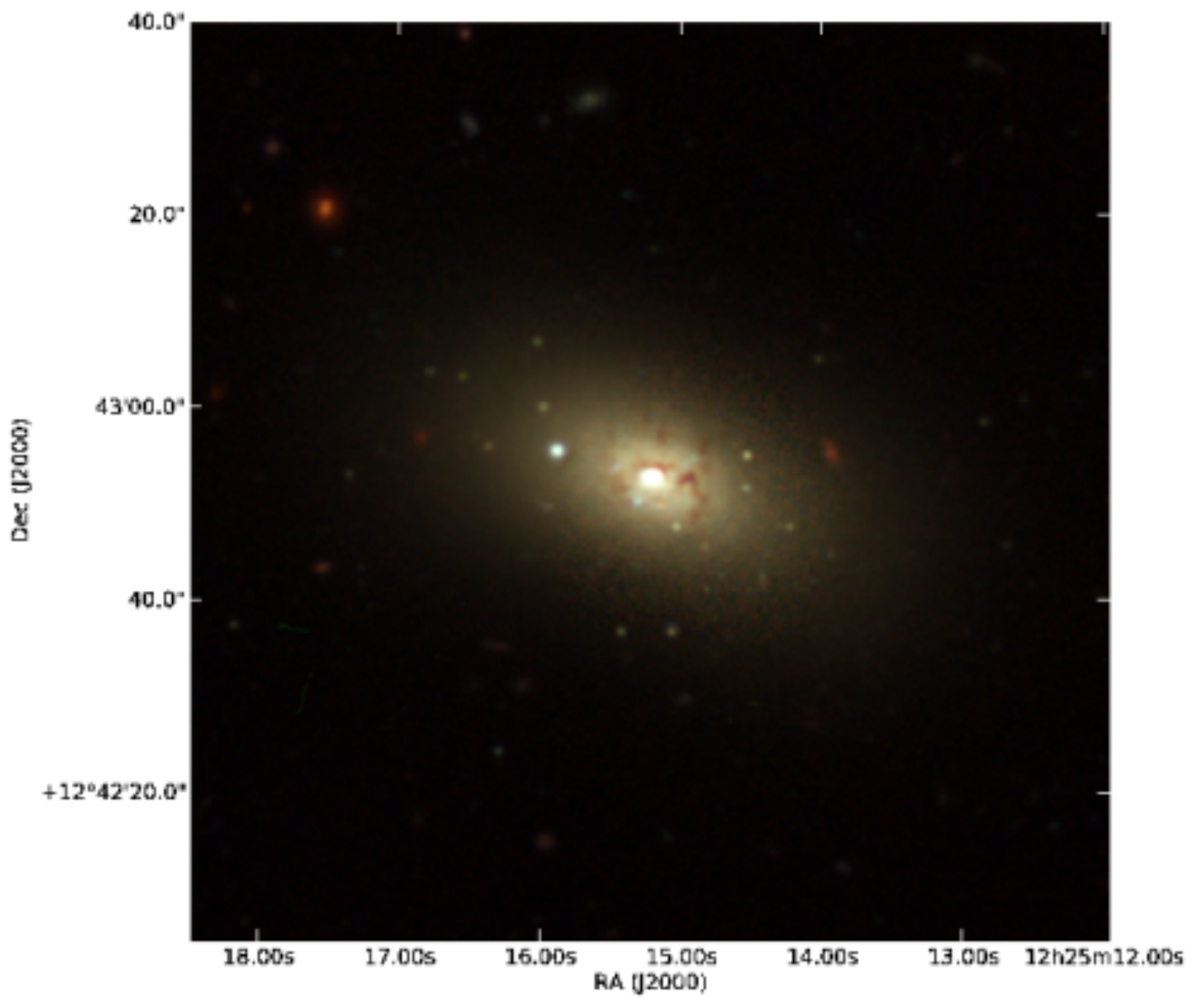}
\includegraphics[width=0.5\textwidth]{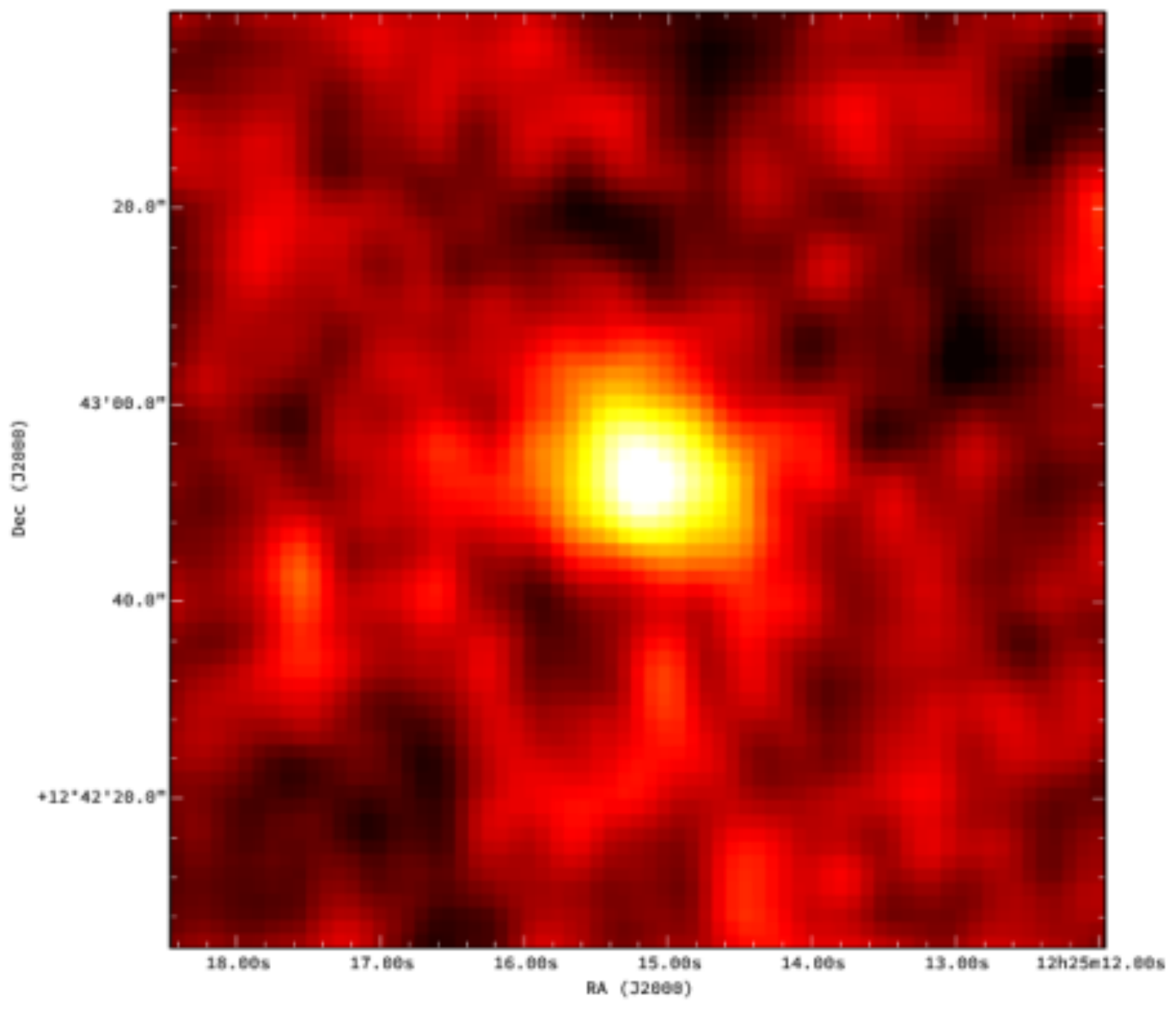}
\caption{The NGVS optical color ({\it top}) and the
\textit{WISE} 12~$\mu$m ({\it bottom}) images of the dwarf
elliptical VCC 781 in the Virgo cluster. The dust lane present in
the optical image is detected in the core of the galaxy in emission
in the mid-infrared by \textit{WISE}}
\label{VCC781}
\end{figure}

The HI blind surveys of the nearby universe have also revealed the
presence in clusters of HI clouds not associated with any stellar
component (Kent et al. 2007, 2009; Kent 2010; Haynes et al. 2007 in
ALFALFA; Taylor et al. 2012, 2013 in AGES). If some of the HI clouds
are probably tidal debris harassed from massive Virgo cluster
objects, such as the case of VIRGOHI21 and NGC 4254
(Haynes et al. 2007), others might have been formed from the stripped
gas of spirals entering the cluster for the first time
(Keent et al. 2009). Interferometric observations have revealed the
presence of spectacular tails of HI gas in the Virgo cluster
(Oosterloo \& van Gorkom 2005; Chung et al. 2007). The typical cometary shape of these
features strongly suggests that ram pressure stripping is the
responsible process. The main results of these observations are that
ram pressure stripping is acting well outside the virial radius of
clusters, making this process the most efficient gas stripping
process also in regions of relatively low density of the
intergalactic medium.

The lack of evident associated regions of star formation also
indicate that, contrary to what predicted by models
(Kepferer et al. 2009), the stripped gas can be hardly transformed into
new stars and, eventually, give birth to dwarf galaxies
(Boissier et al. 2012; see, however Yagi et al. 2013).

Similar long tails of hot gas have been observed in X-rays with
\textit{Chandra} in two spiral galaxies in the cluster A3627
(Sun et al. 2007, 2010). These observations indicate that also the
hot halo gas of galaxies can be stripped by ram pressure in
high-density environments.

\subsection{Dust content}

The \textit{Spitzer} and \textit{Herschel} missions allowed a
detailed analysis of the mid- and far-infrared properties of local
galaxies in high-density environments. The spectral coverage
(5--500~$\mu\hbox{m}$), combined with the high angular resolution
(from a few to 36~arcsec at $500\,\mu\hbox{m}$) and the sensitivity
of the different instruments were crucial for resolving the
different dust components (PAHs, hot and cold dust) in galaxies. The
analysis of the brightest Virgo cluster spirals has clearly shown
that the cold dust component distributed over the disk is removed
with the gaseous component during the interaction with the hot
intracluster medium. Indeed, HI-deficient spiral galaxies have
truncated dust disks compared to unperturbed field objects
(Cortese et al. 2010). Their total dust content is also reduced with
respect to normal, field galaxies (Cortese et al. 2012a). Concerning
the dwarf galaxy population, de Looze et al. (2010, 2013) have shown the
existence of a significant number of dE with presence of dust in
their inner regions (Fig. \ref{VCC781}). 36~\% of the dwarf galaxies
belonging to the green valley, identified by Boselli et al. (2008a) as
galaxies migrating from the blue cloud to the red sequence
(transition type galaxies), have been detected by \textit{Herschel}.
There is also evidence of the presence of a residual dust content in
several dwarf ETGs located outside the diffuse X-ray emitting gas
permeating the cluster (Boselli et al. 2014a). These observations are
consistent with the idea that the dust associated with the gaseous
phase is removed outside-in during the interaction with the
intergalactic medium of galaxies infalling for the first time into
the cluster. Their interstellar medium can be retained only in the
inner regions, where the gravitational potential well of the galaxy
is at its maximum.

\subsection{Star formation}

The statistical studies of the star-formation properties of cluster
galaxies in the past years have been mainly focused on the bright
galaxy population. Using a sample of 79 nearby clusters with
available multifrequency data, Popesso et al. (2007) have shown that the
cluster integrated star-forming properties do not change as a
function of the cluster properties. They see, however, that the
fraction of blue galaxies depends on the total X-ray luminosity of
the clusters, suggesting that environmental processes linked to the
presence of the hot X-ray emitting intracluster gas might affect the
star-formation history of cluster galaxies. Using a sample of nine
nearby clusters, including Coma and A1367, with new spectroscopic
data, Rines et al. (2005) have shown that the fraction of star-forming
bright galaxies, as determined from the presence of the Balmer
$\hbox{H}\alpha$ emission line, increases with clustercentric
distance and reaches that of the field at $\sim$2 to 3 virial radii.
This work mainly confirms the first results obtained by
Gomez et al. (2003) and Lewis et al. (2002) using the SDSS and 2dF surveys.

The combination of $\hbox{H}\alpha$ and HI data has been crucial for
understanding the very nature of the underlying physical process
responsible for the quenching of the star-formation activity of
cluster galaxies. This has been possible in the Virgo cluster and in
the Coma/A1367 supercluster (Gavazzi et al. 2013a,b). These works
have shown that the quenching of the star-formation activity follows
the stripping of the atomic gas as indicated by the tight relation
between the {\it SSFR} and the HI-deficiency parameter.\footnote{The
HI-deficiency parameter is defined as the difference, on logarithmic
scale, between the expected and the observed HI gas mass of each
single galaxy. The expected atomic gas mass is the mean HI mass of a
galaxy of a given optical size and morphological type determined in
a complete sample of isolated galaxies taken as reference (Haynes \& Giovanelli 1984).} Indeed, the specific star-formation rate
decreases with the clustercentric distance exactly as the
HI-deficiency, making LTGs redder (Gavazzi et al. 2013a,b).\footnote{Early claims of enhanced radio-continuum emission from
LTGs in the A1367 and Coma clusters (Gavazzi \& Jaffe 1985, 1986) are
insufficient to infer a significant star-formation enhancement in
clusters. It would not be surprising, however, if on a short time
scale this would take place from the shocked material during the
early phases of a ram pressure event.}

\begin{figure}
\includegraphics[width=0.7\textwidth]{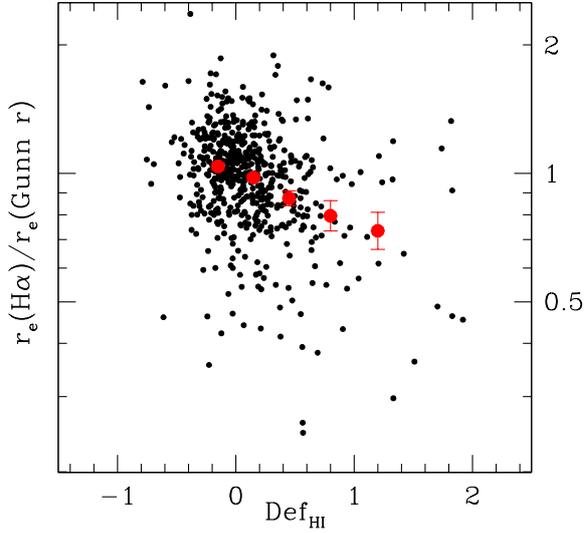}
\caption{The relation between the ratio of the
$\hbox{H}\alpha$ (star forming disk) to $r$-band (old stellar
population) effective radii and the HI-deficiency parameter. {\it
Big red dots} are average values along the $y$-axis in different
bins of HI deficiency, from Fossati et al. (2013). The star-forming disk is
truncated once the atomic gas is removed in cluster galaxies. Courtesy of ESO}
\label{Fossati}
\end{figure}

An accurate study of the star-formation properties of low-luminosity
cluster galaxies has been made possible owing to the narrow-band
$\hbox{H}\alpha$ imaging data that our team has obtained in these
past years. By comparing the $\hbox{H}\alpha$ morphology of galaxies
in Virgo, Coma and A1367 to those of field galaxies selected
according to similar criteria, Fossati et al. (2013) have shown that the
star-forming disk of LTGs, including dwarf systems, is truncated
once the galaxies are devoid of their gas (Fig. \ref{Fossati}). The
star-formation process is thus quenched outside-in, confirming
previous results obtained for the bright galaxy population
(Koopmann et al. 2006; Boselli \& Gavazzi 2006; Boselli et al. 2006; Cortese et al. 2012b). The $\hbox{H}\alpha$ imaging
of a few dE galaxies have also revealed the presence of an emitting
nucleus, suggesting that, after a stripping process, there is still
some retention of gas where the gravitational potential well of the
galaxy is at its maximum.\footnote{Examples of ETGs with nuclear or
circumnuclear $\hbox{H}\alpha$ emission in the Virgo cluster are
given in Boselli et al. (2008a): VCC 450, 597, 710, 1175, 1617, 1855, or in Toloba et al. (2014b): VCC 170, 781, 1304, 1684.}
This gas is able to feed a nuclear star-formation activity
(Boselli et al. 2008a). More in general, multizone
chemo-spectrophotometric models of galaxy evolution especially
tailored to reproduce the effects due to ram pressure stripping in
cluster environments have shown that the observed outside-in
truncation of the star formation in gas stripped galaxies can be
responsible for the inversion of the color gradient observed in
massive galaxies (Boselli et al. 2006) or in dwarf ellipticals
(Boselli et al. 2008a). There is also evidence of some galaxies with a
disturbed $\hbox{H}\alpha$ or UV morphology, witnessing and
undergoing interaction with the hostile cluster environment
(Smith et al. 2010). The $\hbox{H}\alpha$ and UV data have been also
crucial for studying extraplanar star-formation events in galaxies
with clear signs of an undergoing perturbation (see
Sects.~\ref{INDIVIDUAL} and \ref{EXTRAPLANAR}).

\begin{figure}
\includegraphics[width=1.0\textwidth]{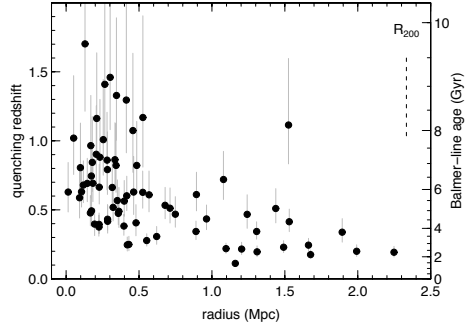}
\caption{The dependence of the quenching age, defined as the
redshift for which the look-back time is equal to the age of dwarf
galaxies determined using the Balmer lines, on the projected
distance from the core of the Coma cluster, from Smith et al. (2008).
Dwarf elliptical galaxies in the periphery of the Coma cluster
stopped their activity of star formation only at recent epochs. Reproduced with permission of Oxford University Press}
\label{Smithspsage}
\end{figure}

A systematic study of the star-formation properties of galaxies in
23 nearby ($z \sim 0.06$) groups based on \textit{GALEX} data has
been presented in Rasmussen et al. (2012a). This work has shown that, as in
clusters, the fraction of star forming galaxies within groups is
suppressed with respect to the field in the inner ${\sim} 2$ virial
radii (${\sim}1.5$~Mpc). The same work has shown a suppression of
the specific star-formation rate by a factor of ${\sim} 40$~\% with
respect to the field, quantifying the impact of the group
environment on quenching the activity of star formation in infalling
galaxies. At fixed galaxy density and stellar mass, this suppression
is stronger in more massive groups. Rasmussen et al. (2012a) concluded that
the average time scale for quenching the star-formation activity is
$\gtrsim$2~Gyr and identified a combination of tidal encounters and
starvation as the responsible process.

\subsection{Stellar populations}

\begin{figure}
\includegraphics[width=0.5\textwidth]{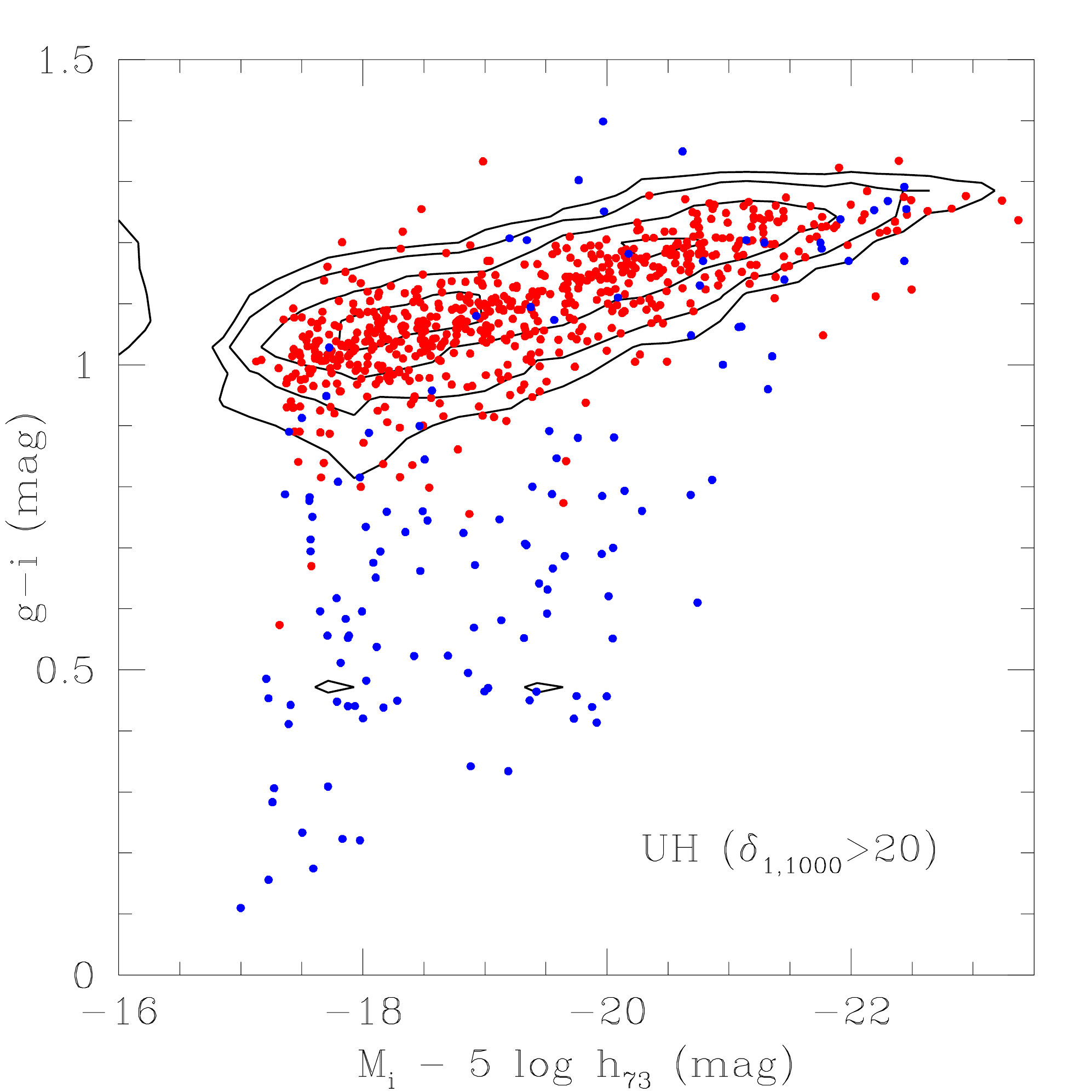}
\includegraphics[width=0.5\textwidth]{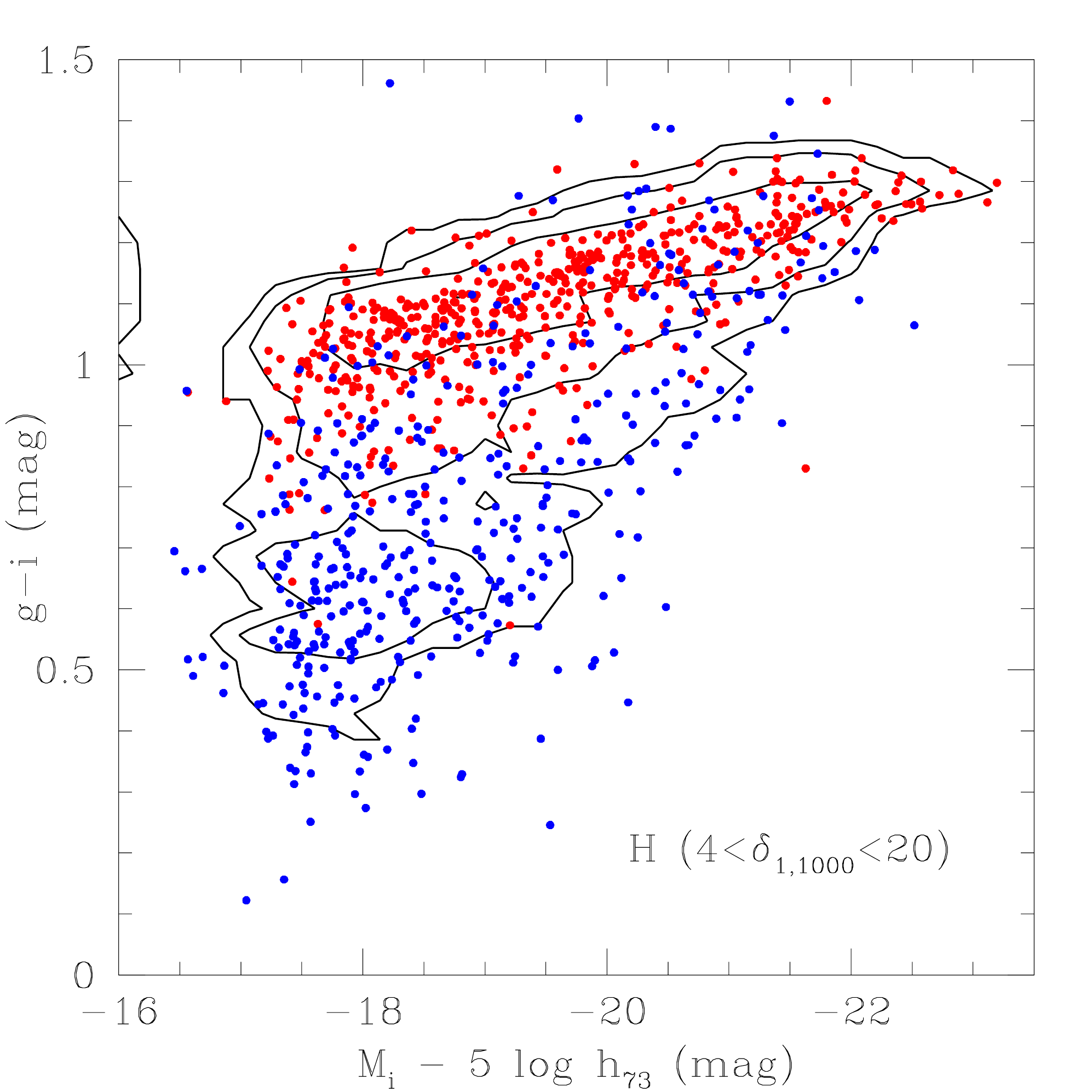}\\
\includegraphics[width=0.5\textwidth]{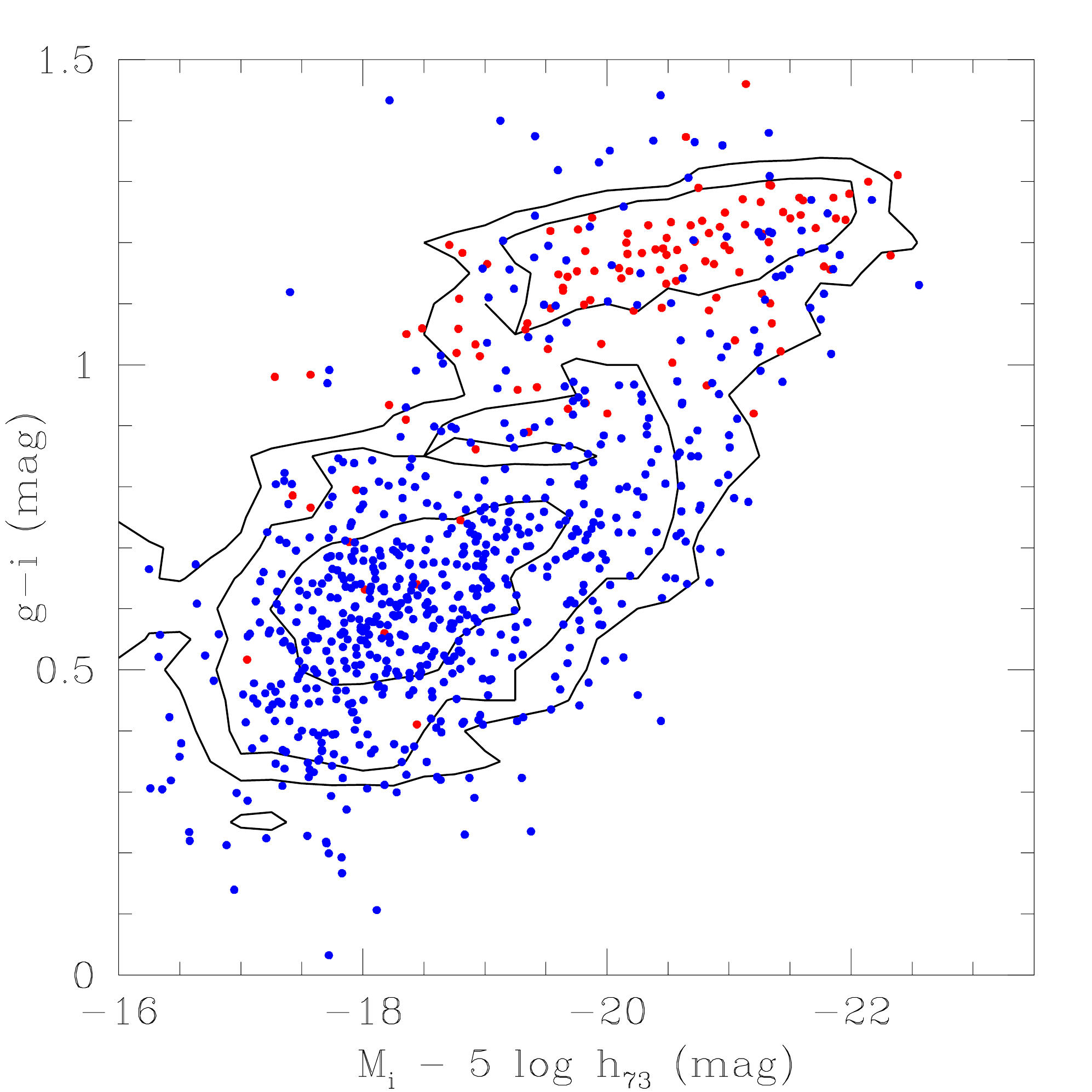}
\includegraphics[width=0.5\textwidth]{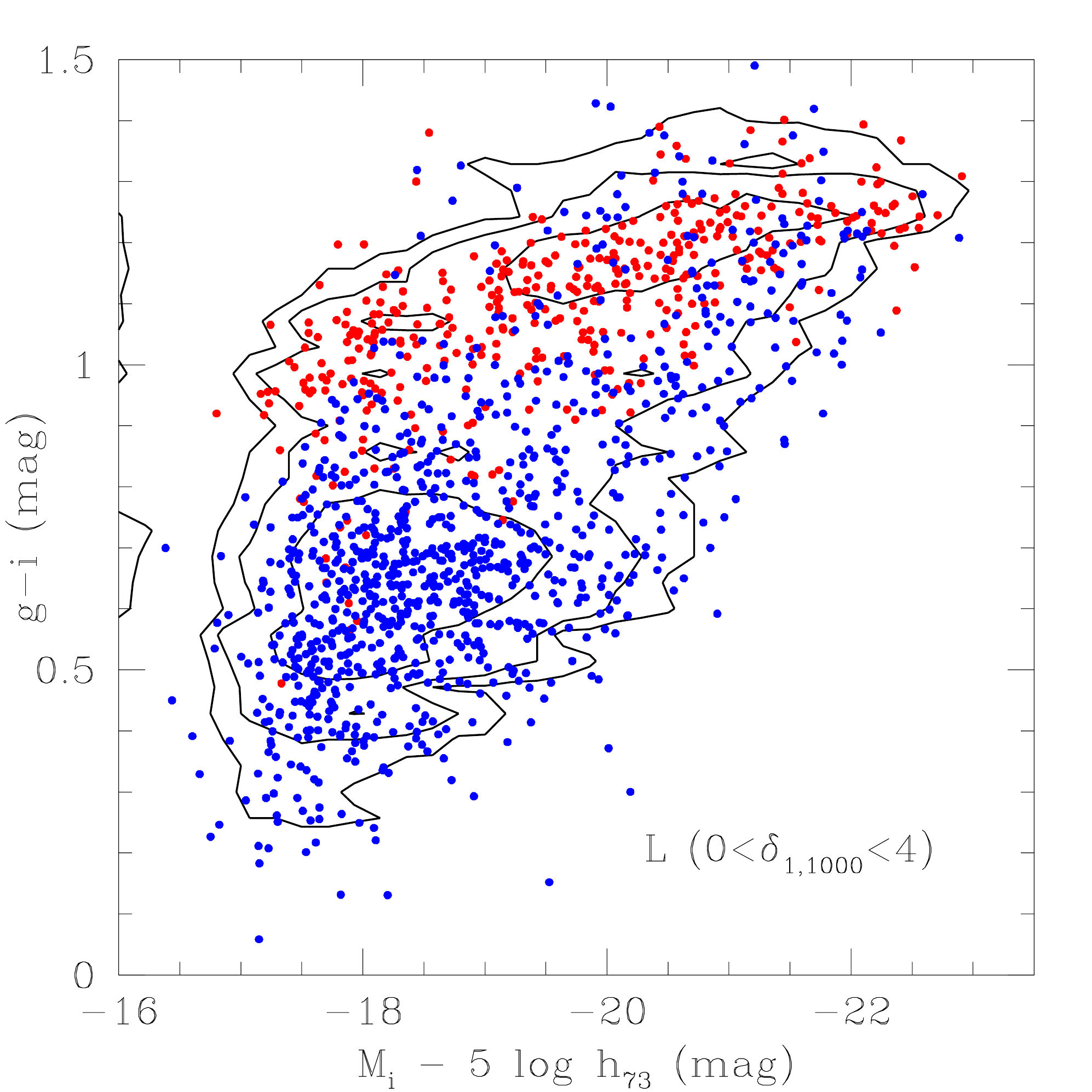}\\
\caption{The $g-i$ vs. $M_i$ CMR for galaxies in the
Coma/A1367 supercluster from very high density regions in the core
of Coma and A1367 to very low density regions in the voids
surrounding the clusters, clockwise from upper left. \textit{Red symbols}
indicate early-type galaxies in the red sequence, \textit{blue symbols}
late-type systems in the blue cloud, from Gavazzi et al. (2010). Star
forming systems are virtually lacking in the densest regions in the
core of the clusters, while the faint end of the red sequence is not
present only in the lowest density regions dominated by star forming
systems. Courtesy of ESO}
\label{Gavazzispsdensity}
\end{figure}

The study of the nature of the various stellar populations of dwarf
galaxies in different environments is one of the topics that has
improved the most in the past years owing to the advent of large
photometric and spectroscopic surveys of nearby clusters. The most
recent works have been mainly focused either on the optical
(Haines et al. 2006b; Lisker et al. 2006a, 2008; Misgeld et al. 2009; Janz \& Lisker 2009; Gavazzi et al. 2010;
Scott et al. 2013; den Brok et al. 2011; McDonald et al. 2011) or on
the optical-UV CMR (Boselli et al. 2005a,b, 2008a, 2014a; Lisker \& Han 2008; Haines et al. 2008; Kim et al. 2010; Smith et al. 2012b), or based
on the analysis of absorption line indices obtained from deep
spectroscopic observations
(Smith et al. 2006, 2008, 2009, 2012a; Haines et al. 2006, 2007; Michielsen et al. 2008; Toloba et al. 2009; Paudel et al. 2010a,b, 2011; Boselli et al. 2014a).
The first works of Haines et al. (2006a,b) on the cluster A2199 and
on the Shapley supercluster, along with that of Smith et al. (2006) on 94
clusters with spectroscopic data from the SDSS, have consistently
shown a different evolution of massive and dwarf galaxies in
high-density regions. While the fraction of massive ($M_{\rm r} <
{-}20$) red galaxies dominated by an old population gradually
decreases from $\sim$80\,\% in the core to $\sim$40\,\% in the
periphery (3--4 virial radii) of the cluster, the radial variation
for dwarfs (${-}19<M_{\rm r}<17.8$~mag) is much more pronounced,
from $\sim$90\,\% in the core to $\sim$20\,\% at one virial radius
(Haines et al. 2006a,b; Boselli et al. 2014a). Consistently, by
studying the dispersion of the fundamental plane relation of
early-type systems, Smith et al. (2006) have shown that, for a given
velocity dispersion---thus total mass---galaxies in the periphery of
clusters have stronger Balmer absorption lines, indicative of
younger ages than those located in the cluster core (see Fig.
\ref{Smithspsage}). Consistent results have been obtained using new
spectroscopic observations in the Coma cluster by
Smith et al. (2008, 2009, 2012a). All these studies have indicated
strong Balmer absorption lines and enhanced $\alpha/{\rm Fe}$ ratios
in the red dwarf galaxy population in the outskirts of Coma compared
to the core, where only the oldest galaxies reside, consistent with
a relatively recent formation of the red sequence ($0.4 <z< 0.7$).
The SW substructure of Coma is also composed of younger dwarf
quiescent galaxies, probably formed between $0.1< z < 0.2$
(Smith et al. 2009). The mean age of the underlying stellar
population is mainly related to the total mass of galaxies in
massive objects (${M}_{\rm star}
> 10^{10}\;\hbox{M}_{\odot}$) and only marginally on the environment,
while the reverse holds for dwarf systems. Indeed, the mean age of
the stellar population of dE galaxies is about a factor of two
younger at one virial radius than in the core of the cluster
(Smith et al. 2012a). The earlier work of Poggianti et al. (2004) has shown the
presence of low luminosity post-starburst galaxies in the Coma
cluster, consistently with the picture where dwarf galaxies abruptly
truncated their star-formation activity and became red objects.
These post-starburst galaxies are mainly low mass systems situated
around massive clusters, as indicated by the recent spectroscopic
survey of 48 nearby clusters ($0.04<z<0.07$) of Fritz et al. (2014).
Similar trends between the mean age of the stellar populations and
the clustercentric distance have been also observed in the Virgo
cluster by Michielsen et al. (2008), Toloba et al. (2009, 2014b), and Paudel et al. (2010a, 2011).
In particular, Toloba et al. (2009) have shown that dE galaxies
characterized by the youngest stellar population not only are
located in the periphery of the cluster, but also are generally
rotationally supported systems. Thanks to the proximity of the Virgo
cluster, spectroscopic studies have been used to investigate stellar
population gradients within the disk of dwarf elliptical galaxies.
The available works have consistently shown that, on average, the
mean age of the stellar populations increases from the center to the
outskirts of galaxies (Chilingarian et al. 2009; Koleva et al. 2009, 2011; Paudel et al. 2011)
contrary to what generally happens in massive systems (Fig.
\ref{Gavazzispsdensity}). All these
evidences are consistent with a scenario where low-luminosity
star-forming galaxies recently entered the cluster environment,
losing their gas content and quenching their star-formation
activity, thus becoming dwarf ellipticals
(Boselli et al. 2008a; Toloba et al. 2009; Gavazzi et al. 2010; Koleva et al. 2013).

The study of the photometric properties of cluster galaxies has been
primarily aimed at understanding whether genuine giant ellipticals
and dwarf systems follow the same CMR, an indication that would
suggest a common origin for the two families of objects. The works
done so far do not bring fully consistent results. A unique CMR has
been invoked by Misgeld et al. (2009) in the Centaurus cluster, by
Hammer et al. (2010b) in Coma and more recently by Smith Castelli et al. (2013) by
combining SDSS data of Virgo galaxies with those obtained by the
ACS Virgo Cluster Survey. Conversely, variations of the CMR have
been found using SDSS data of more than 400 Virgo objects by
Janz \& Lisker( 2009). These authors, however, concluded that this
observational result does not rule out a common origin for the two
populations. Lisker et al. (2008) looked for any systematic difference in
the optical CMR of the different subclasses of dwarf elliptical
galaxies. Their analysis has shown that, at relatively high
luminosities, nucleated dwarf ellipticals have redder colors than
normal dE, while the variations with local galaxy density, if
present, are minor.

The works of Haines et al. (2008) and of Gavazzi et al. (2010) have been
fundamental to extend these interesting results to other
environments such as the periphery or rich clusters or intermediate
mass groups. Using the $NUV-i$ color index, much more sensitive to
age variations than the optical colors, Haines et al. (2008) have shown
that the red sequence is not formed in the field at low
luminosities. The faint end of the red sequence, however, is already
formed within the different substructures of the Virgo cluster
characterized by a wide range in galaxy density (Boselli et al. 2014a).
Consistently, Gavazzi et al. (2010) have shown that both the shape of the
luminosity function determined for galaxies selected according to
their color, and the $g-i$ CMR change as a function of galaxy
density, indicating that the processes that gave birth to the faint
end of the red sequence have been efficient also in relatively
low-density environments. These results are fully consistent with
those determined by analyzing statistically significant samples
extracted from the SDSS indicating that red dwarf galaxies are
extremely rare in the field (Geha et al. 2012). The work of
Gavazzi et al. (2010) have also revealed the presence of low-luminosity
galaxies with clear signs of a recent activity of star formation
(post-starburst) in Coma and A1367. These galaxies are
analogous to those found in the Virgo cluster by Boselli et al. (2008a) and
originally defined as transition type galaxies as their
spectrophotometric properties indicate a recent abrupt truncation of
their star-formation activity (Koleva et al. 2013).

The works of van den Bosch et al. (2008), Wilman et al. (2010), and Cibinel et al. (2013)
studied the relationship between galaxy color and density in nearby
groups using SDSS photometric data. These works have shown that the
perturbation induced by the environment affects more the color than
the morphology of galaxies (van den Bosch et al. 2008). They have also shown
that galaxies become red only once they have been accreted on to
halos of a certain mass (Wilman et al. 2010). The quenching of the
star-formation activity is primarily in the outer disk of galaxies
entering groups (Cibinel et al. 2013).

\begin{figure}
\includegraphics[width=0.7\textwidth]{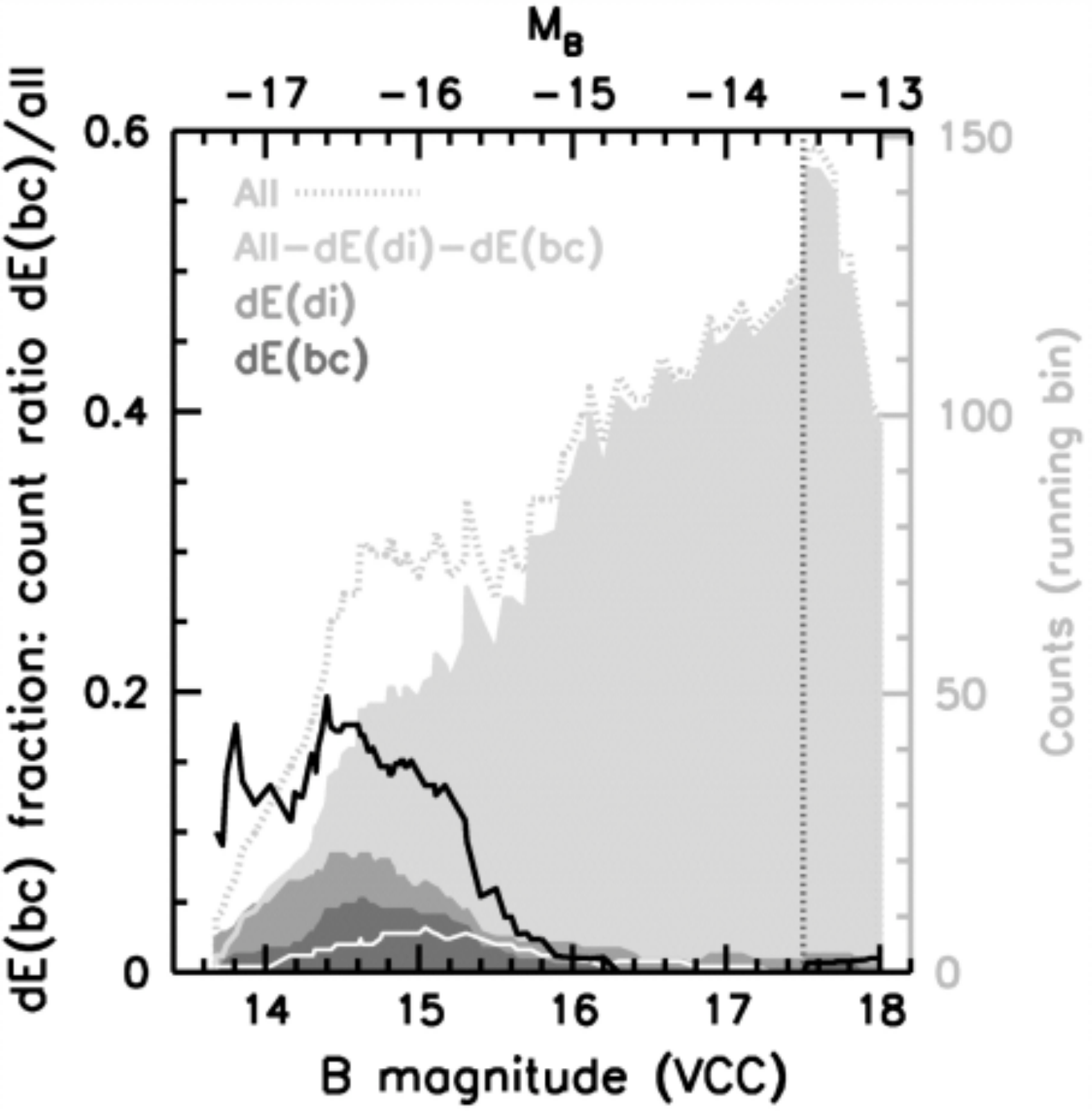}
\caption{Histogram of the number of dwarf elliptical
galaxies in the Virgo cluster in bins of absolute B band magnitude
(from the VCC; right y-axis). The \textit{light gray dashed line} shows the
distribution of all dEs, the dark gray area that of dEs with blue
nuclei, the medium grey that of dEs with an underlying disky
structure, the light gray that of all dEs galaxies after excluding
those with disky structures and blue nuclei. The \textit{black solid line}
shows the fraction of the dEs with blue nuclei (left y-axis), from
Lisker et al. (2006b). \copyright\ AAS. Reproduced with permission}
\label{Lisker}
\end{figure}

\subsection{Structural parameters}\label{STRUCTURAL}

The study of the structural properties of dwarf galaxies in nearby
clusters has enormously benefited from the SDSS. The first
systematic study of the structural properties of dwarf elliptical
and spheroidal galaxies in the Virgo cluster has been done by Lisker
and collaborators
(Lisker et al. 2006a,b, 2007, 2009; Janz \& Lisker 2008; Lisker 2009). These works
basically extended previous analyses done on the photographic plates
taken at the Du Pont telescope at Las Campanas in the eighties by
Binggeli and collaborators (Binggeli et al. 1985; Binggeli \& Cameron 1991). Lisker et al. (2006a)
have looked for unseen disky features such as spiral arms, bars, and
edge-on disks in dwarf ellipticals by applying unsharp masks or
subtracting the axisymmetric light distribution of each galaxy from
the image in a complete sample of Virgo galaxies extracted from the
SDSS. They have shown that dE galaxies with typical spiral features,
as the one first discovered by Jerjen et al. (2000), are rather common
objects. Among the brightest dwarf ellipticals, $\sim$50\,\% of the
objects possess these features, while their fraction decreases with
decreasing luminosity (Fig. \ref{Lisker}). Such galaxies have been
later discovered in other clusters like Coma using the exquisite
quality images obtained with the HST (Graham et al. 2003; den Brok et al. 2011; Marinova et al. 2012). Their
spiral structure is generally grand design and not flocculent, and
they have a flat shape suggesting that they are genuine disk
galaxies (Lisker et al. 2006a; Lisker \& Fuchs 2009). The best quality imaging material
gathered thanks to the ACSVCS survey of Virgo has shown that these
kind of features are very frequent in quiescent systems of
intermediate luminosity (Ferrarese et al. 2006). The same survey has also
indicated that dusty features observed in absorption are also quite
frequent but mainly in the most massive objects (42\,\%). There is
also a population of dwarf ellipticals with blue nuclei probably due
to a recent episode of star formation. The work of Lisker et al. (2007) has
shown that galaxies with disky structures, including those with blue
nuclei, are not distributed near the cluster core as are the bright
ellipticals and lenticulars, or the other ordinary spheroidal dwarf
ellipticals, but are rather distributed uniformly all over the
cluster. Thanks to HST images it has also been shown that dwarf ETGs
in the periphery of the Peresus cluster have more disturbed
morphologies than those located in the core of the cluster
(Penny et al. 2011). Lisker et al. (2009) have studied the properties of the
nucleated dwarf elliptical galaxies without any sign of recent star
formation (blue nuclei) or disky structure (spiral arms, bars,
disks) located close to the bright central elliptical galaxies. By
dividing the sample into fast- and slow-moving objects according to
their velocity with respect to the mean recessional velocity of the
cluster, they have shown that fast-moving objects have a projected
axial ratio consistent with a flatter shape, while the slow-moving
are roundish objects. Deep near-infrared images of dwarf elliptical
galaxies have also revealed different structures in their 2D-stellar
distribution (Janz et al. 2012, 2014). All these results have been interpreted as an
evidence that dwarf elliptical galaxies have different origins.
Those presenting disky structures, or having a flat shape (thus
having a high-velocity with respect to the cluster) have been
recently accreted as star-forming systems and have been transformed
during their interaction with the cluster environment, while the
roundish objects with a low-velocity dispersion have been in the
cluster since a very early epoch, or might have even been formed
within the cluster.

In the recent years a huge effort has also been made at
understanding whether dwarf elliptical galaxies are just the low
luminosity extension of bright ellipticals or rather are a totally
independent category of objects. Systematic differences in the two
galaxy populations, and possible similarities with the properties
observed in spiral galaxies, could be interpreted as a clear
indication of a different formation scenario
(Kormendy et al. 2009; Kormendy \& Bender 2012). The debate is principally motivated by the
possible existence of a few dwarf elliptical galaxies similar to M32
that, contrary to the general dE galaxy population, have a very
compact structure characterized by a very high surface brightness.
These objects might not follow the standard scaling relations
depicted by the other quiescent systems. The seminal work of
Graham \& Guzman (2003) has clearly shown that the observed change in slope in
the surface brightness vs. absolute magnitude relation observed
between massive and dwarf ellipticals is naturally due to the shape
of their Sersic light profile, with an index $n$ increasing with the
luminosity. Continuity between E and dE in different scaling
relations has been also reported by Gavazzi et al. (2005), Ferrarese et al. (2006),
Misgeld et al. (2008, 2009), Misgeld \& Hilker (2011), and Smith Castelli et al. (2013). On the
contrary, there is evidence that the size--luminosity relation
depicted by bright ellipticals is not followed by dwarf systems,
that rather have an almost constant extension ($R_{\rm eff} \simeq
1$~kpc) regardless of their luminosity (Janz \& Lisker 2009; Smith Castelli et al. 2008; Misgeld \& Hilker 2011).
This dispute will certainly come to an end once the NGVS survey
(Ferrarese et al. 2012), that has covered homogeneously the whole Virgo
cluster, thus including several hundreds of early-type systems in
four ($u^*,g',i',z'$) photometric bands, will be fully exploited.

Thanks to its exquisite angular resolution, the ACSVCS
(Cote et al. 2004) allowed the detailed analysis of the structural
properties of the nuclei of ETG. The HST data have shown that  ground
based optical surveys generally underestimate the presence of nuclei
in ETGs probably because of their limited angular resolution
(Cote et al. 2006). The nuclei are generally resolved even in dwarf
systems, ruling out any possible low-level AGN nature
(Cote et al. 2006). The same data have also shown the lack of any
strong evidence that nucleated galaxies are more centrally clustered
than other non-nucleated objects. By fitting their radial light
distribution, Cote et al. (2007) have shown that a Sersic profile
generally overestimates the nuclear emission in bright and massive
galaxies while it underestimates it in low-luminosity objects, with
a difference that steadily varies with galaxy luminosity. They
discussed the observed nuclear properties of ETGs in the context of
gas infall in various formation scenarii.

The comparison of the structural properties of red sequence and blue
cloud galaxies in nearby groups has been carried out by
van den Bosch et al. (2008) using a large sample of galaxy groups extracted from
the SDSS. Studying their color and light concentration, their
analysis has shown that the transformation mechanism operating on
satellites affects more colors than morphology. They have also shown
that the observed differences between satellite and central galaxies
do not depend on the halo mass, suggesting that the process at the
origin of the observed perturbation is the same in different
environments. They thus concluded that the most probable process is
starvation since they considered ram pressure stripping and galaxy
harassment efficient only in high-mass systems such as clusters.

\begin{figure}
\includegraphics[width=0.5\textwidth, angle=-90]{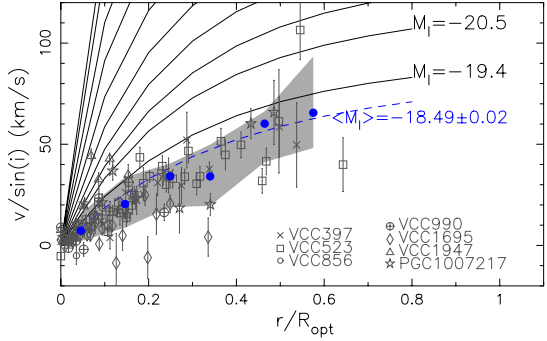}  
\caption{The observed rotation curves of rotationally
supported systems dEs (\textit{gray symbols}) are compared to the mean
rotation curves of observed (\textit{black solid lines}) and expected (\textit{blue
dashed line}) late-type galaxies of different absolute $i$-band
magnitude. \textit{Blue filled dots} represent the median observed rotation
curve of rotationally supported dEs, while the gray area the
rotation velocities within 1 $\sigma$ from the median, from
Toloba et al. (2011). The rotation curves of rotationally supported dEs are
similar to those of star forming systems of comparable luminosity. Courtesy of ESO}
\label{Toloba}\end{figure}

\subsection{Kinematics}

The kinematical properties of galaxies and their relations with the
environment have been the subject of various works done by the
SAURON (Bacon et al. 2001), $\hbox{ATLAS}^{\rm 3D}$
(Cappellari et al. 2011a), CALIFA (Sanchez et al. 2012) and SMAKCED
(Toloba et al. 2014a) teams. While the first three projects were
mainly focused on bright galaxies (see, however, Rys et al. 2013),
with SAURON and $\hbox{ATLAS}^{\rm 3D}$ primarily on early-type
systems, SMAKCED was devoted to the study of dwarf elliptical
galaxies in the Virgo cluster. The final purpose of this project was
that of understanding whether dE are rotationally supported systems,
as first claimed by Pedraz et al. (2002), Geha et al. (2002, 2003) and
van Zee et al. (2004). Based on very small samples, these pioneering works
were mainly limited by the spectral resolution of the adopted
instruments, barely sufficient to observe velocity dispersions of
the order of 20--$30\,\hbox{km\,s}^{-1}$ typical of dwarf systems.
Using a sample of 21 dwarf elliptical galaxies mainly located in the
Virgo cluster, Toloba et al. (2009, 2011) have shown the existence of a
large fraction of rotationally supported dE. Their analysis
indicated that pressure-supported systems, generally characterized
by old stellar populations, are located preferentially in the inner
regions of the cluster, while rotationally supported objects,
composed of younger stellar populations, are mainly located at the
periphery of the cluster (Toloba et al. 2009). Furthermore, these
works have shown that rotationally supported dE have rotation curves
similar to those of LTGs of similar luminosity and follow the same
Tully--Fisher relation as star forming systems (Toloba et al. 2011;
Fig. \ref{Toloba}). The analysis of the main scaling relations, such
as the Faber--Jackson and the Fundamental Plane relations,
consistently revealed that dE are different from massive ellipticals
since have structural and kinematical properties closer to those of
star-forming systems rather than to those of E and S0
(Toloba et al. 2012). Although the kinematical properties of dE might
change from object to object (Rys et al. 2013), all these
observational evidences are consistent with the picture where
gas-rich star-forming systems entering the hostile cluster
environment lose their gaseous component on relatively short time
scales, stopping their activity of star formation and becoming
quiescent systems. The angular momentum of the galaxy is conserved
on relatively long time scales if the galaxy interacts with the hot
intergalactic medium (ram pressure stripping), while the system is
heated if the dominant process is gravitational (galaxy harassment).
The observations indicate that the dynamical interactions with the
IGM are dominant at the present epoch, producing the rotationally
supported systems at the periphery of the cluster, while the
gravitational one, dominant in the past, produced the hot systems
mainly located in the virialized inner region of the cluster
(Toloba et al. 2014b; Boselli et al. 2014a). Independent evidence of gravitational
interactions comes also from the presence of dE with counter
rotating cores observed by Toloba et al. (2014c) close to the Virgo cluster
core. This evolutionary picture is consistent with that proposed for
more massive galaxies by the $\hbox{ATLAS}^{\rm 3D}$ survey, where
the most massive slow rotators are formed by major dry merging
events, while fast rotators by the quenching of the star formation
of spiral galaxies (Cappellari et al. 2011b; Cappellari 2013). The only difference
with the fading model proposed for low luminosity systems is that,
associated with the quenching of the star-formation activity of
spiral disks in cluster, there is also a growth of the bulge
(Cappellari 2013).

\subsection{Individual galaxies}\label{INDIVIDUAL}

As already introduced in Sect.~\ref{s1}, a typical example of
galaxies undergoing environmental transformations is the galaxy FGC
1287 at the periphery of the cluster Abell 1367, showing a cometary
HI tail of 250~kpc projected length. Although the origin of this
feature is still uncertain (Scott et al. 2012), its typical shape
suggests that it might have been created by the interaction of its
ISM with the hot and dense ICM of the cluster during a ram pressure
stripping event. We recall that this result agrees with the presence
of massive ram pressure stripped galaxies up to more than 1~Mpc away
from the cluster core observed in several nearby clusters as
summarized in Boselli \& Gavazzi (2006), including a recent discovery in the
Shapley supercluster (Merluzzi et al. 2013). It also agrees with the
results of the most recent hydrodynamical simulations of gas
stripping in cluster galaxies (see Sect. \ref{HYDRODYNAMICAL}).

Consistent with this scenario is also the discovery of cluster
galaxies with long tails of ionized gas, first observed in A1367 by
Gavazzi et al. (2001) and later in Virgo (Yoshida et al. 2004; Kenney et al. 2008), in
Coma (Yagi et al. 2007, 2010; Yoshida et al. 2008, 2012; Fossati et al. 2012), and in A3627
(Zhang et al. 2013). If the $\hbox{H}\alpha$ filamentary structure
associated with NGC 4388 in Virgo is located close to the cluster
core (Yoshida et al. 2004; kenney et al. 2008), those relative to the galaxies in
A1367 and Coma are much more in the periphery of the clusters, again
indicating that ram pressure stripping is efficient not only in the
cluster core. The observation of these peculiar objects gives also
other important information on the physical process acting on
cluster galaxies. They first indicate that the stripping process is
also able to remove the ionized phase of the gas (Gavazzi et al. 2001).
There are, however, other indications that part of the gas can be
ionized in situ by shocks (Yoshida et al. 2012). They indicate that
star formation in the stripped gas, if present, is generally very
modest and located in small ($\sim$200 to 300~pc) defined knots of
stellar mass $10^6$--$10^7\,\hbox{M}_{\odot}$ (Yoshida et al. 2008).
The few star-forming blobs associated with NGC 4388 or with the
interacting systems VCC 1249--M49 in Virgo are even smaller, with
stellar masses ${\lesssim}10^{4.5}\,\hbox{M}_{\odot}$
(Arrigoni Battaia et al. 2012; Yagi et al. 2013). This result is consistent with the
evidence that the efficiency of star formation in the stripped
material is lower than the standard efficiency observed over spiral
disks generally described by the Schmidt law (Boissier et al. 2012).

This last result is of particular importance in constraining models
of star formation from the stripped gas (see Sect.
\ref{EXTRAPLANAR}). Indeed, it indicates that the formation of dwarf
galaxies in the stripped material, if possible, is not frequent and
cannot produce any significant steepening of the faint end of the
luminosity function. There are, however, a few cases where the
star-formation process in the stripped gas is important. First
discovered by Cortese et al. (2007) in two clusters at $z \sim 0.2$, there exist a few objects with long tails of star-forming regions
prominent in UV images generally named ``fireballs'' or
``jellyfish'' (Smith et al. 2010). They are more frequent in massive
clusters such as Coma than in small systems as Virgo. An exception
to this rule is the spectacular IC 3418 (VCC 1217) close to the core
of the Virgo cluster (Hester et al. 2010; Fumagalli et al. 2011; Jachym et al. 2013; Kenney et al. 2014). The
galaxy, totally stripped of its gas, quenched its activity of star
formation $\sim$300 to 400~Myr ago probably after a starburst
activity. In the tail, the observed $\hbox{H}\alpha$ peaks are
displaced from the UV emitting knots, suggesting that the gas clumps
are continuously accelerated by ram pressure, leaving behind new
stars decoupled from the gas (Kenney et al. 2014). The typical mass of
these fireballs is of ${\sim} 10^5\,\hbox{M}_{\odot}$
(Fumagalli et al. 2011), thus still too small to make this process
relevant for the formation of dwarf galaxies or for modifying the
shape of the luminosity function. Another spectacular example with
similar characteristics is the galaxy ESO 137-001 in the Norma
cluster (A3627, Sun et al. 2007; Jachym et al. 2014).

Multifrequency observations have been critical to demonstrate that
also the hot X-ray emitting gas in the halo of galaxies can be
stripped by ram pressure in cluster
(Sun et al. 2007, 2010; Ehlert et al. 2013; Zhang et al. 2013) and group galaxies
(Rasmussen et al. 2012b). The most evident case is the galaxy ESO 137-001
in A3627 (Sun et al. 2007; Fumagalli et al. 2014). While in clusters the hot and
cold gas removal is primarily due to ram pressure stripping, in
groups both ram pressure and tidal interactions contribute, as
clearly indicated by the recent \textit{Chandra} and VLA
observations of the nearby group of NGC 2563 (Rasmussen et al. 2012b).
The importance of these results resides in the fact that they are the
first observational justification that the gaseous halo of galaxies
is removed in high-density environments, as generally assumed in
cosmological and semi-analytic models of galaxy evolution. This
assumption in models and simulations is crucial since it makes the
feedback process of supernovae extremely efficient in expelling the
disk gas, thus quenching the activity of star formation and
transforming on very short time scales gas-rich systems into
quiescent, red galaxies (see Sect.~\ref{MODELLIspsCOSMOLOGICI}).

Although not frequent because of the high-velocity dispersion within
clusters, tidal interactions can also be related to the formation
and evolution of dwarf systems in high-density regions. There
exists, indeed, a few representative and interesting cases in nearby
clusters. One example is NGC 4254 in Virgo. Entering the cluster at
high velocity for the first time, the galaxy was harassed of a
fraction of its gas that is now forming a cloud (VIRGOHI21;
Davies et al. 2004) apparently not associated with any other optical
counterpart (Haynes et al. 2007; Duc \& Burnaud 2008; Wezgowiec et al. 2012). The wide field and the
high sensitivity to low surface brightness features of the NGVS
survey allowed the identification of three dwarfs galaxies satellite
of the bright NGC 4216 in Virgo undergoing a tidal disruption
process (Paudel et al. 2013). The data obtained by NGVS and GUViCS
have shown the presence of small star-forming complexes produced
during the interaction of the dwarf, gas-rich Im galaxy VCC 1249
(UGC 7636) with the bright elliptical M49 (Arrigoni Battaia et al. 2012). Although the stellar mass of these objects is
still very small $(10^4-10^5\,\hbox{M}_{\odot}$), it has been
suggested that a similar process might have been at the origin of
Ultra Compact Dwarf (UCD) galaxies, a population recently discovered
in high-density regions (Drinkwater et al. 2004). We can also mention the
discovery of a blue infalling group in A1367, the prototypal example
of pre-processing in the nearby universe
(Sakai et al. 2002; Gavazzi et al. 2003b; Cortese et al. 2006b). The low-velocity dispersion
within a group infalling into the main cluster makes gravitational
interactions very efficient in perturbing galaxies before they
become real members of A1367. These perturbations produce small
condensations of matter that might later evolve as independent
entities and thus be progenitors of dwarf cluster galaxies.

\section{Modeling}\label{s4}

\subsection{Cosmological simulations}\label{MODELLIspsCOSMOLOGICI}

Cosmological models of galaxy evolution indicate that galaxies are
formed from the condensation of gas within dark matter halos. In a
hierarchical formation scenario, small structures are formed first
and later merged to give birth to massive objects. By cooling, the
gas conserves its angular momentum leading to the formation of
rotating systems. The violent interactions associated with merging
events heat the systems, forming bulges and ellipticals. In this
bottom-up formation scenario, galaxies now belonging to rich
clusters might have suffered the effects of different environments
during their life (pre-processing; Dressler 2004). They might have
been members of small groups where gravitational interactions with
other members were important before entering the evolved cluster
(Boselli \& Gavazzi 2006).

As described in De Lucia (2011), in cosmological simulations the
physical evolution of galaxies is reproduced using various
techniques. Hydrodynamical simulations mimic the evolution of the
gaseous component modeling different physical processes such as gas
cooling, star formation, and feedback. Because of this complex
approach, these simulations are limited by spatial and mass
resolution (e.g. Berlind et al. 2005). In semi-analytic models of
galaxy formation (SAM), the link between the evolution of dark
matter halos traced by high-resolution N-body simulations and the
barionic matter is accomplished using simple physical prescriptions,
without following explicitly the coupled evolution between structure
formation and gas physics through the numerical integration of
hydro/gravity equations. Nevertheless this (computationally cheap)
approach has the advantage of covering a significantly larger range
in stellar mass and spatial resolution and thus is well suited for
the study of the evolution of dwarf systems in different
environments as those analyzed in this work.

Beside ``harassment'' (Moore et al. 1996), other investigations of
the effects of the cluster environment in cosmological
hydrodynamical simulations have been undertaken by McCarthy et al. (2008). In
this work the authors simulated the stripping\footnote{There is a
clear inconsistency in the definition of ram pressure and starvation
in cosmological simulations and semi-analytic models with respect to
the observation and the simulations of single galaxies in local
clusters. Cosmologists indistinctly define ram pressure, starvation
or strangulation the removal of the hot gaseous halo of satellite
galaxies entering the extended gas halo surrounding groups and
clusters. In the study of nearby objects, ram pressure stripping is
generally referred to the stripping of the cold gas component
exerted by the hot cluster intergalactic medium on galaxies moving
at high velocity. In the same studies, starvation or strangulation
refers to the gas consumption via star formation of galaxies once
the infall of pristine cold gas is stopped.} of the hot halo of
satellite galaxies entering massive halos typical of groups and
clusters. These simulations indicated that an important fraction of
the hot gas of satellite galaxies ($\sim$30\,\%) is still at place
10~Gyr after the beginning of the interaction. More recently, Bahe et al. (2012) and Wetzel et al. (2013) concluded that the confinement
pressure exerted by the intracluster medium is not sufficient to
significantly decrease the impact of the ram pressure exerted on the
hot gas halo of satellite galaxies. Using the same cosmological
simulations, the same team has also shown that ram pressure
stripping exerted by the extended gas halo surrounding groups and
clusters is sufficiently strong to strip the hot gas atmosphere of
infalling galaxies up to $\sim$5 times the virial radius
(Bahe et al. 2013). These results are consistent with the
simulations of Cen et al. (2014).

The most recent semi-analytic models of galaxy evolution  arrive to
reproduce massive and dwarf galaxies down to stellar masses of
$\sim$10$^{7.5}$ (Guo et al. 2011, 2013) and sample a large
range in environments, from the field to massive clusters analog to
Coma in the local universe (some $10^{15}\,\hbox{M}_{\odot}$). Only
some of the most recent models, however, have implemented tuned
recipes such as those proposed by McCarthy et al. (2008) to accurately
reproduce the effects induced by the group and cluster environment
on galaxy evolution
(Font et al. 2008; Kang \& van den Bosch 2008; Kimm et al. 2009; Taranu et al. 2014; Weinmann et al. 2011; Lisker et al. 2013, the last two
especially tailored to mimic the dwarf galaxy populations in local
clusters such as Virgo and Coma). Indeed, in previous studies the
effects of the environments were simulated just by instantaneously
removing the whole hot-gas halo of galaxies once they became
satellites. This assumption makes supernova feedback sufficient to
totally sweep the cold gas disk that, in the lack of a surrounding
halo, is permanently ejected in the intracluster medium. The lack of
gas quenches the activity of star formation on very short time
scales producing a large fraction of red galaxies
(Kang \& van den Bosch 2008; Font et al. 2008; Kimm et al. 2009; De Lucia 2011). For this reason Okamoto \& Nagashima (2003)
and Lanzoni et al. (2005) identified starvation as the main physical process
responsible for the quenching of the star-formation activity of
cluster galaxies.

\begin{figure}
\includegraphics[width=1.0\textwidth]{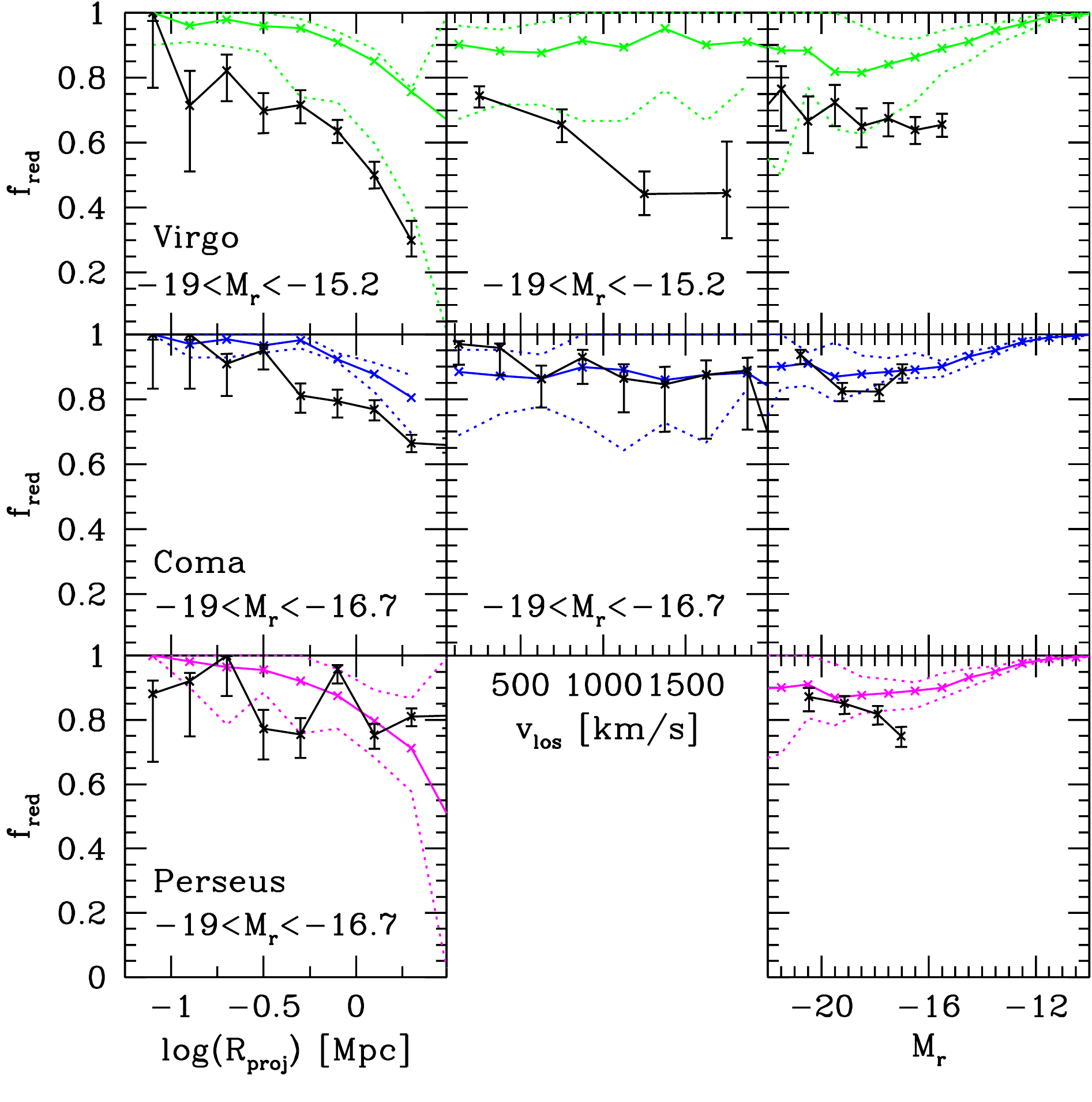} 
\caption{Fraction of red galaxies as a function of the
projected distance from the cluster center (\textit{left panels}),
line-of-sight velocity with respect to the cluster center (\textit{central
panels}), and absolute magnitude (\textit{right panels}), from Weinmann et al. (2011).
Observations of the Virgo cluster (\textit{first row}), of the Coma cluster
(\textit{central row}), and of the Perseus cluster (\textit{lower row}) are shown with
\textit{black lines} and \textit{relative errorbars}, while the results of the
semi-analytic models are the colored lines. The semi-analytic models
of galaxy evolution overpredict the number of red, quiescent
galaxies especially at low stellar masses or in clusters still under
formation (Virgo). This suggests that the gas stripping process and
the subsequent quenching of the star formation activity as predicted
by models, which happens mainly in groups during the pre-processing
of galaxies, is too rapid (Weinmann et al. 2011). Reproduced with permission of Oxford University Press}
\label{Weinmann}
\end{figure}

Despite the implementation of tuned recipes for mimicking the
stripping of the hot gas, the results of the most recent
semi-analytic models still overpredict the number and the colors of
red objects (Fontanot et al. 2009; Wang et al. 2012), suggesting that the effects
induced by the environments are overestimated (Weinmann et al. 2011;
Fig. \ref{Weinmann}).

The same models, based on the MS-II simulations of Guo et al. (2011)
including a physical prescription for the stripping of the hot gas
and supernova feedback, reproduce well the velocity dispersion and
the luminosity functions of nearby clusters, but overpredict the
dwarf to giant fraction probably because of an incorrect
prescription for tidal disruption (see however Henriques et al. 2013).

Several works have also indicated that the evolution of cluster
galaxies might have been affected during their previous membership
to groups (pre-processing)
(Book \& Benson 2010; De Lucia et al. 2012; Bahe et al. 2013; Lisker et al. 2013; Taranu et al. 2014). For this reason, gas
stripping and quenching of the star formation might have happened at
earlier epochs and outside the virial radius of the evolved cluster
(Bahe et al. 2013). The models also indicate that pre-processing
might have been more important in low-mass systems than in massive
galaxies (De Lucia et al. 2012). Models thus suggest that dwarf
elliptical galaxies in local clusters might have been processed
early and continuously in groups and cluster halos instead of being
late-type objects recently transformed in quiescent systems
(Lisker et al. 2013; Taranu et al. 2014).

\subsection{Hydrodynamical simulations of gas stripping}\label{HYDRODYNAMICAL}

Different teams developed their own hydrodynamical simulations to
reproduce the effects induced by the hostile cluster or group
environment on galaxy evolution. Tonnesen et al. (2007) tried to identify,
using cosmological simulations, which, among the different processes
acting on galaxies in rich cluster, is the dominant one. Their work
has indicated that the interaction with the hot intergalactic medium
(ram pressure stripping) is the most important at the present epoch.
Models and simulations indeed show that, although dominant in the
core of the cluster where the velocity of the galaxy and the density
of the ICM are maximal, ram pressure stripping is an ongoing process
eroding the gaseous component all over the orbit of the galaxy
within the cluster (Roediger \& Hensler 2005; Roediger \& Bruggen 2006, 2007; Bruggen \& De Lucia 2008; Kapferer et al. 2008). The simulations
of Tonnesen et al. (2007) indeed indicate that ram pressure is an important
process out to the virial radius, able to remove all the gas on time
scales of the order of $\geq$1~Gyr. The limit on the cluster region
where ram pressure stripping is active and efficient has been later
extended to $\sim$3 virial radii by Cen et al. (2014), consistent with
the most recent observations of head-tail galaxies in the periphery
of nearby clusters (see Sect. \ref{INDIVIDUAL}). Timescales of the
order of 1.5~Gyr are also obtained using 3D hydrodynamical
simulations by Roediger \& Bruggen (2007) and Cen et al. (2014) for a substantial
reduction of the total gas content of the perturbed galaxy via ram
pressure. As mentioned before, however, these hydrodynamical
simulations based on a single and homogeneous gas phase for the ISM
generally underestimate the efficiency of ram pressure stripping
(Tonnesen \& Bryan 2009). Multiphase hydrodynamical simulations indicate
that gas ablation in all its phases (from diffuse atomic to dense
molecular gas) can take place at all galactic radii if ram pressure
stripping is sufficiently strong, as it is the case in rich clusters
of galaxies such as Virgo (Tonnesen \& Bryan 2009) and as seen in the few
extreme cases of ram-pressure known to date. Modeling the effects of
ram pressure on a multiphase gas disk is still challenging since it
requires to take into account several physical effects as to resolve
hydrodynamic and thermal instabilities across a large range of
physical scales, such as heating and cooling of the ISM,
self-gravity, star formation, stellar feedback, and magnetic field.
Given the fractal distribution of the gaseous component within the
ISM, models must also predict simultaneously the gas distribution on
different scales, from giant molecular clouds to the tails whose
size can exceed the size of galaxies (Roediger 2009). Tonnesen \& Bryan (2009)
performed high-resolution (40~pc) 3D hydrodynamical simulations of
galaxies undergoing a ram pressure stripping event. They have
included in their code radiative cooling on a multiphase medium.
This naturally produces a clumpy ISM with densities spanning six
orders of magnitude, thus quite representative of the physical
conditions encountered in normal, LTGs. Their simulations show that
under these conditions the gas is stripped more efficiently up to
the inner regions with respect to an homogeneous gas. They also show
that all the low-density, diffuse gas is quickly stripped at all
radii. When the ram pressure stripping is strong there is also less
gas at high densities. The deficiency in high-density regions
results from the lack of the diffuse component feeding giant
molecular clouds (Tonnesen \& Bryan 2009). The recent work of Ruszkowski et al. (2014)
indicates that an accurate description of magnetic fields in models
does not significantly change the efficiency of gas stripping, but
only explains the formation of the filamentary structures observed
in the stripped material.

Smith et al. (2012c) studied the effects induced by ram pressure stripping
on the stellar disk and dark matter halo of cluster galaxies. Their
simulations indicate that, although the ISM--IGM interaction acts
only on the gaseous component, thanks to their mutual gravitational
interaction, the gas displacement perturbs the potential well of the
galaxy, dragging the stellar disk and the cusp of the dark matter
halo off center. The perturbation can also mildly deform and heat
the stellar disk. The same team has also simulated the effects of
ram pressure stripping on newly formed tidal dwarf galaxies
(Smith et al. 2013). Because of their low dark matter content, tidal
dwarfs are very fragile systems that can be easily perturbed and
even totally destroyed through gas and stellar loss. Consistently
with these works, Kronberger et al. (2008a) have shown that a ram pressure
stripping event can affect the 2D velocity field of galaxies
determined from emission lines. The perturbation is symmetric in a
face-on interaction, while can displace the dynamical center of the
galaxy in edge-on interactions. The interaction can also increase
the activity of star formation in the inner regions where the
compressed gas is located (Kronberger et al. 2008b).

Bekki (2009) simulated the effects of halo gas stripping in galaxies
of different mass belonging to different environments. This work has
shown that halo gas stripping on Milky Way type galaxies is very
efficient not only in massive clusters but also in small and compact
groups. The removal of gas happens outside-in, producing truncated
star forming radial profiles. The stripping process is more rapid in
dense environments than in groups, and in low-mass systems with
respect to massive galaxies. The same team has also shown that
repeated slow encounters within groups are able to transform star
forming systems into S0 (Bekki \& Couch 2011). These gravitational
interactions can trigger the formation of new stars through
repetitive starbursts, and at the same time consume the gas
reservoir producing gas-poor objects. The resulting systems have
lower velocity rotations and higher velocity dispersions than their
progenitors. The ram pressure stripping event can either enhance or
reduce the activity of star formation depending on the mass of the
galaxy, on the inclination of its disk with respect to the orbit and
the environment in which the galaxy resides (Bekki 2014).
Kawata \& Mulchaey (2008) have simulated the effects of gas stripping in groups of
virial mass $8\times10^{12}\,\hbox{M}_{\odot}$ and total X-ray
luminosity $L_{\rm X} \simeq 10^{41}\,\hbox{erg s}^{-1}$ on a
galaxy of $M_{\rm star}
\simeq 3.4\times 10^{10}\,\hbox{ M}_{\odot}$. Their N-body/smoothed
particle hydrodynamic simulations show that ram pressure stripping
is not able to remove the cold gas over the disk of the galaxy, but
is sufficient to remove the hot gas located in the halo on time
scales of $\sim$1~Gyr. Because of the lack of new gas feeding star
formation, the galaxy quenches its activity in $\sim$4~Gyr. These
conclusions, however, are not confirmed by the simulations of
Hester (2006) who identify ram pressure as the most efficient process
in stripping the cold gas component even inside groups of galaxies.
Stripping of the hot gaseous halos in clusters and groups galaxies
have also been modeled using 3D-hydrodynamical simulations by
McCarthy et al. (2008). These simulations indicate that a significant
fraction of the hot gas, $\simeq$30\,\%, can remain in place after a
10~Gyr interaction, thus in contradiction with other works.
McCarthy et al. (2008) developed simple analytic models ideally constructed
to be included in semi-analytic models of galaxy evolution to
reproduce the gas stripping process in high-density environments.

\subsection{Formation of dwarf elliptical galaxies via galaxy harassment}

Early N-body simulations specially designed to follow the
hierarchical growth of clusters and galaxy harassment and to study
the possible formation of dwarf elliptical galaxies through the
transformation of star forming disks were carried out by
Mastropietro et al. (2005). These simulations are designed for a
$\Lambda\hbox{CDM}$ cluster of $10^7$ particles with a total mass
similar to the Virgo cluster. The simulations indicate that most of
the galaxies undergo major structural modifications even at the
outskirts of the cluster, with a large fraction of them transforming
from late-type rotating systems into dwarf spheroidal hot systems on
time scales of a few Gyrs. The effects are, however, most important
in the inner 100~kpc of the cluster. The harassed galaxies are more
compact and have comparable or higher surface brightness than
unperturbed objects, probably because of the formation of bars or
grand design spiral arms. The mass loss is important and induces the
formation of round galaxies. Harassment also heats the systems,
decreasing the rotational velocity-to-velocity dispersion $v/\sigma$
ratio. Rotation is totally lost only in the most perturbed objects.

\begin{figure}
\includegraphics[width=0.7\textwidth]{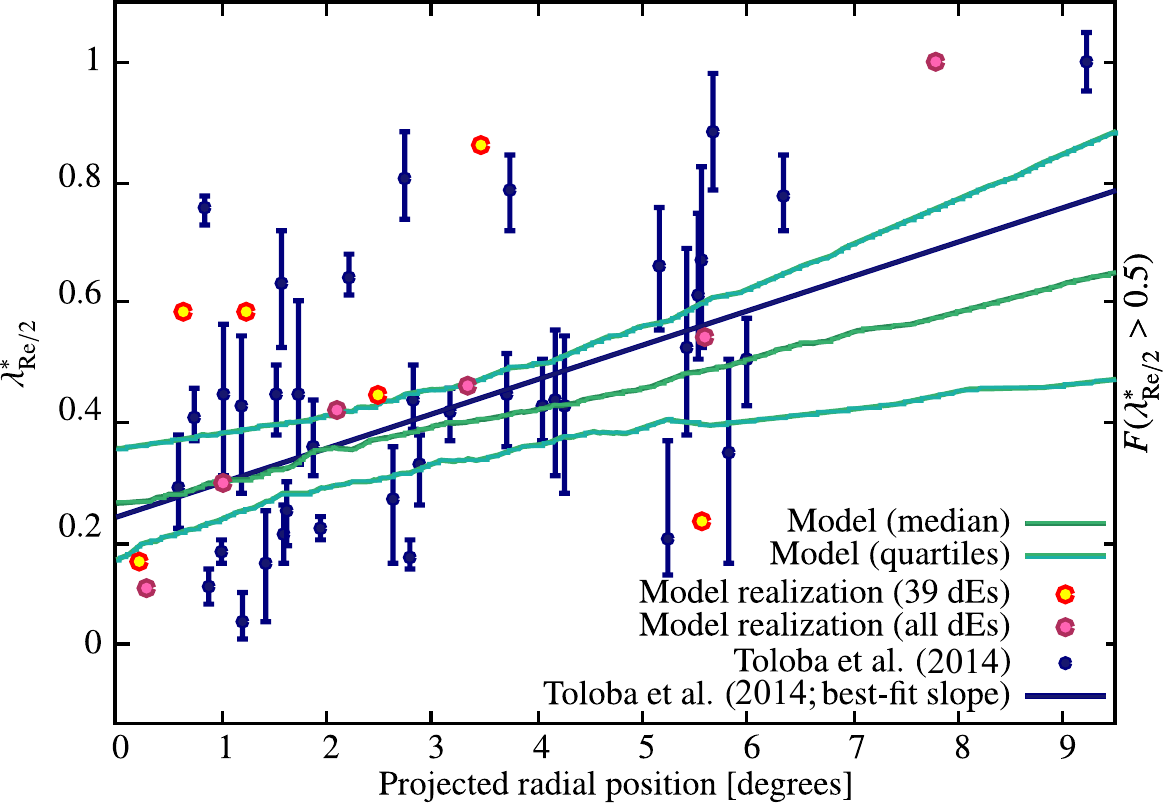}
\caption{The distribution of the angular momentum of dEs as
a function of the projected distance from M87 (from Benson et al. 2014). \copyright\ AAS. Reproduced with permission}
\label{benson}
\end{figure}

More recently, Aguerri \& Gonzalez-Garcia (2009) have developed high-resolution N-body
simulations to test the tidal stripping scenario for the formation
of dE. These simulations studied the perturbation induced to disk
galaxies with different bulge-to-disk ratios. They show that, while
the bulge is only marginally affected, the disk and the dark matter
halo are efficiently perturbed in the outer parts. The scale length
of the stellar disk, for instance, can be reduced by 40--50\,\%.
After several fast encounters galaxies can lose up to 50--80\,\% of
their mass and 30--60\,\% of their luminous matter. Prograde
interactions produce stable bars, while retrograde encounters do
not. The formation of bars is more important in the absence of
bulges. The interaction heats the system decreasing its $v/\sigma$
ratio. In a recent study, Benson et al. (2014) simulated the effects of
different environmental processes (starvation, ram pressure
stripping, tidal stirring) on the evolution of dwarf galaxies in
Virgo-like clusters. These simulations were principally focused in
reproducing the galaxy kinematical properties. Benson et al. (2014)
identified tidal stirring induced by the cluster halo on disk
galaxies as the most probable process able to reproduce the observed
gradient of the angular momentum of dEs with the clustercentric
distance to M87 at the center of the Virgo cluster (see Fig.
\ref{benson}).

\subsection{Spectro-photometric models of galaxy transformation}\label{MODELspsSPECTRO}

Boselli et al. (2006, 2008a,b) have developed 2D
chemo-spectrophotometric models of galaxy evolution especially
designed to reproduce the perturbations induced by the hostile
environment on star forming disks infalling for the first time in
rich clusters. These models reproduce two different kinds of
perturbations: ram pressure stripping and starvation. The models are
based on the chemo-spectrophotometric models of galaxy evolution of
Boissier \& Prantzos (2000) where disk galaxies are formed in a
dark matter halo and form stars following a Schmidt law modulated by
the rotation of the galaxy (Boissier et al. 2003). Calibrated on the
Milky Way, the models have two free parameters, the spin parameter
$\lambda$, which is a dimensionless measurement of the specific
angular momentum, and the rotational velocity $V_{\rm C}$.
Starvation is simulated simply stopping the infall of fresh gas,
that the model requires for unperturbed systems to reproduce the
observed color and metallicity gradients of nearby
galaxies.\footnote{We stress that this definition of starvation
differs from the one originally proposed by Larson et al. (1980), where the
galaxy quenches its activity of star formation once the gas of its
halo, generally feeding the disk in unperturbed systems, is removed
during the interaction with the hostile environment. In the Boselli
et al. models, starvation is a passive process where star formation
decreases after gas consumption because of the lack of the infall of
pristine gas from the surrounding environment.} A stripping event is
modeled considering that ram pressure has an intensity varying along
the orbit of the galaxy within the cluster, as predicted by the
scenario of Vollmer et al. (2001) explicitly designed for Virgo.

\begin{figure}
\includegraphics[width=0.7\textwidth]{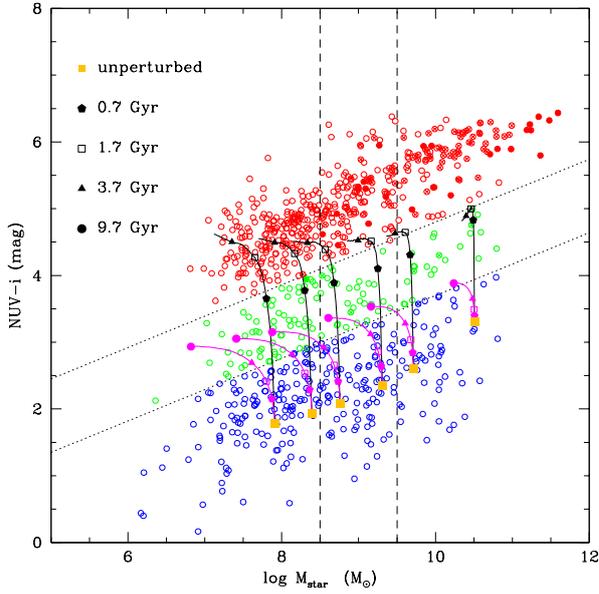}  
\caption{The extinction corrected $NUV-i$ (AB system) vs.
$M_{star}$ relation for all galaxies of the sample, from
Boselli et al. (2014a). Filled symbols are for slow rotators, crosses for
fast rotators. \textit{Symbols} are color coded for dividing objects in the
blue cloud form the green valley and the red sequence. The large
\textit{orange filled squares} indicate the models of unperturbed galaxies of
spin parameter $\lambda=0.05$ and rotational velocity 40, 55, 70,
100, 130, 170, and $220\;\hbox{km s}^{-1}$. The \textit{magenta lines}
indicate the starvation models. The \textit{black lines} show the ram
pressure stripping models. \textit{Different symbols along the models}
indicate the position of the model galaxies at a given look-back
time from the beginning of the interaction. Courtesy of ESO}
\label{modellisam}
\end{figure}

These models have been first tuned on a well-known anemic Virgo
cluster galaxy undergoing a ram pressure stripping event, NGC 4569
(M90, Boselli et al. 2006), and then extended to dwarf galaxies in the
Virgo cluster (Boselli et al. 2008a,b). They show that only ram
pressure stripping can reproduce the truncated radial profiles
observed in the gaseous component and in the young stellar
populations of NGC 4569, ruling out the starvation scenario. They
also show that ram pressure stripping can efficiently remove, on
very short time scales ($\simeq$150~Myr) all the gas in low-mass
star forming systems. The lack of gas induces a quenching of the
star-formation activity, transforming gas-rich star-forming systems
into gas-poor quiescent objects (Fig. \ref{modellisam}). This
transformation happens on time scales of the order of 0.8--1.7~Gyr.
The chemical, structural, spectrophotometric properties of the
transformed galaxies are very similar to those of dwarf elliptical
galaxies (Boselli et al. 2008a,b). This transformation process
would imply that the overall cluster and field luminosity functions
are similar, with a mutual inversion of the faint end slope of blue,
star forming systems, frequent in low-density regions, and red,
quiescent objects, abundant in the core of the cluster, as indeed
observed in Virgo and Coma. The models predictions are also
consistent with the observed properties of the globular clusters of
dwarf ellipticals and their expected progenitors (Boselli et al. 2008a;
see however Sanchez-Janssen \& Aguerri 2012). These results consistently indicate
that ram pressure stripping is able to explain the formation of the
faint end of the red sequence characterizing rich clusters of
galaxies.

\begin{figure}
\includegraphics[width=0.3\textwidth] {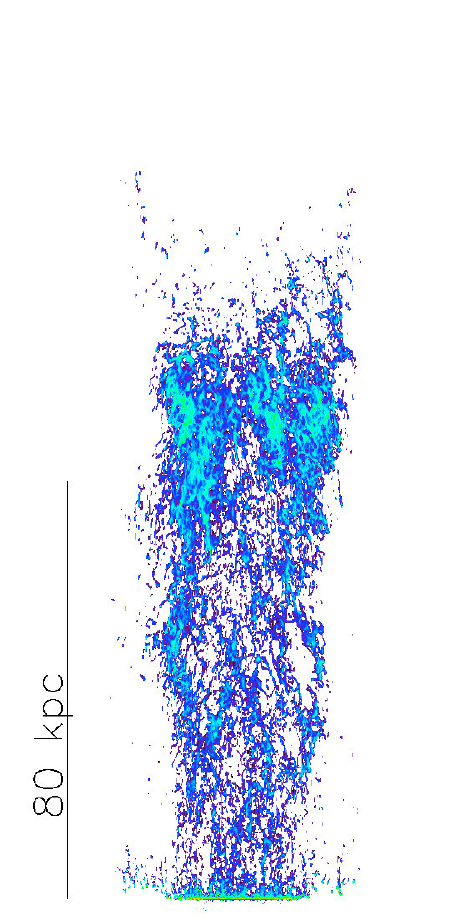}    
\includegraphics[width=0.17\textwidth]{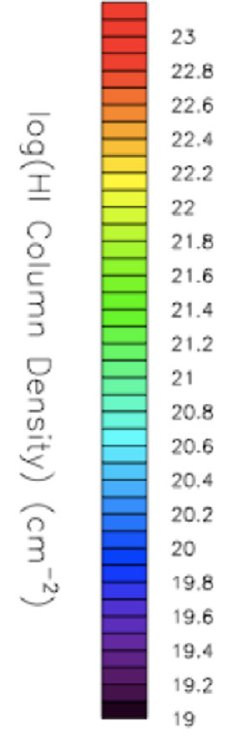}    
\includegraphics[width=0.3\textwidth] {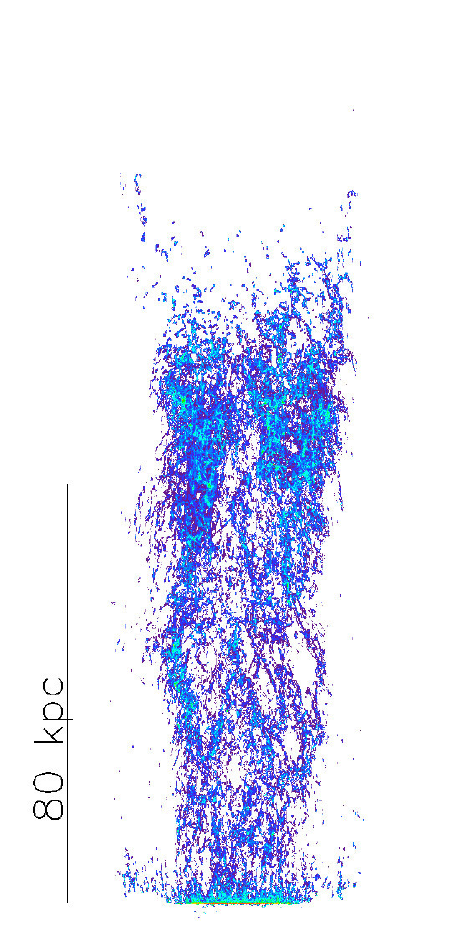}    
\includegraphics[width=0.17\textwidth]{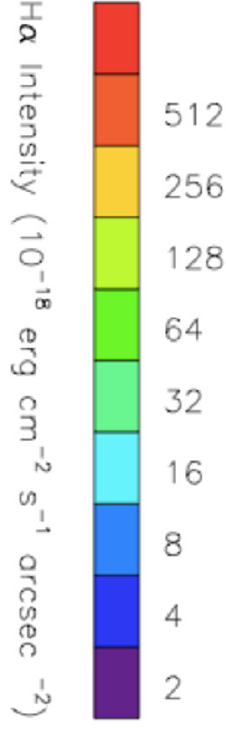}\\  
\caption{Projections of the HI column density (left) and of
the $\hbox{H}\alpha$ surface density (right) in the ram pressure
stripped gas for a model galaxy with star formation and feedback,
adapted from Tonnesen \& Bryan (2012). A small amount of star formation can take
place in the highest column density regions of the HI gas stripped
during the interaction out to 80 kpc from the disk of the galaxy. Reproduced with permission of Oxford University Press}
\label{SFstrippedgas}
\end{figure}

\subsection{Formation of dwarf galaxies in stripped material }\label{EXTRAPLANAR}

N-body/hydrodynamic simulations of ram pressure stripping events on
star forming galaxies in rich clusters show the formation of
extended gas tails (350~kpc) in the direction opposite to the motion
of the galaxy within the intergalactic medium
(Kronberger et al. 2008b; Kapferer et al. 2009). These gas tails are similar to those
observed in the periphery of A1367 and Coma by Gavazzi et al. (2001),
Yoshida et al. (2008); Yagi et al. (2010), Fossati et al. (2012) in $\hbox{H}\alpha$, by
Scott et al. (2012) in HI and in X-rays in A3627 (Sun et al. 2010). The
same simulations also predict an increase of the total
star-formation activity of the galaxy by an order of magnitude,
95\,\% of which produced in the tail of diffuse gas
(Kapferer et al. 2009). These results were later questioned by Tonnesen \& Bryan (2012) who showed, using high-resolution adaptive mesh
simulations, that gas stripping produces a truncation of the star
forming disk on time scales of a few hundreds million years. They
show that there is a moderate increase of the star formation in the
bulge but without a starburst phase. They also show that the star
formation in the tail is low, and any contribution to the
intercluster light, if present, is likely to be very small (see Fig.
\ref{SFstrippedgas}). The difference in the results between Tonnesen \& Bryan (2012) and Kapferer et al. (2009) comes mainly from the nature of
the two different codes and only marginally on the adopted
conditions (Tonnesen \& Bryan 2012). Consistently with Tonnesen \& Bryan (2012), Yamagami \& Fujita (2011) predict that in the absence of magnetic
field, Kelvin--Helmholtz instabilities destroy molecular clouds,
preventing the formation of new stars in the tails of stripped gas.

Observations seem more consistent with the prediction of Tonnesen \& Bryan (2012) and Yamagami \& Fujita (2011). Indeed, there are only a few
examples of cluster galaxies with major ongoing star-formation
events in the stripped material far from the galaxy disk. Typical
examples are IC 3418 in Virgo (Hester et al. 2010; Fumagalli et al. 2011; Kenney et al. 2014),
ESO137-001 in A3627 (Sun et al. 2007; Jachym et al. 2014; Fumagalli et al. 2014), and two
massive objects in two clusters at $z \simeq 0.2$
(Cortese et al. 2007). The extended tails of ionized gas observed in
the $\hbox{H}\alpha$ images of galaxies in nearby clusters
(Gavazzi et al. 2001; Kenney et al. 2008; Yoshida et al. 2008; Yagi et al. 2010; Fossati et al. 2012; Jachym et al. 2014) are only
marginally associated with major star-formation events in the tails.
In the Virgo cluster there are also evident cases of galaxies with
extended HI tails but without any associated star-formation event
(Boissier et al. 2012), or objects where the outplanar star-formation
process is happening only very close to the galaxy disk
(Abramson et al. 2011).

\section{Observations vs. models: a new evolutionary picture}\label{s5}

\subsection{The Virgo cluster}

The comparison of the observational results obtained so far with
model predictions can be used to reconstruct the evolutionary
picture that gave birth to the dwarf quiescent galaxies inhabiting
nearby clusters that form the faint end of the red sequence.
Multifrequency data are critical for this purpose since they provide
us with a complete and coherent view of the undergoing process and
of its effect on galaxy evolution (e.g. Boselli 2011). We can
first apply the exercise to the Virgo cluster. Here, indeed,
multifrequency data spanning the whole range of frequencies are
available for galaxies down to stellar masses of ${\sim}
10^7\,\hbox{M}_{\odot}$. At the same time, most of the available
models and simulations have been carried out with the aim of
reproducing the physical conditions undergone by galaxies located in
clusters with characteristics similar to those of Virgo.

Observations and models consistently indicate that dwarf elliptical
galaxies might have been formed by the transformation of
low-luminosity late-type systems recently entered the Virgo cluster
once they have lost their gaseous content during the interaction
with the hostile cluster environment, as first proposed by
Boselli et al. (2008a,b). The most plausible physical process
responsible for this transformation is the ram pressure $\rho_{\rm
ICM} V^2$ exerted by the hot and dense intracluster medium (where
$\rho_{\rm ICM}$ is its density) on galaxies moving at high velocity
($V$) within the cluster. Both models and observations indicate that
this ram pressure is able to overcome the gravitational forces
keeping the gas anchored to the potential well of the galaxy, $G
\Sigma_{\rm gas} \Sigma_{\rm star}$, where $G$ is the gravitational
constant and $\Sigma_{\rm gas}$ and $\Sigma_{\rm star}$ are the
gaseous and stellar surface density, respectively (Gunn \& Gott 1972).
This is particularly true in dwarf systems, where the gravitational
potential well is shallower than in massive galaxies. Ram pressure
stripping easily removes both the atomic (e.g.
Cayatte et al. 1990; Solanes et al. 2001; Gavazzi et al. 2005, 2013a) and in the strongest cases
the molecular gas phases (Fumagalli et al. 2009; Boselli et al. 2014b), as well as the
associated dust (Cortese et al. 2010, 2012a,b). Models and
observations consistently indicate that ram pressure stripping is
efficient even outside the virial radius of the cluster
(Roediger \& Hensler 2005; Roediger \& Bruggen 2006, 2007; 
Boselli \& Gavazzi 2006; Tonnesen \& Bryan 2009; Cen et al. 2014). The higher dispersion
in the velocity distribution of star-forming systems with respect to
that of massive ellipticals definitely shows the presence of gas
rich systems infalling in the Virgo cluster
(Boselli et al. 2008a, 2014a).

The stripping process is quite rapid since it is able to remove most
of the gas content on time scales of $\sim$1.5~Gyr
(Roediger \& Bruggen 2007; Tonnesen \& Bryan 2009). This time scale even reduces to 100--200~Myr
in dwarf systems because of their weak restoring forces
(Boselli et al. 2008a). Some gas retention, with the associated dust
(de Looze et al. 2010, 2013), might be present in the core of the
stripped galaxies where the gravitational potential well is at its
maximum. This gas might feed star formation up to more recent
epochs, and thus be at the origin of the blue centers observed in
the most massive dwarf ellipticals (${\sim} 10^{9.5 - 10} M_\odot$) 
(Lisker et al. 2006b; Boselli et al. 2008a; Michielsen et al. 2008; Paudel et al. 2011). Indeed, gas
removal is an outside-in process able to perturb only the gaseous
component not affecting the stellar component, if not indirectly via
the star-formation process (Boselli et al. 2006). The stripped galaxies
often show in their morphology several remnants of their past spiral
origin, such as grand design spiral arms, disks, and bars
(Jerjen et al. 2000; Geha et al. 2003; Lisker et al 2006a,b; Lisker \& Fuchs 2009). They also conserve
their angular momentum, showing rotation curves similar to those
observed in spiral galaxies of similar luminosity
(Toloba et al. 2009, 2011, 2014b). Because
of the lack of gas, the star formation stops, making galaxies redder
and redder (Boselli et al. 2008a, 2014a; Cortese \& Hughes 2009; Hughes \& Cortese 2009; Gavazzi et al. 2013a). When
plotted in a color magnitude diagram, they leave the blue sequence,
cross the green valley and become red, quiescent systems. The time
scale for this transformation is only slightly longer than the time
scale for the gas stripping (Boselli et al. 2008a, 2014a). A rapid
quenching of the star-formation activity is also indicated by the
presence of several dwarf ellipticals such as VCC 1499 (see
Sect.~\ref{s1}) in a post-starburst phase.\footnote{The definition
of post-starburst does not necessarily imply there was a
particularly acute starburst phase, but that a normal star-formation
phase was abruptly interrupted.} The structural properties of the
newly formed dwarf elliptical galaxies, as well as those of their
relative globular cluster content, are fully consistent with those
of their star forming progeny (Boselli et al. 2008a,b; see,
however, Sanchez-Janssen \& Aguerri 2012). There is strong and consistent evidence
that the kinematical, structural, and spectrophotometric properties
of dwarf ellipticals within Virgo tightly depend on their position
within the cluster. Indeed, the most roundish (Lisker et al. 2009),
pressure-supported systems (Toloba et al. 2009, 2011) dominated by
old stellar populations in their center
(Michielsen et al. 2008; Paudel et al. 2010a, 2011) are located close to the cluster
core, while the disky, rotationally supported systems, dominated by
younger stellar populations in their center, are predominantly
situated in the outskirts of the Virgo cluster.

The statistical properties of the Virgo cluster galaxies are also
consistent with this picture. The shape of the optical and UV
luminosity functions, and in particular their faint end slope, are
very similar to those observed in the field once low surface
brightness galaxies are considered (Rines \& Geller 2008; Boselli et al. 2011; Ferrarese
et al., in prep.). There is just an inversion of the relative
contribution of star forming and quiescent galaxies, the former
dominating in low-density regions, the latter typical of rich
environments, as indeed expected in such a scenario. We recall,
however, that both the GUViCS UV (Boselli et al. 2011) and the NGVS
optical (Ferrarese et al., in prep.) luminosity functions have been
determined for the central regions of the cluster and thus do not
sample any possible radial variation (steepening of the faint end in
the cluster periphery) already observed in other nearby clusters
(Popesso et al. 2006; Barkhouse et al. 2007, 2009).

The comparison of models with observations strongly favors a soft
interaction of galaxies with the hot intergalactic medium (ram
pressure) rather than more violent phenomena such as tidal
interactions or galaxy harassment for several reasons. The most
important one is that gravitational interactions remove a
significant fraction of the stellar component, producing lower
luminosity objects with systematically truncated stellar disks
(Mastropietro et al. 2005; Aguerri \& Gonzalez-Garcia 2009). The interaction
would also significantly increase stochastic motions and reduce
ordered motions, thus finally decreasing $v/\sigma$
(Mastropietro et al. 2005; Aguerri \& Gonzalez-Garcia 2009). The observed structural
properties of dwarf galaxies are more consistent with those of
star-forming systems stripped of their gas by ram pressure stripping
rather than those of harassed galaxies (Boselli et al. 2008b).
Furthermore, efficient gravitational perturbations are quite rare
given the high-velocity dispersion of the cluster. We thus expect
that harassment requires long time scales for galaxy transformation,
longer than the observed evolution with redshift of the faint end of
the CMR (see Sect. \ref{EVOLUZIONEz}). We can also expect that
strong gravitational interactions, through tidal disruption and
tidal galaxy formation, should significantly modify the shape of the
luminosity function (Popesso et al. 2006; Barkhouse et al. 2007; de Filippis et al. 2011). As previously
mentioned, the observations do not show any systematic difference in
the luminosity function of Virgo and the field. At the same time,
the detailed observations of the few cluster galaxies with clear
signs of an undergoing perturbation do not show the formation of
intermediate mass objects, but rather the formation in the stripped
material of very small, compact blobs of stellar mass ${\sim}
10^3{-}10^6\,\hbox{M}_{\odot}$
(Yoshida et al. 2008; Fumagalli et al. 2011; Arrigoni Battaia et al. 2012; Yagi et al. 2013). This phenomenon of galaxy
disruption/formation does not seem to be very frequent and is thus
quite unlikely that it is able to modify the shape of the luminosity
function in the observed stellar mass range. Furthermore, the output
of this process are not low surface brightness, extended systems as
those dominating the faint end of the luminosity function and of the
CMR, but rather compact sources much more similar to UCD galaxies
(yet of lower mass; Arrigoni Battaia et al. 2012).

All the evidence described above, however, does not necessarily rule
out the hypothesis that gravitational perturbations might play an
important role on those galaxies that entered the cluster long time
ago or have been even formed within it. On long time scales, galaxy
harassment can heat the perturbed systems, producing roundish,
pressure-supported objects such as those observed in the core of the
cluster by Lisker et al. (2007, 2009). These galaxies, that have on
average a velocity distribution similar to the Gaussian distribution
drawn by the massive virialized ellipticals, are indeed
characterized by older stellar populations
(Michielsen et al. 2008; Paudel et al. 2010a, 2011) and lower $v/\sigma$
(Toloba et al. 2009, 2011; Boselli et al. 2014a) than their disky dominated
counterparts at the periphery of the cluster. Furthermore, the
beautiful systematic work of Lisker, Janz and collaborators, have
undoubtedly shown that dwarf elliptical galaxies are not an
homogeneous class of objects, but rather present different
structures remnants of different possible origins
(Lisker 2009; Janz et al. 2012, 2014). We can also add that
chemo-spectrophotometric models of galaxy evolution expressly
conceived to take into account the perturbations induced by the
cluster environment ruled out the hypothesis that starvation, or
strangulation, is at the origin of the observed properties of dwarf
(and giant) gas-poor galaxies in clusters
(Boselli et al. 2006, 2008a). We recall, however, that  these models
simulate starvation by stopping the infall of pristine gas, as
already noted in Sect. \ref{MODELspsSPECTRO}. In the original
definition of Larson et al. (1980) starvation is a more aggressive process,
where the interaction of galaxies with the hot and dense
intracluster medium removes the hot gaseous halos of galaxies, a
process that is much more similar to the ram pressure stripping
considered here.

A still open question is the range of stellar mass where this
transformation process is efficient. Given the tight relation
between stellar mass and restoring forces on galaxy disks, we expect
the gas stripping process to be less efficient in massive objects
(Boselli et al. 2008a). Here an important fraction of the atomic and
molecular gas can still be retained on the inner disk, feeding star
formation. Total gas stripping might occur only after several
crossings of the cluster (Boselli et al. 2014b). It is thus possible in
this scenario that the most massive ETGs inhabiting the cluster have
been formed through major merging events, as indeed suggested by
kinematical arguments (Cappellari et al. 2011a,b; Boselli et al. 2014a). It would
be interesting to extend the studies of the main scaling relations
and see whether the possible presence of strong discontinuities, as
claimed by Kormendy et al. (2009) and Kormendy \& Bender (2012), or continuity
(Graham \& Guzman 2003; Gavazzi et al. 2005; Ferrarese et al. 2006; Misgeld et al. 2008, 2009; Misgeld \& Hilker 20011; Smith Castelli et al. 2013) can shade
light on this specific point.

Within this evolutionary picture we see only one apparent
inconsistency. The cosmological simulations by Weinmann et al. (2011) and
Lisker et al. (2013) especially tuned to reproduce the properties of the
Virgo cluster seem to indicate that dwarf elliptical galaxies in
local clusters such as Virgo might have been processed early and
continuously in groups and in the cluster halo and are thus not LTGs
recently transformed in quiescent systems. There are, however, a few
indications that these semi-analytic models fails to reproduce
several statistical properties of local clusters. Among these, the
most evident is that they overestimate the color and the number of
red objects, suggesting that in semi-analytic models the effects of
the environment are still not optimized (Weinmann et al. 2011).

\subsection{Other clusters}

Are these results consistent with what observed in other nearby
clusters? The multifrequency observations of Coma, A1367, and the
Shapley supercluster confirm this scenario. All observations
indicate that the fraction of gas-rich star-forming galaxies
continuously increases with clustercentric distance up to a few
virial radii (Rines et al. 2005; Gavazzi et al. 2013b), and that, on average,
gas-poor late-type systems typical of high-density environments
populate the green valley in between the blue cloud and the red
sequence. The observations also indicate that dwarf elliptical
galaxies in the infalling regions at the periphery of these rich
clusters have, on average, younger stellar populations than those
located close to the cluster core
(Smith et al. 2006, 2008, 2009, 2012a), often with characteristic
spectra indicating a recent abrupt truncation of their
star-formation activity (Poggianti et al. 2004; Gavazzi et al. 2010). There is also
evidence that some of these low-mass quiescent systems have spiral
arms (Graham et al. 2003) and residual star formation in their center
(Haines et al. 2008). The physical properties of star-forming and
quiescent dwarf galaxies in other nearby clusters are thus similar
to those observed in Virgo.

As for Virgo, we expect that ram pressure stripping is the most
probable process responsible for the observed trends. In rich
clusters such as Coma the density of the intracluster medium is, on
average, a factor of ten higher than in less relaxed clusters such
as Virgo. The velocity dispersion of galaxies also increases with
the mass of the cluster; it is thus natural that the efficiency of
ram pressure stripping, which varies as $\rho_{\rm ICM} V^2$, is
higher in these environments than in Virgo. On the contrary, the
increase of the velocity dispersion of the cluster makes
gravitational perturbations less efficient just because the time
during which two galaxies can interact becomes shorter
(Boselli \& Gavazzi 2006). There are also a few spectacular observations that
confirm this result. Besides the aforementioned FGC1287 in the outskirts of A1367 (Scott et al. 2012), 
the detection of several gas-rich galaxies at $\sim$ 1 virial radius with extended H$\alpha$ tails in Coma and A1367
(Gavazzi et al. 2001; Yagi et al. 2007, 2010; Yoshida et al. 2008, 2012; Fossati et al. 2012) has indeed
shown that gas stripping via ram pressure can also remove the
ionized gas component. At the same time, the X-ray and CO
observations of ESO 137-001 in A3627 have indicated that the hot and
the molecular gas components can also be removed (Sun et al. 2007;
Jachym et al. 2014).  Ram pressure stripping seems thus more
important than previously thought. This statement is consistent with
the most recent hydrodynamical simulations indicating that the ram
pressure stripping process is more efficient whenever a multiphase
ISM medium (atomic plus molecular gas) is considered
(Tonnesen \& Bryan 2009). Other simulations also indicate that ram pressure
stripping is efficient  to remove gas in galaxies up to $\sim$3
virial radii. Altogether these results have proven that ram pressure
stripping is still active well outside the virial radius, making it
the most plausible process responsible for the radial variation of
the gas stripping and following quenching of the star-formation
activity observed in nearby clusters.

Not all pieces of evidence, however, rule out gravitational
interactions as a possible process able to modify the evolution of
cluster galaxies. In a recent work, for instance, Poggianti et al. (2013)
discovered a population of compact objects analogue to those found
at high redshift, three times more frequent in clusters than in the
field, accounting for $\sim$12\,\% of the massive ($M_{\rm star}
> 10^{10}\,\hbox{M}_{\odot}$) galaxy population in nearby clusters.
These objects, mainly lenticulars or ellipticals, have been probably
shaped by gravitational interactions able to truncate stellar disks.
There are also some inconsistencies with the observation of the
optical luminosity function of nearby clusters. If the results of
Popesso et al. (2005) and Cortese et al. (2008b), indicating that the cluster
luminosity function is significantly steeper than in the field,
(with a faint end slope in the fitted Schechter function of ${-}2.1
\leq \alpha \leq {-}1.6$) are confirmed, other gravitational
processes such as harassment and tidal disruption should be invoked.
We recall, however, that these results should still be confirmed
observationally, in particular because they might suffer from the
quite uncertain background subtraction technique (Rines \& Geller 2008).
In any case, it is also conceivable that, given the very different
nature of Virgo and Coma or other massive clusters, the former being
spiral rich and still under formation, the latter quite relaxed and
spiral poor, the relative weight of the different physical process
acting on galaxies might significantly change. If observations and
simulations mainly suggest that the dynamical interactions between
galaxies and the hot intracluster medium are probably modulating the
evolution of galaxies in the present epoch, this might not have been
the case in the past, when pre-processing was probably more
important (e.g. Dressler 2004; Boselli \& Gavazzi 2006). Gravitational perturbations in
infalling groups are rare in the present epoch: the blue infalling
group in A1367 is indeed the only known case in the local universe
(Sakai et al. 2002; Gavazzi et al. 2003b; Cortese et al. 2006b). Furthermore, the luminosity
function, in particular in the optical bands, gives a view of the
cumulative evolution of galaxies and thus might be not
representative of a recent evolution.

\subsection{Lower density environments}

It is now interesting to determine the range of galaxy density
within which this evolutionary process holds. The works of
Haines et al. (2008), Gavazzi et al. (2010, 2013b),  and Rasmussen et al. (2012a) are
crucial for this purpose. In particular, Gavazzi et al. (2010) have shown
(see Fig. \ref{Gavazzispsdensity}) that while the bright end of the
red sequence is already fully defined in all kinds of environments,
the faint end is present only for densities $\delta_{1,1{,}000} >4,$
where $\delta_{1,1{,}000}$ is the 3D density contrast defined as
\begin{equation}
{\delta_{1,1{,}000} = \frac{\rho -\langle  \rho\rangle }{{\langle
\rho\rangle }}}
\end{equation}
\noindent $\rho$ is the local number density and $\langle
\rho\rangle  = 0.05$~gal~${\rm h}^{-1}\,\hbox{Mpc}^{-3}$ is the mean
number density measured in the Coma/A1367 supercluster
region.\footnote{The local number density $\rho$ around each galaxy
is measured within a cylinder of radius 1~$h^{-1}$~Mpc and
half-length $1{,}000\,\hbox{km\,s}^{-1}$.} This threshold in local
number density of $\delta_{1,1{,}000} > 4$ roughly corresponds to
the density observed in groups with more than 20 objects and
velocity dispersion of ${\sim }200\,\hbox{km s}^{-1}$. The lack of
red sequence faint quiescent galaxies in the field has been later
confirmed on strong statistical basis by Wilman et al. (2010) and
Geha et al. 2012. Kilborn et al. (2009), Fabello et al. (2012), Gavazzi et al. (2013b), and
Catinella et al. (2013) have shown that LTGs in medium-density environments
such as groups also suffer from gas deficiency. The lack of gas
quenches the activity of star formation, making, on average, redder
galaxies (Gavazzi et al. 2013b). Wilman et al. (2010), using a large sample
of SDSS galaxies in different density environments, have shown that
galaxies become red only once they have been accreted into halos of
a certain mass. Rasmussen et al. (2012a), on the other hand, have clearly
shown that the quenching of the star-formation activity is stronger
in galaxies with stellar mass $M_{\rm star} <
10^{9.2}\,\hbox{M}_{\odot}$.

\textit{ROSAT} and \textit{Chandra} observations have shown that
about half of the optically selected nearby groups are characterized
by a diffuse X-ray emission, similar to clusters of galaxies
(Mulchaey \& Zabuldoff 1988; Mulchaey 2000; Mulchaey et al. 2003; Osmond \& Ponman 2004; 
Rasmussen et al. 2008; Sun et al. 2009). The X-ray emission,
which is typical of those groups dominated by an ETG, generally
extends out to less than half of the virial radius. They are
characterized by gas densities of $n_0 \sim
10^{-2}{-}10^{-3}\,\hbox{cm}^{-3}$ and velocity dispersions of
$\sigma \sim 150{-}400\,\hbox{km\,s}^{-1}$ (Rasmussen et al. 2008). It is
thus conceivable that ram pressure stripping is also acting on
galaxies in these medium density systems
(Kantharia et al. 2005; Verdes-Montenegro et al. 2007; Sengupta et al. 2007; McConnachie et al. 2007; Jeltema et al. 2008). Its efficiency,
however, significantly drops since both the mean density of the
intergalactic medium and the velocity dispersion are significantly
smaller than in cluster galaxies (Kawata \& Mulchaey 2008). In dwarf systems,
however, the ram pressure stripping process can still be active
given their shallow gravitational potential well, as suggested by
the results of Rasmussen et al. (2012a). Indeed, through a statistical study
of the spectroscopic properties of a large sample of nearby, ETGs,
Thomas et al. (2010) have shown that the impact of the environment on the
star-formation history of galaxies increases with decreasing galaxy
mass. At the same time, in groups the role of gravitational
perturbations can be more important than in rich systems just
because the duration of the interactions is longer
(Kilborn et al. 2009). The hydrodynamical simulations of Bekki \& Couch (2011),
indeed, have shown that repeated slow encounters within groups are
able to transform star forming systems into lenticular galaxies. It
is thus possible that the same process modulates the evolution of
lower mass systems. On the other hand, the analysis of a large
sample of galaxies in groups using SDSS data done by van den
van den Bosch et al. (2008) seems to indicate starvation or strangulation as the
most probable process quenching the activity of star formation of
late-type systems and thus forming the red sequence. This last
conclusion, however, is under debate since it is based on the
assumption that harassment and ram pressure stripping are efficient
only in massive systems, an assumption not fully supported by the
results presented in this review. We recall, however, that the
simulations of Kawata \& Mulchaey (2008) indicate that the total hot halo gas of
galaxies of intermediate mass can be removed on time scales of
$\sim$1~Gyr. The lack of new gas feeding the star-formation process
quenches the activity of these objects. We also remind that other
processes have been proposed in the literature to remove the gas and
quench the activity of star formation of dwarf systems in small
groups of galaxies similar to the Local group. Among these, we can
mention tidal stirring (Mayer et al. 2001a,b), or the
combination of tidal interactions and ram pressure stripping exerted
on dwarf satellites during their crossing of the halo of massive
galaxies (Mayer et al. 2006). Regardless of the very nature of the
gas stripping process, we expect that the lack of gas feeding star
formation quenches the activity and makes galaxies in groups redder,
thus populating the faint end of the red sequence.

\subsection{Dependence on redshift}\label{EVOLUZIONEz}

The identification of the process at the origin of the faint end of
the red sequence can be further constrained by comparing the time
scale for quenching the activity of star formation, transforming
blue galaxies into red systems, and the infall rate in nearby
clusters, with the evolution of the red sequence as a function of
redshift. This exercise has been done for the first time by
Boselli et al. (2008a) and later revisited by Gavazzi et al. (2013a,b).
Boselli et al. (2008a) used their own chemo-spectrophotometric models of
galaxy evolution to calibrate different indicators necessary to
quantify the lookback time since the beginning of the interaction.
In this way they identified four different indices: the H$\alpha$ emission
line equivalent width and the HI-deficiency parameter are sensitive
to short time scales ($\sim$200 Myr), while the equivalent width
of the H$\beta$ absorption line and the FUV-H colour index are
sensitive to significantly longer lookback times ($\sim$1 Gyr). By determining the fraction of dwarf
elliptical galaxies still undergoing the transformation process as
indicated by these four indices in the whole dE Virgo population,
Boselli et al. (2008a) deduced that the infall rate of dwarf galaxies in
Virgo is of the order of 300 objects $\hbox{Gyr}^{{-}1}$. Following
similar arguments, Gavazzi et al. (2013a) determined infall rates of
$\sim$300 to 400~$\hbox{Gyr}^{-1}$ galaxies in Virgo and of 100
galaxies $\hbox{Gyr}^{-1}$ of mass $M_{\rm star} >
10^9\,\hbox{M}_{\odot}$ in Coma. These are important rates
considering that the total number of dwarf Virgo members in the same
luminosity range is $\sim$650 and implies that the faint end of the
red sequence has been formed in the last $\sim$2~Gyr (corresponding
to $z = 0.16$ in a $H_0 = 70$\,km\,s$^{-1}\,\hbox{Mpc}^{-1},\Omega_m
= 0.3$, and $\Omega_{\lambda} = 0.7$ cosmology) if we assume that
this rate did not change significantly with time
(Boselli et al. 2008a).

\begin{figure}
\includegraphics[width=0.8\textwidth]{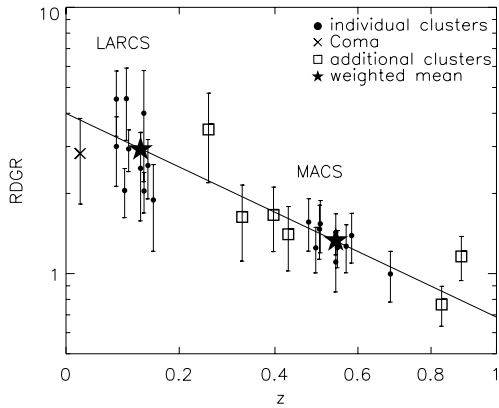} 
\caption{The variation in the red sequence of the dwarf to
giant number ratio as a function of the redshift, from
Stott et al. (2007). Dwarf elliptical galaxies have been formed mainly at
recent epochs. \copyright\ AAS. Reproduced with permission}
\label{StottspsRDTG}
\end{figure}

The most recent studies of the CMR and of the luminosity function in
clusters at different redshifts can be used for a direct comparison
with this result. Since the work of Kodama et al. (2004) and De Lucia et al. (2004)
there is a growing evidence that the fraction of luminous-to-faint
galaxies on the red sequence significantly decreased since $z = 0.8$
(Fig. \ref{StottspsRDTG};
De Lucia et al. 2007, 2009; Stott et al. 2007, 2009; Gilbank \& Balogh 2008; see however
Andreon 2008). Stott et al. (2007) determined that the number of dwarf
galaxies on the red sequence increased by a factor of 2.2 since $z =
0.5$ (4~Gyr), a number roughly consistent with that determined from
the infall rate of galaxies in Virgo (Boselli et al. 2008a; Gavazzi et al. 2013a).
Jaffe et al. (2011), by analyzing the shape of the CMR traced by galaxies
selected according to morphological criteria, have shown that the
CMR was already formed in clusters at redshift $0.4 < z < 0.8$. They
noticed, however, the presence of several low luminosity objects
with bluer colors than those of the mean CMR, indicating that these
galaxies have reached the red sequence later in time than more
massive galaxies. By comparing the fraction of red dwarf galaxies
with clear signs of a post-starburst activity (`$k+a$') with the
infalling rate, De Lucia et al. (2009) have shown that not all dwarf galaxies
have moved from the blue cloud to the red sequence after having
passed a post-starburst phase. They have thus concluded that the
transformation process that gave birth to the faint end of the red
sequence is not always rapid, but can mildly and continuously change
the galaxies properties with time. Using a different set of data of
high redshift galaxies, Bolzanella et al. (2010) suggested that the
environmental mechanisms of galaxy transformation start to be
effective only below $z = 1$. Furthermore, they indicated that the
migration from the blue cloud to the red sequence occurs on a
shorter timescale than the transformation from disk-like
morphologies to ellipticals, consistently with that observed in the
local universe (presence of spiral arms in Virgo and Coma dE, see
Sect. \ref{STRUCTURAL}) (Giodini et al. 2012; Mok et al. 2013; Vulcani et al. 2013). It has
been shown that the fraction of galaxies in dense environments at
$z= 0.7$ located in between the red and the blue sequence of the
color--magnitude relation, thus of objects probably undergoing the
transformation process, increases with decreasing luminosity
(Cassata et al. 2007), as expected in the proposed scenario. There are
also several indications that the faint end of the luminosity
function of red cluster galaxies has been formed only after $z =
0.6$ (Harsono \& de Propris 2007; Gilbank et al. 2008; Rudnick et al. 2009, 2012; Lemaux et al. 2012; see, however
Crawford et al. 2009; de Propris et al. 2013). More locally, the work of Hansen et al. (2009)
based on the photometric properties of galaxies in group and
clusters at redshift $0.1 < z < 0.3$ has shown that, while the
luminosity function of blue and red satellites is only weakly
dependent on the cluster richness for masses above $3 \times
10^{13}\,\hbox{ M}_{\odot}$, the mix of faint red and blue galaxies
changes dramatically.

Overall, although controversial results have been reported, most of
the observational evidence suggests that the faint end of the red
sequence has been formed only at relatively recent epochs. It might
be  more challenging to see whether the physical process at the
origin of this transformation in the past was the same as the one
identified in the local universe (ram pressure stripping).
Simulations indicate that the present day rich clusters have been
formed by the aggregation of smaller structures and that
$\sim$40\,\% of the galaxies in local clusters have been accreted
through groups (Gnedin 2003; McGee et al. 2009; De Lucia et al. 2012). It is thus
conceivable that galaxies have been processed during their
membership to these systems before entering in the cluster
(pre-processing, Dressler 2004; Vijayaraghavan \& Ricker 2013; Dressler et al. 2013). There is also some
evidence that clusters were denser at high $z$ than in the local
universe (Poggianti et al. 2010). It is thus possible that
gravitational interactions played a more important role in shaping
dwarf galaxy evolution in the past than in the local universe. The
short time scale for the assembly of the faint end of the red
sequence ($\sim$4~Gyr), however, favors rapid processes such as ram
pressure stripping rather than harassment, whose time scale is
longer given that multiple encounters are required. The interesting
works of Poggianti and collaborators also favor this scenario. By
studying the spectral properties of cluster galaxies at redshift
$0.4< z < 0.8$, Poggianti et al. (2009) have shown that the fraction of
post-starburst galaxies increases with the velocity dispersion of
the cluster, suggesting that the process at the origin of their
transformation is related to the intracluster medium. Consistently
with this scenario, Dressler and collaborators concluded that
massive ellipticals have been formed by early major merging
events within the groups accreting the cluster, while lenticulars
more recently by less violent processes. This view is also
consistent with the most recent multifrequency observations of the
Virgo cluster (Boselli et al. 2014a).

\begin{figure}
\includegraphics[width=1\textwidth]{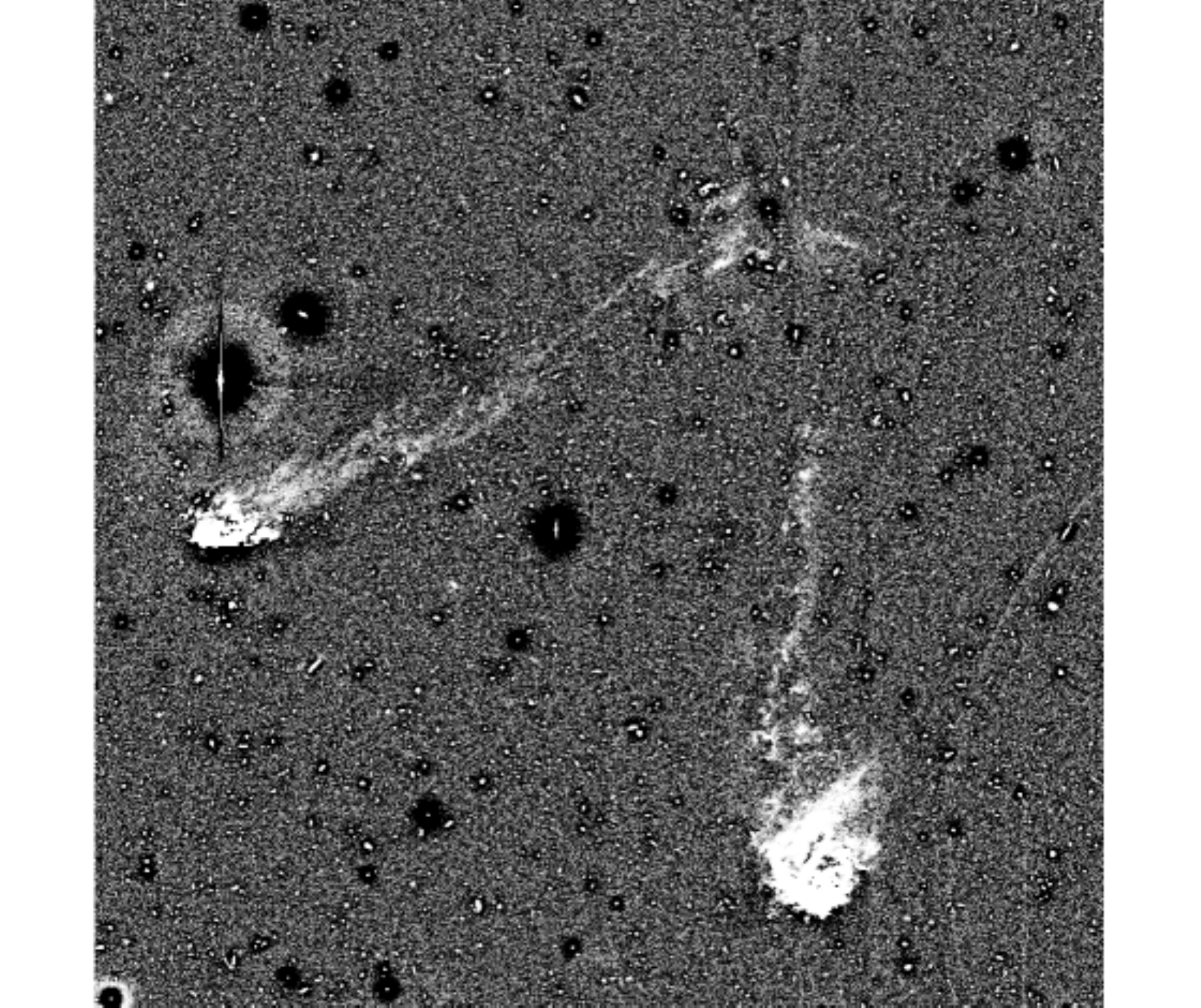} 
\caption{High contrast representation of the
$\hbox{H}\alpha$ NET  frame obtained with the Suprime-Cam at the
Subaru telescope (4 hours exposure) zoomed on a 10x10 $\rm arcmin^2$
region containing galaxies 97073 and 97079 (compare with Figure
\ref{f1}), showing tails of ionized gas exceeding 100 kpc in
length} 
\label{subaru}
\end{figure}

\section{Concluding remark}\label{s6}

In spring 2014 we took a $1\,{\rm deg}^2$ deep (4~h exposure)
$\hbox{H}\alpha$ image of A1367 using Suprime-Cam at the Subaru
telescope, in collaboration with Michintoshi Yosida and Masafumi
Yagi. Moreover, we obtained one field at the NW periphery of the
Coma cluster which, added to the two fields previously obtained by
Yagi et al. (2010) with the same instrument, brings to 1.5~sq. deg. the
area surveyed in this cluster at similar depth. Figure \ref{subaru}
speaks for itself. It contains a detail of the A1367 field, zoomed
on two of the galaxies highlighted in Fig.~\ref{f1}. The gray scale
represents the intensity of the  $\hbox{H}\alpha+\hbox{N[II]}$ lines
after a preliminary subtraction of the stellar continuum. The figure
shows ionized gas (no stars) trailing behind the two low-mass
galaxies (${\textit{M}_{\rm star}} \ 10^{9.3, 9.5} \ \hbox{M}_{\odot}$) with approximately 75 (97073)
and 100~kpc (97079) projected length. It cannot be excluded that the
two galaxies suffered from a close encounter at the location where
the two tails seem to cross each other. Nevertheless they
dramatically witness gas loss due to ram-pressure.

In total we covered 36 LTGs in A1367 and 28 in the Coma cluster. The
preliminary result of the survey is that at least 24 (perhaps 27) of
them, i.e. 40\,\%, present a cometary $\hbox{H}\alpha$ tail. In
other words, roughly one out of two of the star forming galaxies in
the two clusters shows signs of an ongoing ram pressure interaction.
This confirms that ram pressure is a quick phenomenon, that
replenishment of fresh gas-rich galaxies is currently taking place
in the clusters belonging to the Great Wall and that environmental
transformations are indeed ubiquitous. If this rate is
representative of other clusters, we predict that future
$\hbox{H}\alpha$ surveys of similar depths (few $10^{-18} \,{\rm
erg\,cm}^{-2}\,\hbox{s}^{-1}\,\hbox{arcsec}^{-2}$) of the Virgo
cluster would lead to the detection of several hundred LTGs showing
cometary structures, especially in dwarf systems. The recent
availability of $\hbox{H}\alpha$ filters matching the field of view
of MegaCam at the CFHT will make such survey feasible in the near
future. Another major breakthrough for unveiling at the same time
the kinematics and the chemistry of these systems will certainly be
provided by the wide-field integral-field units that are becoming
available at 10~m class telescopes (e.g. MUSE at VLT:
Fumagalli et al. 2014 and KCWI at Keck) and by ALMA for disclosing the
study of the gas in the molecular phase.

\begin{acknowledgements}
We wish to thank Massimo Dotti, Matteo Fossati, Michele
Fumagalli, and Elisa Toloba for their comments on the manuscript and
Yannick Roehlly for his help in the preparation of the
illustrations. G.G wishes to thank Michitoshi Yoshida and Masafumi
Yagi for their permission to use the Halpha map in
Fig.\,\ref{subaru} prior to publication. The authors would like to thank 
L. Cortese, K. Rines, R. Smith, T. Lisker, E. Toloba, S. Tonnesen, and J. Stott for allowing 
reproducing their published figures. During the writing of this review we made extensive use of the GOLDMine database
(Gavazzi et al. 2003a) and of the NASA/IPAC Extragalactic
Database (NED) which is operated by the Jet Propulsion Laboratory,
California Institute of Technology, under contract with the National
Aeronautics and Space Administration. Funding for the SDSS and
SDSS-II has been provided by the Alfred P. Sloan Foundation, the
Participating Institutions, the National Science Foundation, the
U.S. Department of Energy, the National Aeronautics and Space
Administration, the Japanese Monbukagakusho, and the Max Planck
Society, and the Higher Education Funding Council for England. The
SDSS Web site is \emph{http://www.sdss.org/}. The SDSS is managed by
the Astrophysical Research Consortium (ARC) for the Participating
Institutions. The Participating Institutions are the American Museum
of Natural History, Astrophysical Institute Potsdam, University of
Basel, University of Cambridge, Case Western Reserve University, The
University of Chicago, Drexel University, Fermilab, the Institute
for Advanced Study, the Japan Participation Group, The Johns Hopkins
University, the Joint Institute for Nuclear Astrophysics, the Kavli
Institute for Particle Astrophysics and Cosmology, the Korean
Scientist Group, the Chinese Academy of Sciences (LAMOST), Los
Alamos National Laboratory, the Max-Planck-Institute for Astronomy
(MPIA), the Max-Planck-Institute for Astrophysics (MPA), New Mexico
State University, Ohio State University, University of Pittsburgh,
University of Portsmouth, Princeton University, the United States
Naval Observatory, and the University of Washington.
\end{acknowledgements}

\bibliographystyle{spbasic}

\begin{thebibliography}{}

\bibitem{} Abazajian KN et al (2009) ApJS 182:543

\bibitem{} Abramson A, Kenney JDP, Crowl HH et al (2011) AJ 141:164

\bibitem{AG09} Aguerri J-A-L, Gonz{\'a}lez-Garc{\'{\i}}a A-C (2009) A\&A 494:891

\bibitem{Ahet12} Ahn CP, Alexandroff R, Allende Prieto C et al (2012) ApJS 203:21

\bibitem{Ahet13} Ahn CP, Alexandroff R, Allende Prieto C et al (2014) ApJS 211:17

\bibitem{Aket03} Akimoto F, Kondou K, Furuzawa A, Tawara Y, Yamashita K (2003) ApJ 596:170

\bibitem{Anet13} Andrade-Santos F, Nulsen PEJ, Kraft RP et al (2013) ApJ 766:107

\bibitem{An08} Andreon S (2008) MNRAS 386:1045

\bibitem{Aret01} Arnaud M, Aghanim N, Gastaud R et al (2001) A\&A 365:L67

\bibitem{Aret12} Arrigoni Battaia F, Gavazzi G, Fumagalli M et al (2012) A\&A 543:A112

\bibitem{Baet01} Bacon R, Copin Y, Monnet G et al (2001) MNRAS 326:23

\bibitem{Ba12} Bah\'{e} Y-M, McCarthy I-G, Crain R-A, Theuns T (2012) MNRAS 424:1179

\bibitem{Baet13} Bah\'{e} YM, McCarthy IG, Balogh ML, Font AS (2013) MNRAS 430:3017

\bibitem{Baiet06} Bai L, Rieke GH, Rieke MJ et al (2006) ApJ 639:827

\bibitem{Baiet09} Bai L, Rieke GH, Rieke MJ, Christlein D, Zabludoff AI (2009) ApJ 693:1840

\bibitem{Baet00} Balogh ML, Navarro JF, Morris SL (2000) ApJ 540:113

\bibitem{Baet04} Balogh ML, Baldry IK, Nichol R et al (2004) ApJl 615:L101

\bibitem{Baet07} Barkhouse WA, Yee HKC, L{\'o}pez-Cruz O (2007) ApJ 671:1471

\bibitem{Baet09} Barkhouse WA, Yee HKC, L{\'o}pez-Cruz O (2009) ApJ 703:2024

\bibitem{Be09} Bekki K (2009) MNRAS 399:2221

\bibitem{Be14} Bekki K (2014) MNRAS 438:444

\bibitem{BC11} Bekki K, Couch WJ (2011) MNRAS 415:1783

\bibitem{Beet14} Benson A, Toloba E, Mayer L, Simon J, Guhathakurta P (2014) ApJ. (in press)

\bibitem{Beet05} Berlind AA, Blanton MR, Hogg DW et al (2005) ApJ 629:625

\bibitem{BC91} Binggeli B, Cameron LM (1991) A\&A 252:27

\bibitem{Biet85} Binggeli B, Sandage A, Tammann GA (1985) AJ 90:1681

\bibitem{BM09} Blanton MR, Moustakas J (2009) ARA\&A 47:159

\bibitem{Blet05a} Blanton MR, Schlegel DJ, Strauss MA et al (2005a) AJ 129:2562

\bibitem{Blet05b} Blanton MR, Lupton RH, Schlegel DJ et al (2005b) ApJ 631:208

\bibitem{Boet94} B{\"o}hringer H, Briel UG, Schwarz RA et al (1994) Nature 368:828

\bibitem{BP00} Boissier S, Prantzos N (2000) MNRAS 312:398

\bibitem{Boi03} Boissier S, Prantzos N, Boselli A, Gavazzi G (2003) MNRAS 346:1215

\bibitem{Boet12} Boissier S, Boselli A, Duc P-A et al (2012) A\&A 545:A142

\bibitem{Bolet10} Bolzonella M, Kova\v{c} K, Pozzetti L et al (2010) A\&A 524:A76

\bibitem{Bonet01} Bonamente M, Lieu R, Nevalainen J, Kaastra JS (2001) ApJ 552:L7

\bibitem{BB10} Book LG, Benson AJ (2010) ApJ 716:810

\bibitem{Bo11} Boselli A (2011) A panchromatic view of galaxies. In: Practical approach book, vol XVI. Wiley, New York 2011. ISBN-10: 3-527-40991-2. ISBN-13: 978-3-527-40991-4

\bibitem{BG02} Boselli A, Gavazzi G (2002) A\&A 386:124

\bibitem{BG06} Boselli A, Gavazzi G (2006) PASP 118:517

\bibitem{Boet01} Boselli A, Gavazzi G, Donas J, Scodeggio M (2001) AJ 121:753

\bibitem{Boet02} Boselli A, Iglesias-P{\'a}ramo J, V{\'{\i}}lchez JM, Gavazzi G (2002) A\&A 386:134

\bibitem{Boet05a} Boselli A, Cortese L, Deharveng JM et al (2005a) ApJ 629:L29

\bibitem{Boet05b} Boselli A, Boissier S, Cortese L et al (2005b) ApJ 623:L13

\bibitem{Boet06} Boselli A, Boissier S, Cortese L et al (2006) ApJ 651:811

\bibitem{Boet08a} Boselli A, Boissier S, Cortese L, Gavazzi G (2008a) ApJ 674:742

\bibitem{Boet08b} Boselli A, Boissier S, Cortese L, Gavazzi G (2008b) A\&A 489:1015

\bibitem{Boet09} Boselli A, Boissier S, Cortese L et al (2009) ApJ 706:1527

\bibitem{Boset10} Boselli A, Eales S, Cortese L et al (2010) PASP 122:261

\bibitem{Boet11} Boselli A, Boissier S, Heinis S et al (2011) A\&A 528:A107

\bibitem{Boet14a} Boselli A, Voyer E, Boissier S et al (2014a) A\&A 570:A69

\bibitem{Boet14b} Boselli A, Cortese L, Boquien M et al (2014b) A\&A 564:A67

\bibitem{Bret09} Braglia FG, Pierini D, Biviano A, Boheringer H (2009) A\&A 500:947

\bibitem{Bret92} Briel UG, Henry JP, Boehringer H (1992) A\&A 259:L31

\bibitem{Bret01} Briel UG, Henry JP, Lumb DH et al (2001) A\&A 365:L60

\bibitem{BL08} Br{\"u}ggen M, De Lucia G (2008) MNRAS 383:1336

\bibitem{Ca13} Cappellari M (2013) ApJ 778:L2

\bibitem{Caet11a} Cappellari M, Emsellem E, Krajnovi{\'c} D et al (2011a) MNRAS 413:813

\bibitem{Caet11b} Cappellari M, Emsellem E, Krajnovi{\'c} D et al (2011b) MNRAS 416:1680

\bibitem{Caet08} Carter D, Goudfrooij P, Mobasher B et al (2008) ApJS 176:424

\bibitem{Caet07} Cassata P, Guzzo L, Franceschini A et al (2007) ApJS 172:270

\bibitem{Caet13} Catinella B, Schiminovich D, Cortese L et al (2013) MNRAS 436:34

\bibitem{Caet90} Cayatte V, van Gorkom JH, Balkowski C, Kotanyi C (1990) AJ 100:604

\bibitem{Ceet14} Cen R, Pop A, Bachall N (2014) Proc Natl Acad Sci USA. (arXiv:1405.0537)
 
\bibitem{Chi09} Chilingarian IV (2009) MNRAS 394:1229

\bibitem{Chet07} Chung A, van Gorkom JH, Kenney JDP, Vollmer B (2007) ApJ 659:L115

\bibitem{Chet09} Chung A, van Gorkom JH, Kenney JDP, Crowl H, Vollmer B (2009) AJ 138:1741

\bibitem{Chet12} Churazov E, Vikhlinin A, Zhuravleva I et al (2012) MNRAS 421:1123

\bibitem{Ciet13} Cibinel A, Carollo CM, Lilly SJ et al (2013) ApJ 777:116

\bibitem{CH09} Cortese L, Hughes TM (2009) MNRAS 400:1225

\bibitem{Coet05} Cortese L, Boselli A, Gavazzi G et al (2005) ApJ 623:L17

\bibitem{Coet06a} Cortese L, Boselli A, Buat V et al (2006a) ApJ 637:242

\bibitem{Coet06b} Cortese L, Gavazzi G, Boselli A et al (2006b) A\&A 453:847

\bibitem{Coet07} Cortese L, Marcillac D, Richard J et al (2007) MNRAS 376:157

\bibitem{Coet08a} Cortese L, Boselli A, Franzetti P et al (2008a) MNRAS 386:1157

\bibitem{Coet08b} Cortese L, Gavazzi G, Boselli A (2008b) MNRAS 390:1282

\bibitem{Coet08c} Cortese L, Minchin RF, Auld RR et al (2008c) MNRAS 383:1519

\bibitem{Coet10} Cortese L, Davies JI, Pohlen M et al (2010) A\&A 518:L49

\bibitem{Coet12a} Cortese L, Ciesla L, Boselli A et al (2012a) A\&A 540:A52

\bibitem{Coet12b} Cortese L, Boissier S, Boselli A et al (2012b) A\&A 544:A101

\bibitem{Cotet04} C{\^o}t\'{e} P, Blakeslee JP, Ferrarese L et al (2004) ApJS 153:223

\bibitem{Cotet06} C{\^o}t\'{e} P, Piatek S, Ferrarese L et al (2006) ApJS 165:57

\bibitem{Cotet07} C{\^o}t\'{e} P, Ferrarese L, Jord{\'a}n A et al (2007) ApJ 671:1456

\bibitem{Craet09} Crawford SM, Bershady MA, Hoessel JG (2009) ApJ 690:1158

\bibitem{Daet04} Davies J, Minchin R, Sabatini S et al (2004) MNRAS 349:922

\bibitem{Daet10} Davies JI, Baes M, Bendo GJ et al (2010) A\&A 518:L48

\bibitem{Daet12} Davies JI, Bianchi S, Cortese L et al (2012) MNRAS 419:3505

\bibitem{Daet13} Davies JI, Bianchi S, Baes M et al (2013) MNRAS 428:834

\bibitem{Fiet11} de Filippis E, Paolillo M, Longo G et al (2011) MNRAS 414:2771

\bibitem{Loet10} de Looze I, Baes M, Zibetti S et al (2010) A\&A 518:L54

\bibitem{Loet13} de Looze I, Baes M, Boselli A, et al. (2013) MNRAS, 2280

\bibitem{Lu11} De Lucia G (2011) Environment and the formation of galaxies: 30 years later. pp 203--210

\bibitem{Luet04} De Lucia G, Poggianti BM, Arag{\'o}n-Salamanca A et al (2004) ApJ 610:L77

\bibitem{Luet07} De Lucia G, Poggianti BM, Arag{\'o}n-Salamanca A et al (2007) MNRAS 374:809

\bibitem{Luet09} De Lucia G, Poggianti BM, Halliday C et al (2009) MNRAS 400:68

\bibitem{Luet12} De Lucia G, Weinmann S, Poggianti BM, Arag{\'o}n-Salamanca A, Zaritsky D (2012) MNRAS 423:1277

\bibitem{Bret11} den Brok M, Peletier RF, Valentijn EA et al (2011) MNRAS 414:3052

\bibitem{Pret13} De Propris R, Phillipps S, Bremer MN (2013) MNRAS 434:3469

\bibitem{Seret07} di Serego Alighieri S, Gavazzi G, Giovanardi C et al (2007) A\&A 474:851

\bibitem{Dr04} Dressler A (2004) Clusters of galaxies: probes of cosmological structure and galaxy. Evolution 206 

\bibitem{Dret13} Dressler A, Oemler A Jr, Poggianti BM et al (2013) ApJ 770:62

\bibitem{Driet04} Drinkwater MJ, Gregg MD, Couch WJ et al (2004) PASA 21:375

\bibitem{DB08} Duc P-A, Bournaud F (2008) ApJ 673:787

\bibitem{Eaet10} Eales S, Dunne L, Clements D et al (2010) PASP 122:499

\bibitem{Eket07} Eckert D, Neronov A, Courvoisier TJ-L, Produit N (2007) A\&A 470:835

\bibitem{Ehet13} Ehlert S, Werner N, Simionescu A et al (2013) MNRAS 430:2401

\bibitem{Faet12} Fabello S, Kauffmann G, Catinella B et al (2012) MNRAS 427:2841

\bibitem{Feet06} Ferrarese L, C{\^o}t\'{e} P, Jord{\'a}n A et al (2006) ApJS 164:334

\bibitem{Feet12} Ferrarese L, C{\^o}t\'{e} P, Cuillandre J-C et al (2012) ApJS 200:4

\bibitem{Fiet04a} Finoguenov A, Briel UG, Henry JP et al (2004a) A\&A 419:47

\bibitem{Fiet04b} Finoguenov A, Henriksen MJ, Briel UG, de Plaa J, Kaastra JS (2004b) ApJ 611:811

\bibitem{Foet08} Font AS, Bower RG, McCarthy IG et al (2008) MNRAS 389:1619

\bibitem{Foet09} Fontanot F, De Lucia G, Monaco P, Somerville RS, Santini P (2009) MNRAS 397:1776

\bibitem{Foet12} Fossati M, Gavazzi G, Boselli A, Fumagalli M (2012) A\&A 544:A128

\bibitem{Foet13} Fossati M, Gavazzi G, Savorgnan G et al (2013) A\&A 553:A91

\bibitem{Fre10} Freeland E, Sengupta C, Croston J-H (2010) MNRAS 409:1518

\bibitem{Fret14} Fritz J, Poggianti BM, Cava A et al (2014) A\&A 566:A32

\bibitem{Fuet09} Fumagalli M, Krumholz MR, Prochaska JX, Gavazzi G, Boselli A (2009) ApJ 697:1811

\bibitem{Fuet11} Fumagalli M, Gavazzi G, Scaramella R, Franzetti P (2011) A\&A 528:A46

\bibitem{Fuet14} Fumagalli M, Fossati M, Hau JKT, Gavazzi G, Bower R, Sun M, Boselli A (2014) (arXiv:1407.7527)

\bibitem{GJ85} Gavazzi G, Jaffe W (1985) ApJ 294:L89

\bibitem{GJ86} Gavazzi G, Jaffe W (1986) ApJ 310:53

\bibitem{Gav95} Gavazzi G, Contursi A, Carrasco L et al (1995) A\&A 304:325

\bibitem{Gaet96} Gavazzi G, Pierini D, Boselli A (1996) A\&A 312:397

\bibitem{Gaet98} Gavazzi G, Catinella B, Carrasco L, Boselli A, Contursi A (1998) AJ 115:1745

\bibitem{Gaet99} Gavazzi G, Boselli A, Scodeggio M, Pierini D, Belsole E (1999) MNRAS 304:595

\bibitem{Gaet01} Gavazzi G, Boselli A, Mayer L et al (2001) ApJ 563:L23

\bibitem{Gaet02} Gavazzi G, Boselli A, Pedotti P, Gallazzi A, Carrasco L (2002) A\&A 386:114

\bibitem{Gaet03a} Gavazzi G, Boselli A, Donati A, Franzetti P, Scodeggio M (2003a) A\&A 400:451

\bibitem{Gaet03b} Gavazzi G, Cortese L, Boselli A et al (2003b) ApJ 597:210

\bibitem{Gaet05} Gavazzi G, Boselli A, van Driel W, O'Neil K (2005) A\&A 429:439

\bibitem{Gaet06a} Gavazzi G, Boselli A, Cortese L et al (2006a) A\&A 446:839

\bibitem{Gaet06b} Gavazzi G, O'Neil K, Boselli A, van Driel W (2006b) A\&A 449:929

\bibitem{Gaet08} Gavazzi G, Giovanelli R, Haynes MP et al (2008) A\&A 482:43

\bibitem{Gaet10} Gavazzi G, Fumagalli M, Cucciati O, Boselli A (2010) A\&A 517:A73

\bibitem{Gaet12} Gavazzi G, Fumagalli M, Galardo V et al (2012) A\&A 545:A16

\bibitem{Gaet13a} Gavazzi G, Fumagalli M, Fossati M et al (2013a) A\&A 553:A89

\bibitem{Gaet13b} Gavazzi G, Savorgnan G, Fossati M et al (2013b) A\&A 553:A90

\bibitem{Geet02} Geha M, Guhathakurta P, van der Marel RP (2002) AJ 124:3073

\bibitem{Geet03} Geha M, Guhathakurta P, van der Marel RP (2003) AJ 126:1794

\bibitem{Geet12} Geha M, Blanton MR, Yan R, Tinker JL (2012) ApJ 757:85

\bibitem{Gilet07} Gil de Paz A, Boissier S, Madore BF et al (2007) ApJS 173:185

\bibitem{GB08} Gilbank DG, Balogh ML (2008) MNRAS 385:L116

\bibitem{Giet08} Gilbank DG, Yee HKC, Ellingson E et al (2008) ApJ 673:742

\bibitem{Giet12} Giodini S, Finoguenov A, Pierini D et al (2012) A\&A 538:A104

\bibitem{Giet05} Giovanelli R, Haynes MP, Kent BR et al (2005) AJ 130:2598

\bibitem{Gioet07} Giovanelli R, Haynes MP, Kent BR et al (2007) AJ 133:2569

\bibitem{Gn03} Gnedin OY (2003) ApJ 582:141

\bibitem{Goet03} G{\'o}mez PL, Nichol RC, Miller CJ et al (2003) ApJ 584:210

\bibitem{GG03} Graham AW, Guzm{\'a}n R (2003) AJ 125:2936

\bibitem{Graet03} Graham AW, Jerjen H, Guzm{\'a}n R (2003) AJ 126:1787

\bibitem{GG72} Gunn JE, Gott JR III (1972) ApJ 176:1

\bibitem{Guet11} Guo Q, White S, Boylan-Kolchin M et al (2011) MNRAS 413:101

\bibitem{Guet13} Guo Q, Cole S, Eke V, Frenk C, Helly J (2013) MNRAS 434:1838

\bibitem{Haet06a} Haines CP, La Barbera F, Mercurio A, Merluzzi P, Busarello G (2006a) ApJ 647:L21

\bibitem{Haet06b} Haines CP, Merluzzi P, Mercurio A et al (2006b) MNRAS 371:55

\bibitem{Haet07} Haines CP, Gargiulo A, La Barbera F et al (2007) MNRAS 381:7

\bibitem{Haet08} Haines CP, Gargiulo A, Merluzzi P (2008) MNRAS 385:1201

\bibitem{Haet11} Haines CP, Busarello G, Merluzzi P et al (2011) MNRAS 412:127

\bibitem{Haet10a} Hammer D, Hornschemeier AE, Mobasher B et al (2010a) ApJS 190:43

\bibitem{Haet10b} Hammer D, Verdoes Kleijn G, Hoyos C et al (2010b) ApJS 191:143

\bibitem{Haet12} Hammer DM, Hornschemeier AE, Salim S et al (2012) ApJ 745:177

\bibitem{Haet09} Hansen SM, Sheldon ES, Wechsler RH, Koester BP (2009) ApJ 699:1333

\bibitem{Haoet11} Hao C-N, Kennicutt RC, Johnson BD et al (2011) ApJ 741:124

\bibitem{HP07} Harsono D, de Propris R (2007) MNRAS 380:1036

\bibitem{HG84} Haynes M-P, Giovanelli R (1984) AJ 89:758

\bibitem{Hayet07} Haynes MP, Giovanelli R, Kent BR (2007) ApJ 665:L19

\bibitem{Hayet11} Haynes MP, Giovanelli R, Martin AM et al (2011) AJ 142:170

\bibitem{Hanet13} Henriques BMB, White SDM, Thomas PA et al (2013) MNRAS 431:3373

\bibitem{He06} Hester JA (2006) ApJ 647:910

\bibitem{Heet10} Hester JA, Seibert M, Neill JD et al (2010) ApJ 716:L14

\bibitem{Hiet14} Hickinbottom S, Simpson C, James P et al (2014) MNRAS 442:1286

\bibitem{Hoet04} Hogg DW, Blanton MR, Brinchmann J et al (2004) ApJ 601:L29

\bibitem{Huet12} Huang S, Haynes MP, Giovanelli R, Brinchmann J (2012) ApJ 756:113

\bibitem{HC09} Hughes TM, Cortese L (2009) MNRAS 396:L41

\bibitem{Jaet13} J{\'a}chym P, Kenney JDP, R\v{z}ui\v{c}ka A et al (2013) A\&A 556:A99

\bibitem{Jacet14} J{\'a}chym P, Combes F, Cortese L, Kenney JDP (2014) ApJ 792:11

\bibitem{JG86} Jaffe W, Gavazzi G (1986) AJ 91:204

\bibitem{Jaet11} Jaff\'{e} YL, Arag{\'o}n-Salamanca A, De Lucia G et al (2011) MNRAS 410:280

\bibitem{JL08} Janz J, Lisker T (2008) ApJ 689:L25

\bibitem{JL09} Janz J, Lisker T (2009) ApJ 696:L102

\bibitem{Jaet12} Janz J, Laurikainen E, Lisker T et al (2012) ApJ 745:L24

\bibitem{Jaet14} Janz J, Laurikainen E, Lisker T et al (2014) ApJ 786:105

\bibitem{Jeet08} Jeltema TE, Binder B, Mulchaey JS (2008) ApJ 679:1162

\bibitem{Jeet00} Jerjen H, Kalnajs A, Binggeli B (2000) A\&A 358:845

\bibitem{Joet07a} Jord{\'a}n A, Blakeslee JP, C{\^o}t\'{e} P et al (2007a) ApJS 169:213

\bibitem{Joet07b} Jord{\'a}n A, McLaughlin DE, C{\^o}t\'{e} P et al (2007b) ApJS 171:101

\bibitem{KB08} Kang X, van den Bosch FC (2008) ApJ 676:L101

\bibitem{Kaet05} Kantharia NG, Ananthakrishnan S, Nityananda R, Hota A (2005) A\&A 435:483

\bibitem{Kaet08} Kapferer W, Kronberger T, Ferrari C, Riser T, Schindler S (2008) MNRAS 389:1405

\bibitem{Kaet09} Kapferer W, Sluka C, Schindler S, Ferrari C, Ziegler B (2009) A\&A 499:87

\bibitem{Kaet04} Kauffmann G, White SDM, Heckman TM et al (2004) MNRAS 353:713

\bibitem{KM08} Kawata D, Mulchaey JS (2008) ApJ 672:L103

\bibitem{Kenet08} Kenney JDP, Tal T, Crowl HH, Feldmeier J, Jacoby GH (2008) ApJ 687:L69

\bibitem{Keet14} Kenney JDP, Geha M, J{\'a}chym P et al (2014) ApJ 780:119

\bibitem{Kilet09} Kilborn VA, Forbes DA, Barnes DG et al (2009) MNRAS 400:1962

\bibitem{Ke98} Kennicutt RC Jr (1998) ARA\&A 36:189

\bibitem{Ke10} Kent BR (2010) ApJ 725:2333

\bibitem{Keet07} Kent BR, Giovanelli R, Haynes MP et al (2007) ApJ 665:L15

\bibitem{Kentet08} Kent BR, Giovanelli R, Haynes MP et al (2008) AJ 136:713

\bibitem{Keet09} Kent BR, Spekkens K, Giovanelli R et al (2009) ApJ 691:1595

\bibitem{Kiet10} Kim S, Rey S-C, Lisker T, Sohn ST (2010) ApJ 721:L72

\bibitem{Kimet09} Kimm T, Somerville RS, Yi SK et al (2009) MNRAS 394:1131

\bibitem{Koet04} Kodama T, Balogh ML, Smail I, Bower RG, Nakata F (2004) MNRAS 354:1103

\bibitem{Kolet09} Koleva M, de Rijcke S, Prugniel P, Zeilinger WW, Michielsen D (2009) MNRAS 396:2133

\bibitem{Koet11} Koleva M, Prugniel P, de Rijcke S, Zeilinger WW (2011) MNRAS 417:1643

\bibitem{Koet13} Koleva M, Bouchard A, Prugniel P, De Rijcke S, Vauglin I (2013) MNRAS 428:2949

\bibitem{Koet01} Koopmann RA, Kenney JDP, Young J (2001) ApJS 135:125

\bibitem{Koet06} Koopmann RA, Haynes MP, Catinella B (2006) AJ 131:716

\bibitem{KB12} Kormendy J, Bender R (2012) ApJS 198:2

\bibitem{Koret09} Kormendy J, Fisher DB, Cornell ME, Bender R (2009) ApJS 182:216

\bibitem{Kret08a} Kronberger T, Kapferer W, Unterguggenberger S, Schindler S, Ziegler BL (2008a) A\&A 483:783

\bibitem{Kret08b} Kronberger T, Kapferer W, Ferrari C, Unterguggenberger S, Schindler S (2008b) A\&A 481:337

\bibitem{Laet05} Lanzoni B, Guiderdoni B, Mamon GA, Devriendt J, Hatton S (2005) MNRAS 361:369

\bibitem{Laet80} Larson RB, Tinsley BM, Caldwell CN (1980) ApJ 237:692

\bibitem{Leet12} Lemaux BC, Gal RR, Lubin LM et al (2012) ApJ 745:106

\bibitem{Leet02} Lewis I, Balogh M, De Propris R et al (2002) MNRAS 334:673

\bibitem{Liet12} Lieder S, Lisker T, Hilker M, Misgeld I, Durrell P (2012) A\&A 538:A69

\bibitem{Li09} Lisker T (2009) Astron Nachr 330:1043

\bibitem{LH08} Lisker T, Han Z (2008) ApJ 680:1042

\bibitem{LF09} Lisker T, Fuchs B (2009) A\&A 501:429

\bibitem{Liet06a} Lisker T, Grebel EK, Binggeli B (2006a) AJ 132:497

\bibitem{Liet06b} Lisker T, Glatt K, Westera P, Grebel EK (2006b) AJ 132:2432

\bibitem{Liet07} Lisker T, Grebel EK, Binggeli B, Glatt K (2007) ApJ 660:1186

\bibitem{Liet08} Lisker T, Grebel EK, Binggeli B (2008) AJ 135:380

\bibitem{Liet09} Lisker T, Janz J, Hensler G et al (2009) ApJ 706:L124

\bibitem{Liet13} Lisker T, Weinmann SM, Janz J, Meyer HT (2013) MNRAS 432:1162

\bibitem{Maret12} Marinova I, Jogee S, Weinzirl T et al (2012) ApJ 746:136

\bibitem{Maret05} Martin DC, Fanson J, Schiminovich D et al (2005) ApJ 619:L1

\bibitem{Maret07} Martin DC, Wyder TK, Schiminovich D et al (2007) ApJS 173:342

\bibitem{Maret10} Martin AM, Papastergis E, Giovanelli R et al (2010) ApJ 723:1359

\bibitem{Maset05} Mastropietro C, Moore B, Mayer L et al (2005) MNRAS 364:607

\bibitem{Mayet01a} Mayer L, Governato F, Colpi M et al (2001a) ApJ 559:754

\bibitem{Mayet01b} Mayer L, Governato F, Colpi M et al (2001b) ApJ 547:L123

\bibitem{Mayet03} Mayer L, Mastropietro C, Wadsley J, Stadel J, Moore B (2006) MNRAS 369:1021

\bibitem{Mcet08} McCarthy IG, Frenk CS, Font AS et al (2008) MNRAS 383:593

\bibitem{Mcet07} McConnachie AW, Venn KA, Irwin MJ, Young LM, Geehan JJ (2007) ApJ 671:L33

\bibitem{Mcet11} McDonald M, Courteau S, Tully RB, Roediger J (2011) MNRAS 414:2055

\bibitem{Mcet09} McGee SL, Balogh ML, Bower RG, Font AS, McCarthy IG (2009) MNRAS 400:937

\bibitem{Meet07} Mei S, Blakeslee JP, C{\^o}t\'{e} P et al (2007) ApJ 655:144

\bibitem{Meret06} Mercurio A, Merluzzi P, Haines CP et al (2006) MNRAS 368:109

\bibitem{Meret10} Merluzzi P, Mercurio A, Haines CP et al (2010) MNRAS 402:753

\bibitem{Meret13} Merluzzi P, Busarello G, Dopita MA et al (2013) MNRAS 429:1747

\bibitem{Micet08} Michielsen D, Boselli A, Conselice CJ et al (2008) MNRAS 385:1374

\bibitem{MH11} Misgeld I, Hilker M (2011) MNRAS 414:3699

\bibitem{Miet08} Misgeld I, Mieske S, Hilker M (2008) A\&A 486:697

\bibitem{Miet09} Misgeld I, Hilker M, Mieske S (2009) A\&A 496:683

\bibitem{Moet13} Mok A, Balogh ML, McGee SL et al (2013) MNRAS 431:1090

\bibitem{Mooet96} Moore B, Katz N, Lake G, Dressler A, Oemler A (1996) Nature 379:613

\bibitem{Mu00} Mulchaey JS (2000) ARA\&A 38:289

\bibitem{MZ98} Mulchaey JS, Zabludoff AI (1998) ApJ 496:73

\bibitem{Muet03} Mulchaey JS, Davis DS, Mushotzky RF, Burstein D (2003) ApJS 145:39

\bibitem{Muet07} Murakami H, Baba H, Barthel P et al (2007) PASJ 59:369

\bibitem{Muet11} Murakami H, Komiyama M, Matsushita K et al (2011) PASJ 63:963

\bibitem{Neet01} Neumann DM, Arnaud M, Gastaud R et al (2001) A\&A 365:L74

\bibitem{Oc99} O'Connell RW (1999) ARA\&A 37:603

\bibitem{ON03} Okamoto T, Nagashima M (2003) ApJ 587:500

\bibitem{OG05} Oosterloo T, van Gorkom J (2005) A\&A 437:L19

\bibitem{OP04} Osmond JPF, Ponman TJ (2004) MNRAS 350:1511

\bibitem{Paet10a} Paudel S, Lisker T, Kuntschner H, Grebel EK, Glatt K (2010a) MNRAS 405:800

\bibitem{Paet10b} Paudel S, Lisker T, Janz J (2010b) ApJ 724:L64

\bibitem{Paet11} Paudel S, Lisker T, Kuntschner H (2011) MNRAS 413:1764

\bibitem{Paet13} Paudel S, Duc P-A, C{\^o}t\'{e} P et al (2013) ApJ 767:133

\bibitem{Peet02} Pedraz S, Gorgas J, Cardiel N, S{\'a}nchez-Bl{\'a}zquez P, Guzm{\'a}n R (2002) MNRAS 332:L59

\bibitem{Peet08} Peng EW, Jord{\'a}n A, C{\^o}t\'{e} P et al (2008) ApJ 681:197

\bibitem{Peet11} Penny SJ, Conselice CJ, de Rijcke S et al (2011) MNRAS 410:1076

\bibitem{Poet04} Poggianti BM, Bridges TJ, Komiyama Y et al (2004) ApJ 601:197

\bibitem{Poet09} Poggianti BM, Arag{\'o}n-Salamanca A, Zaritsky D et al (2009) ApJ 693:112

\bibitem{Pog10} Poggianti B-M, De Lucia G, Varela J et al (2010) MNRAS 405:995

\bibitem{Poet13} Poggianti BM, Calvi R, Bindoni D et al (2013) ApJ 762:77

\bibitem{Poet05} Popesso P, B{\"o}hringer H, Romaniello M, Voges W (2005) A\&A 433:415

\bibitem{Poet06} Popesso P, Biviano A, B{\"o}hringer H, Romaniello M (2006) A\&A 445:29

\bibitem{Poet07} Popesso P, Biviano A, Romaniello M, B{\"o}hringer H (2007) A\&A 461:411

\bibitem{Raet08} Rasmussen J, Ponman TJ, Verdes-Montenegro L, Yun MS, Borthakur S (2008) MNRAS 388:1245

\bibitem{Raet12a} Rasmussen J, Mulchaey JS, Bai L et al (2012a) ApJ 757:122

\bibitem{Raet12b} Rasmussen J, Bai X-N, Mulchaey JS et al (2012b) ApJ 747:31

\bibitem{Reet06} Renaud M, B\'{e}langer G, Paul J, Lebrun F, Terrier R (2006) A\&A 453:L5

\bibitem{RG08} Rines K, Geller MJ (2008) AJ 135:1837

\bibitem{Riet05} Rines K, Geller MJ, Kurtz MJ, Diaferio A (2005) AJ 130:1482

\bibitem{Roet10} Robotham A, Phillipps S, de Propris R (2010) MNRAS 403:1812

\bibitem{Ro09} Roediger E (2009) Astron Nachr 330:888

\bibitem{RH05} Roediger E, Hensler G (2005) A\&A 433:875

\bibitem{RB06} Roediger E, Br{\"u}ggen M (2006) MNRAS 369:567

\bibitem{RB07} Roediger E, Br{\"u}ggen M (2007) MNRAS 380:1399

\bibitem{Ruet09} Rudnick G, von der Linden A, Pell{\'o} R et al (2009) ApJ 700:1559

\bibitem{Ruet12} Rudnick GH, Tran K-V, Papovich C, Momcheva I, Willmer C (2012) ApJ 755:14

\bibitem{Ruet14} Ruszkowski M, Br{\"u}ggen M, Lee D, Shin M-S (2014) ApJ 784:75

\bibitem{Ryet13} Ry\'{s} A, Falc{\'o}n-Barroso J, van de Ven G (2013) MNRAS 428:2980

\bibitem{Saet02} Sakai S, Kennicutt RC Jr, van der Hulst JM, Moss C (2002) ApJ 578:842

\bibitem{Saet12} S{\'a}nchez SF, Kennicutt RC, Gil de Paz A et al (2012) A\&A 538:A8

\bibitem{SA12} S{\'a}nchez-Janssen R, Aguerri JAL (2012) MNRAS 424:2614

\bibitem{Saet85} Sandage A, Binggeli B, Tammann GA (1985) AJ 90:1759

\bibitem{Sa86} Sarazin CL (1986) Rev Mod Phys 58:1

\bibitem{Scet09} Schawinski K, Virani S, Simmons B et al (2009) ApJ 692:L19

\bibitem{Scet14} Schawinski K, Urry CM, Simmons BD et al (2014) MNRAS 440:889

\bibitem{Scet12} Scott TC, Cortese L, Brinks E et al (2012) MNRAS 419:L19

\bibitem{Scet13} Scott TC, Usero A, Brinks E et al (2013) MNRAS 429:221

\bibitem{Senet07} Sengupta C, Balasubramanyam R, Dwarakanath KS (2007) MNRAS 378:137

\bibitem{Shet01} Shibata R, Matsushita K, Yamasaki NY et al (2001) ApJ 549:228

\bibitem{Smet06} Smith RJ, Hudson MJ, Lucey JR, Nelan JE, Wegner GA (2006) MNRAS 369:1419

\bibitem{Smet07} Smith RJ, Lucey JR, Hudson MJ (2007) MNRAS 381:1035

\bibitem{Smet08} Smith RJ, Marzke RO, Hornschemeier AE et al (2008) MNRAS 386:L96

\bibitem{Smet09} Smith RJ, Lucey JR, Hudson MJ et al (2009) MNRAS 392:1265

\bibitem{Smet10} Smith RJ, Lucey JR, Hammer D et al (2010) MNRAS 408:1417

\bibitem{Smet12a} Smith RJ, Lucey JR, Price J, Hudson MJ, Phillipps S (2012a) MNRAS 419:3167

\bibitem{Smet12b} Smith RJ, Lucey JR, Carter D (2012b) MNRAS 421:2982

\bibitem{Smet12c} Smith R, Fellhauer M, Assmann P (2012c) MNRAS 420:1990

\bibitem{Smiet13} Smith R, Duc PA, Candlish GN et al (2013) MNRAS 2255

\bibitem{SCet08} Smith Castelli AV, Bassino LP, Richtler T et al (2008) MNRAS 386:2311

\bibitem{Scaet13} Smith Castelli AV, Gonz{\'a}lez NM, Faifer FR, Forte JC (2013) ApJ 772:68

\bibitem{Soet01} Solanes JM, Manrique A, Garc{\'{\i}}a-G{\'o}mez C et al (2001) ApJ 548:97

\bibitem{Spet05} Springel V, White SDM, Jenkins A et al (2005) Nature 435:629

\bibitem{Stet07} Stott JP, Smail I, Edge AC et al (2007) ApJ 661:95

\bibitem{Stet09} Stott JP, Pimbblet KA, Edge AC, Smith GP, Wardlow JL (2009) MNRAS 394:2098

\bibitem{Stet02} Strauss MA, Weinberg DH, Lupton RH et al (2002) AJ 124:1810

\bibitem{Suet07} Sun M, Donahue M, Voit GM (2007) ApJ 671:190

\bibitem{Suet09} Sun M, Voit GM, Donahue M et al (2009) ApJ 693:1142

\bibitem{Suet10} Sun M, Donahue M, Roediger E et al (2010) ApJ 708:946

\bibitem{Taet14} Taranu DS, Hudson MJ, Balogh ML et al (2014) MNRAS 440:1934

\bibitem{Taet12} Taylor R, Davies JI, Auld R, Minchin RF (2012) MNRAS 423:787

\bibitem{Taet13} Taylor R, Davies JI, Auld R, Minchin RF, Smith R (2013) MNRAS 428:459

\bibitem{Thet10} Thomas D, Maraston C, Schawinski K, Sarzi M, Silk J (2010) MNRAS 404:1775

\bibitem{Toet09} Toloba E, Boselli A, Gorgas J et al (2009) ApJ 707:L17

\bibitem{Toet11} Toloba E, Boselli A, Cenarro AJ et al (2011) A\&A 526:A114

\bibitem{Toet12} Toloba E, Boselli A, Peletier RF et al (2012) A\&A 548:A78

\bibitem{Toet14a} Toloba E, Guhathakurta P, Peletier R et al (2014a) ApJS (in press)

\bibitem{Toet14b} Toloba E, Guhathakurta P, Boselli A et al (2014b) ApJ (in press)

\bibitem{Toet14c} Toloba E, Guhathakurta P, van de Ven G et al (2014c) ApJ 783:120

\bibitem{TB12} Tonnesen S, Bryan G-L (2012) MNRAS 422:1609

\bibitem{Toet07} Tonnesen S, Bryan GL, van Gorkom JH (2007) ApJ 671:1434

\bibitem{TB09} Tonnesen S, Bryan GL (2009) ApJ 694:789

\bibitem{Uret11} Urban O, Werner N, Simionescu A, Allen SW, B{\"o}hringer H (2011) MNRAS 414:2101

\bibitem{Boet08} van den Bosch FC, Aquino D, Yang X et al (2008) MNRAS 387:79

\bibitem{Zeet04} van Zee L, Skillman ED, Haynes MP (2004) AJ 128:121

\bibitem{Veet07} Verdes-Montenegro L, Yun MS, Borthakur S, Rasmussen J, Ponman T (2007) New Astron Rev 51:87

\bibitem{Veet08} Verdugo M, Ziegler BL, Gerken B (2008) A\&A 486:9

\bibitem{Viet01} Vikhlinin A, Markevitch M, Forman W, Jones C (2001) ApJ 555:L87

\bibitem{VR13} Vijayaraghavan R, Ricker PM (2013) MNRAS 435:2713

\bibitem{Voet01} Vollmer B, Cayatte V, Balkowski C, Duschl WJ (2001) ApJ 561:708

\bibitem{Voet14} Voyer E, Boselli A, Boissier S et al (2014) A\&a (in press)

\bibitem{Vuet13} Vulcani B, Poggianti BM, Oemler A et al (2013) A\&A 550:A58

\bibitem{Waet12} Wang L, Weinmann SM, Neistein E (2012) MNRAS 421:3450

\bibitem{Wetet13} Wetzel AR, Tinker JL, Conroy C, van den Bosch FC (2013) MNRAS 432:336

\bibitem{Weiet11} Weinmann SM, Lisker T, Guo Q, Meyer HT, Janz J (2011) MNRAS 416:1197

\bibitem{Wezet12} We\.{z}gowiec M, Bomans DJ, Ehle M et al (2012) A\&A 544:A99

\bibitem{Wilet10} Wilman DJ, Zibetti S, Budav{\'a}ri T (2010) MNRAS 406:1701

\bibitem{Wriet10} Wright EL, Eisenhardt PRM, Mainzer AK et al (2010) AJ 140:1868

\bibitem{Wydet05} Wyder TK, Treyer MA, Milliard B et al (2005) ApJ 619:L15

\bibitem{Yaet07} Yagi M, Komiyama Y, Yoshida M et al (2007) ApJ 660:1209

\bibitem{Yaet10} Yagi M, Yoshida M, Komiyama Y et al (2010) AJ 140:1814

\bibitem{Yaet13} Yagi M, Gu L, Fujita Y et al (2013) ApJ 778:91

\bibitem{YF11} Yamagami T, Fujita Y (2011) PASJ 63:1165

\bibitem{Yoet00} York DG, Adelman J, Anderson JE Jr et al (2000) AJ 120:1579

\bibitem{Yoet04} Yoshida M, Ohyama Y, Iye M et al (2004) AJ 127:90

\bibitem{Yoet08} Yoshida M, Yagi M, Komiyama Y et al (2008) ApJ 688:918

\bibitem{Yoet12} Yoshida M, Yagi M, Komiyama Y et al (2012) ApJ 749:43

\bibitem{ZM11} Zandivarez A, Mart{\'{\i}}nez HJ (2011) MNRAS 415:2553

\bibitem{Zaet06} Zandivarez A, Mart{\'{\i}}nez HJ, Merch{\'a}n ME (2006) ApJ 650:137

\bibitem{Zaet13} Zhang B, Sun M, Ji L et al (2013) ApJ 777:122

\end{thebibliography}

\end{document}